\documentclass[10pt]{article}
\pdfoutput=1
\usepackage{amssymb}
\usepackage{amsmath}
\usepackage{graphicx}
\usepackage{xcolor}
\usepackage{ifthen}
\usepackage{a4wide}
\usepackage{float}
\usepackage{wrapfig}
\usepackage[titles]{tocloft}
\usepackage[english]{babel}

\newcommand{\cenlin}{\centerline}

\newcommand{\ovline}{\overline}

\newcommand{\larr}{\leftarrow}
\newcommand{\rarr}{\rightarrow}

\newcommand{\lesq}{\leqslant}
\newcommand{\greq}{\geqslant}

\newcommand{\vphi}{\varphi}

\newcommand{\forceparindent}{\hskip 1.5em}

\newcommand{\vgap}{\vskip 3mm}

\newcommand{\fzNOT}{\mathop{\mathtt{NOT}}}
\newcommand{\fzAND}{\mathbin{\mathtt{AND}}}
\newcommand{\fzOR}{\mathbin{\mathtt{OR}}}
\newcommand{\fzXOR}{\mathbin{\mathtt{XOR}}}
\newcommand{\fzADD}{\mathbin{\mathtt{ADD}}}
\newcommand{\fzSHL}{\mathbin{\mathtt{SHL}}}

\newcommand{\fzROR}{\mathbin{\mathtt{ROR}}}
\newcommand{\fzLEBE}{\mathbin{\mathtt{LEBE}}}
\newcommand{\atanxy}{\mathop{\mathtt{ATANXY}}}

\newcounter{Formulanum}
\newcounter{Resultitemnum}
\newcommand{\resultitem}{%
\refstepcounter{Resultitemnum}%
\hrule

{\ }

\begin{wrapfigure}{r}{7mm}
\vspace{-4mm}
\fbox{\textbf{\arabic{Resultitemnum}}}
\end{wrapfigure}}

\newcommand{\traintestcharts}[4]{
{\ }

\cenlin{\footnotesize{\begin{tabular}{ccc}
\includegraphics[height=2.7cm]{#1.pdf}
&
\includegraphics[height=2.7cm]{#2.pdf}
&
\includegraphics[height=2.7cm]{#3.pdf}
\\
Loss ($L_1$) & Misses for Predicted & Misses for Random
\end{tabular}}}
\ifthenelse{\equal{#4}{}}{}{
\cenlin{\footnotesize{\begin{tabular}{c}
\includegraphics[height=2.5cm]{#4.pdf}
\\
Miss frequencies per bit
\end{tabular}}}}

{\ }
}

\newcommand{\biauth}[1]{\textsc{#1}} % Author(s)
\newcommand{\bitita}[1]{\textit{#1}} % Article title
\newcommand{\bititb}[1]{\textrm{#1}} % Book title
\newcommand{\bititu}[1]{\texttt{#1}} % URL
\newcommand{\bijour}[1]{\textrm{#1}} % Journal abbrev.
\newcommand{\bicoor}[3]{%
\ifthenelse{\equal{#1}{}}{}{\textbf{#1}}%
\ifthenelse{\equal{#2}{}}{}{:#2}%
\ifthenelse{\equal{#3}{}}{}{, #3}} % Volume[(Number)]:Pages, Year - "Coordinates"
\newcommand{\bipubl}[1]{\textrm{#1}} % Publisher
\newcommand{\bidoid}[1]{\texttt{doi:#1}} % DOId
\newcommand{\binote}[1]{(\textrm{#1})} % Note

\begin{document}

\title{Using fuzzy bits and neural networks to partially invert few rounds of some cryptographic hash functions}

\author{Sergij V. Goncharov\thanks{Faculty of Mechanics and Mathematics, Oles Honchar Dnipro National University, 72 Gagarin Avenue, 49010 Dnipro, Ukraine.
\textit{E-mail: goncharov@mmf.dnulive.dp.ua}}}

\date{January 2019}

\maketitle

\begin{abstract}
We consider fuzzy, or continuous, bits, which take values in $[0;1]$ and $(-1;1]$ instead of $\{0;1\}$,
and operations on them (NOT, XOR etc.) and on their sequences (ADD),
to obtain the generalization of cryptographic hash functions, CHFs, for the messages consisting of fuzzy bits,
so that CHFs become smooth and non-constant functions of each bit of the message.
We then train the neural networks to predict the message that has a given hash,
where the loss function for the hash of predicted message and given true hash is backpropagatable.
The results of the trainings for the standard CHFs --- MD5, SHA1, SHA2-256, and SHA3/Keccak ---
with small number of (optionally weakened) rounds are presented and compared.
\end{abstract}

\cenlin{\small \textit{Keywords:} bit, neural network, hash, fuzzy, CHF, round, inverse, preimage, training, approximation}

\vgap

% \tableofcontents
\makeatletter
\@starttoc{toc}% Omit "Contents"
\makeatother

\section*{Introduction}
\addcontentsline{toc}{section}{Introduction}

\forceparindent
{\color{white} \tiny Mermaids are boring to fishes from below and to humans from above...}

One of the main requirements for a cryptographic hash function (CHF) is the \textit{preimage resistance} (\cite[5.1]{katz2015}):
for the known message/bit string $M = (m_1; ...; m_r)$ it is ``easy'' to calculate the hash $H = h(M) = (h_1; ...; h_k)$,
but for the known $H$ and the unknown $M$ it is ``hard'' to obtain any $M' = (m'_1; ...; m'_s)$ such that $h(M') = h(M)$.
In other words, the calculation of the inverse function $h^{-1}(H)$
(that is, to find any element from the preimage of $H$ with respect to $h$) requires, in average, ``too much'' computation,
even though $h$ is not injective for $s > k$.
The worst case is when only the exhaustive search guarantees the obtaining of $M'$,
which takes time being proportional on the average to $2^k$ (if not applying quantum computers, \cite{bernstein2009}).

In due course many methods, or attacks, were proposed and developed to reduce the average time needed to invert the
specific CHF: see \cite{sasaki2009} for MD5, \cite{canniere2008} for SHA1, \cite{aoki2009} for SHA2,
\cite{morawiecki2013} for SHA3/Keccak.
These attacks usually come after ``easier'' \textit{collision} attacks
(see \cite{mendel2013}, \cite{song2017}, \cite{stevens2017}, \cite{wang2005}),
which require in average $2^{k/2}$ attempts in the worst case.
From the practical side, for instance, there are widely used \textit{SAT(isfiability) solvers}
(\cite{cook1997}, \cite{de2007}, \cite{morawiecki2013}, \cite{nejati2017}).
These approaches work with discrete, binary as a rule, variables,
they search for the combinations of these variables that satisfy certain equations using various techniques to speed up the search.

{\ }

(Artificial) neural networks (NNs) on their own are generally considered to be ineffective,
--- or, to put it stronger, absolutely useless, ---
for CHF inversion due to discreteness of the arguments and the values of CHFs, as well as of inverse CHFs.
See e.g. \cite{crypto.stackexch.q14509}, \cite{datasc.stackexch.q16639}, \cite{sec.stackexch.q135211}. Here's why:

Let the input of NN be $H = h(M)$, and we try to train NN to predict $M'$ such that $h(M') = H$
($M' = M$ isn't necessary). For the predicted message $M'$ we calculate $H'= h(M')$ and compare it with $H$,
obtaining the value of the loss function, which generally decreases when $H'$ ``becomes closer'' to $H$, and increases otherwise.
For the learning/optimization algorithms, based on gradient descent (GD) and back propagation (BP)
(\cite[4.3, 5.9, 6.5]{goodfellow2017}, \cite{rumelhart1986}),
to work, the loss function must satisfy certain properties; in particular, it has to be smooth enough and non-constant
(though some special, less restrictive types of the optimization and GD have been developed).

Clearly, if we use CHFs defined as they are for bit strings, the outputs of (the last layer of)
NN should be discretized to 0 or 1 before we can calculate the CHF of predicted message;
here we cannot treat the outputs as probabilities like we do for classification or pattern recognition problems.
Regardless of how we perform the discretization, the loss function will be piecewise-constant and non-continuous,
because both hashes are discrete then.
Even if we replace it with piecewise-linear one, or something smoother, the dependency of the loss on the weights
will be greatly reduced. However many samples we take and average on them,
the loss remains discrete, hence the small change of any weight either leaves it the same, or makes it leap.

{\ }

Now, if we extend the definition of CHF to the bits from the whole $[0;1]$, so that it becomes smooth, non-constant,
and more dependable on each bit of the message, then these obstacles to GD and BP will be removed.
Of course, other issues appear, --- local minima; rounding errors of the floating point arithmetic;
NN may learn to predict ``essentially fuzzy'' messages
for given usual hashes, such messages when rounded to usual bits change hashes,
--- but at least NN will be able to learn.
And if the learning were ``successful'', then trained NN, perhaps with assistance of other methods,
would invert the CHF without trying a lot of combinations of binary variables.

Such bits are the well-known generalization of usual ones,
they are called \textit{fuzzy} and are the basis of \textit{fuzzy logic}
(see e.g. \cite{hajek1998}, \cite[III]{kaufmann1977}, \cite{zadeh1965}).
However, we do not need to keep all ``logical'' properties, because they may ``stick'' us to the extension
that is not suitable for our task.
Moreover, we can extend the bit domain even further, from $[0;1]$ to $(-1;1]$.

Although we do not rely on logical or probabilistic interpretation of the bit value, cf. \cite[XII]{dadhich2014}.
Perhaps here it would be better to call such bits \textit{continuous} (see \cite[3.2]{coutinho2018}).

{\ }

The task for NN then is rather an approximation than a pattern recognition. To be precise,
we can train NN in two different ways, to be either

a) \textit{general inverter}: at the training NN is presented with hashes of many random messages.
After the training NN should be able to invert arbitrary hash, not only those it saw at the training (at least partially).
Here NN becomes the approximator of the whole inverse CHF; or

b) \textit{single inverter}: at the training phase we repeatedly input the one and the same hash into NN,
which attempts to predict the message whose hash is as ``close'' to the input as possible.
When the training is finished, NN should invert only this hash (again, partially).
Here NN is the optimizer of the ``distance'' between the hash of its output and the fixed input hash.
In principle, such NN doesn't have to be a ``network'' and may consist of the output layer only
(however, we found that letting such NN know the input hash seems to improve the accuracy of inversion).

(a) is more universal, (b) is simpler.

{\ }

\textbf{Disclaimer.}
Because of the small, if any, accuracy of such inversion for a substantial number of CHFs' rounds and message length,
as shown in the Results, we do not consider it to be an ``attack''
or to have any practical cryptanalytic value (e.g. it does not at all help to mine cryptocurrencies), at least on its own.
There may exist theorems, unknown to us, proving that under certain conditions NNs cannot invert better/faster
than discrete methods.
Rather it is the analysis of inverse CHFs from the approximation by composition of elementary functions point of view,
or perhaps a mere ``proof-of-concept'' to make the aforementioned \textsl{absolute} uselessness doubtful.

{\ }

We suppose that similar investigations were done before, in multiple places,
probably some of their outcomes were not published due to various reasons.
We were able to find the researches on resemblant themes,
gathered under Neural Crypto-graphy/-analysis name,
the closest ones being the \textit{CryptoNet} from \cite[3.2]{coutinho2018} and the descriptions from \cite[XII]{dadhich2014}
(presumably, there are others we'll be glad to know about).
Many works are dedicated to the construction of CHF on the basis of NN,
when the NN input is a message, and the output is a hash,
among the others see \cite[7]{guyeux2016}, \cite{lian2007}, \cite{turcanik2016};
such CHFs are cryptanalyzed e.g. in \cite{qin2015};
in \cite{alani2012} NNs assist in known-plaintext attack on DES;
in \cite{kanter2002} the key exchange protocol is proposed, and analyzed in \cite{klimov2002}, based on synchronization of 2 NNs;
in \cite{xie2014} NNs are used in conjunction with homomorphic encryption.
See also \cite{laskari2007}, \cite{rivest1993}, \cite{crypto.stackexch.q2715}.

\section{Bits, hashes, and neural networks}

\subsection{Fuzzy bits and operations}

\forceparindent
While the definition of the fuzzy, or continuous, bit here is standard, as in \cite[1.1]{hajek1998}, \cite[III.31]{kaufmann1977},

{\ }

\hfil Usual, binary bit (binbit): $b = 0, 1$ \hfil Fuzzy bit (fuzbit): $b \in [0;1]$ \hfil

{\ }

\noindent
the ops on such bits can be generalized in many ways, accordingly to the properties we try to keep.
Even when the goal is to build a ``natural fuzzy logic'' (which we do not aim at),
it can be done differently; see \cite[2]{hajek1998}.
We describe some of these generalizations below and select one for each op.
Fuzzy ops are denoted in the same way as the usual ones.

There are 4 basic ops: $\fzNOT$, $\fzAND$, $\fzOR$, $\fzXOR$. We define each of them for a pair of bits,
then extend to the bit strings of equal length bitwisely: if the op on bits $a$ and $b$ is $a * b$, then
$(a_1; ...; a_n) * (b_1; ...; b_n) := (a_1 * b_1; ...; a_n * b_n)$.

{\ }

\textbf{1)} Consistency: for binbits the result of the fuzzy op will be equal to that of the usual op
(then any composition of the fuzzy ops has this property too):

{\ }

\cenlin{
\begin{tabular}{c|c}
$a$ & $\fzNOT(a)$\\
\hline
0 & 1\\
1 & 0
\end{tabular}
\hfil
\begin{tabular}{c|c|c}
$a$ & $b$ & $a \fzAND b$\\
\hline
0 & 0 & 0\\
0 & 1 & 0\\
1 & 0 & 0\\
1 & 1 & 1
\end{tabular}
\hfil
\begin{tabular}{c|c|c}
$a$ & $b$ & $a \fzOR b$\\
\hline
0 & 0 & 0\\
0 & 1 & 1\\
1 & 0 & 1\\
1 & 1 & 1
\end{tabular}
\hfil
\begin{tabular}{c|c|c}
$a$ & $b$ & $a \fzXOR b$\\
\hline
0 & 0 & 0\\
0 & 1 & 1\\
1 & 0 & 1\\
1 & 1 & 0
\end{tabular}
}

{\ }

\textbf{2)} Domain: for fuzbits the result will stay in $[0;1]$.

\textbf{3)} Feedback: the ops must be smooth enough w.r.t. the operands, and they must not block the back propagation.

{\ }

Some, but not all, algebraic and ``logical'' properties will remain (see below).
We cannot say for sure if keeping or losing certain property of this kind,
or a subset of properties, makes NN training easier or harder.
Also, for binbits all these properties are kept due to 1st requirement.

{\ }

\textbf{NOT.} $\fzNOT(a) := 1 - a$

\noindent
is less ambiguous than other ops, though alternatives exist:
$\fzNOT_1(a) = (1-a)^2$, $\fzNOT_2(a) = \cos (\frac{\pi}{2} a)$.

{\ }

\textbf{AND.} $a \fzAND b := ab$

\noindent
called \textit{product t-norm}.
Alternatives are $a \fzAND_1 b = \min \{ a, b\}$ (the ``traditional'' one, as e.g. in \cite{kaufmann1977}),
$a \fzAND_2 b = (ab)^p$ etc. See also \cite[Ex. 2.1.2]{hajek1998}.
It may be non-commutative: $a \fzAND_3 b = a^2 b$.

{\ }

\textbf{OR.} $a \fzOR b := a + b - ab$.

It's associative, and De Morgan's laws hold:

\cenlin{$\fzNOT (a \fzOR b ) = 1 - a - b + ab = (1 - a)(1 - b) = \fzNOT(a) \fzAND \fzNOT(b)$,}

\cenlin{$\fzNOT (a \fzAND b ) = 1 - ab = (1 - a) + (1 - b) - (1 - a)(1 - b) = \fzNOT(a) \fzOR \fzNOT(b)$}

Alternatives: $a \fzOR_1 b = \max \{ a, b \}$, $a \fzOR_2 b = a + b - \min \{ a, b \}$,
$a \fzOR_3 b = \min \{ a + b, 1 \}$.

{\ }

\textbf{XOR.} $a \fzXOR b := a(1 - b) + (1 - a) b$.

It's associative, however
$a \fzXOR b \ne \bigl[ \bigl( [\fzNOT (a)] \fzAND b \bigr) \fzOR \bigl(a \fzAND \fzNOT (b)] \bigr) \bigr]$
(that would be provided by non-associative $a \fzXOR_1 b = a(1 - b) + b(1 - a) - ab(1-a)(1-b)$).
Again, the absence of this property doesn't significantly affect the training results.

Alternatives: $a \fzXOR_2 b = |a - b|^p$, $a \fzXOR_3 b = \min \{ a + b, 2 - a - b \}$,
$a \fzXOR_4 b = \sin (\frac{\pi}{2} (a + b))$.

$a \fzXOR_5 b = (a + b) \pmod 2$, where $x \pmod y = x - y \lfloor \frac{x}{y} \rfloor$ for $x \greq0$, $y > 0$,
although being more natural in some sense, doesn't satisfy 2nd requirement.

{\ }

\textbf{ADD.} The addition (to be precise, the addition modulo $2^n$) is special: when applied to two bit strings of equal length,
it is not reduced to bitwise version like the previous ops, due to carry propagation.
This propagation contributes to \textit{diffusion}, which is an important component of CHF.

Let $A = (a_1; ...; a_n)$ and $B = (b_1; ... ; b_n)$ (``little-endian'': the lower bits have lesser indices).
We obtain the bits of the result $S = (s_1; ...; s_n) = A \fzADD B$ one by one, starting from $s_1$ and calculating
the \textit{digit} $s_i = D(a_i; b_i; c_{i-1})$ and the \textit{carry} $c_i = C(a_i; b_i; c_{i-1})$, where $c_0 = 0$.

For binbits we have

\cenlin{\begin{tabular}{c|c|c|c|c}
$a_i$ & $b_i$ & $c_{i-1}$ & $D$ & $C$\\
\hline
0 & 0 & 0 & 0 & 0\\
0 & 0 & 1 & 1 & 0\\
0 & 1 & 0 & 1 & 0\\
0 & 1 & 1 & 0 & 1\\
1 & 0 & 0 & 1 & 0\\
1 & 0 & 1 & 0 & 1\\
1 & 1 & 0 & 0 & 1\\
1 & 1 & 1 & 1 & 1
\end{tabular}}

{\ }

In usual logical ops: $D = a_i \fzXOR b_i \fzXOR c_{i-1}$,
$C = (a_i \fzAND b_i) \fzOR \bigl( (a_i \fzXOR b_i) \fzAND c_{i-1} \bigr)$.

We use the same expression to define $D$
(since our $\fzXOR$ is associative, it doesn't matter if we take
$(a_i \fzXOR b_i) \fzXOR c_{i-1}$ or $a_i \fzXOR (b_i \fzXOR c_{i-1})$;
in general case we would have to choose), but represent $C$ from the identity $a_i + b_i + c_{i-1} = D + 2C$,
implying $C = \frac{1}{2} (a_i + b_i + c_{i-1} - D)$.

Thus defined $\fzADD$ is commutative, but not associative (except for the 1st bit) on fuzbit strings.

Alternatives ($a = a_i, b = b_i, c = c_{i-1}$):

\cenlin{$D_1 = (a \fzXOR b) \fzXOR c$,
$C_1 = (a \fzAND b) \; \mathtt{[X]OR} \; \bigl( (a \fzXOR b) \fzAND c\bigr)$ --- ``full fuzzy adder''}

\cenlin{$D_2 = \frac{1}{2} \bigl( 1.0 - \cos [\pi (a + b + c)] \bigr)$, $C_2 = \frac{1}{2} (a + b + c  - D_2)$}

\cenlin{$C_3 = \min \{ \max \{ a + b + c - 1, 0 \}, 1 \}$, $D_3 = a + b + c - 2C_3$}

{\ }

\textbf{Interpolation.} Our fuzzy bitwise ops correspond with the following well-known interpolation
of the function $f\colon [0;1]^2 \mapsto \mathbb{R}$ inside the $[0;1]^2$ square, based on its values at corners:

\cenlin{$f(x;y) = (1 - y)\bigl[ (1 - x) f(0; 0) + x f(1; 0) \bigr] + y \bigl[ (1 - x) f(0; 1) + x f(1; 1) \bigr]$}

For $D$ and $C$ we use such interpolation inside the $[0;1]^3$ cube.

{\ }

\textbf{Modular arithmetic on circle.} Another way to fuzzy some ops relies on the domain $(-1;1]$,
viewed as scaled angles of the point on unit circle: $a \in (-1;1]$ maps to
$\bigl( \cos (\pi a); \sin (\pi a) \bigr) \in \mathbb{R}^2$
or to $z = e^{i \pi a} = \cos (\pi a) + i \sin (\pi a) \in \mathbb{C}$.
Then, for instance, we define (cf. \cite[3.2]{coutinho2018})

\cenlin{$\fzNOT(a) := \frac{1}{\pi} \atanxy \bigl( \cos \pi(a + 1); \sin \pi(a + 1) \bigr) = \begin{cases}
a - 1, & a > 0,\\
a + 1, & a \lesq 0
\end{cases}$}

\cenlin{$\fzAND(a, b) := ab$}

\cenlin{$\fzOR(a, b) := \fzNOT \bigl( \fzNOT(a) \fzAND \fzNOT(b) \bigr)$}

\cenlin{$\fzXOR(a, b) := \frac{1}{\pi} \atanxy \bigl( \cos \pi (a + b); \sin \pi (a + b) \bigr) = \begin{cases}
a + b - 2, & a + b > 1,\\
a + b, & a + b \in (-1;1],\\
a + b + 2, & a + b \lesq -1
\end{cases}$}

\cenlin{$\fzADD$: $D := a_i \fzXOR b_i \fzXOR c_{i-1}$,
$C := (a_i \fzAND b_i) \fzXOR \bigl( (a_i \fzXOR b_i) \fzAND c_{i-1} \bigr)$}

\noindent
where $\atanxy(x, y)$ for $x^2 + y^2 = 1$ is the angle $\vphi \in (-\pi; \pi]$ such that $\cos \vphi = x$ and $\sin \vphi = y$.

In particular, thus defined $\fzXOR$ has more ``modular linearity'' than the interpolation-based one.

The distance between two such fuzbits is angular $\rho(a, b) = \min \{ |a - b|, |a - b + 2|, |b - a + 2| \}$ or
Euclidean distance between corresponding points on unit circle. At the end of calculations circular fuzbits
can be mapped to ``standard'' fuzbits, $a \mapsto |a|$, --- note that this mapping provides another ``distance'',
$\rho(a, b) = \bigl| |a| - |b| \bigr|$, --- and then to binbits, by rounding $|a|$.

We use circular fuzbits and ops, up to NN training, only for SHA3/Keccak
(in addition to ``standard'' fuzbits for it).
For other CHFs with ADD op, and for Keccak with long enough messages, they sometimes impede learning:
NN learns to predict an ``essentially fuzzy'' message whose bits are far from 0/1
and thus cannot be rounded to get binbit string with the same hash.

On the other hand, we could add a penalty for the output fuzbits of NN being far from binbits.

\subsection{Fuzzy hashes}

which might be called ``hashes of fuzbit strings'',
because the term ``fuzzy hash'' has certain established meaning already, and it's different from ours;
cf. \cite{oliver2013}, \cite{sec.stackexch.q117905}.

The approach is simply this:
take the algorithm to calculate CHF and consider all ops on bits and bit strings as fuzzy.
To be precise, we fuzzy only the ``change bits'' ops and keep untouched the ``move bits'' ops like
left shift $\fzSHL$/$\ll$, right rotation $\fzROR$/$\ggg$, or any other fixed permutation of bits.
In particular, $\fzLEBE$ reverses byte order, switching between little- and big-endian.

Message length in bits is a usual number, hence the padding is done with binbits.

The algorithms are described in many places, see \cite{fips180.4}, \cite{fips202}, \cite{rfc1321}, \cite{rfc3174}, \cite{rfc6234}.

{\ }

For example, the single ($i$-th for $i=\ovline{0, 63}$) SHA2-256 round in pseudocode:

{\ }

IF $i \greq 16$:

\qquad $S := W_{i - 15}$

\qquad $R_0 :=  (S \ggg 7) \fzXOR (S \ggg 18) \fzXOR (S \gg 3)$

\qquad $S := W_{i - 2}$

\qquad $R_1 :=  (S \ggg 17) \fzXOR (S \ggg 19) \fzXOR (S \gg 10)$

\qquad $W_j := W_{i - 16} \fzADD R_0 \fzADD W_{i-7} \fzADD R_1$

ELSE:

\qquad $W_i \larr \fzLEBE \bigl( \mathrm{CHUNK}[32i, 32i + 1, ..., 32i + 31] \bigr)$

{\ }

$S_1 : = (E \ggg 6) \fzXOR (E \ggg 11) \fzXOR (E \ggg 25)$

$P := (E \fzAND F) \fzOR (\fzNOT(E) \fzAND G)$

$T_1 := H \fzADD S_1 \fzADD P \fzADD K_i \fzADD W_i$

$S_0 := (A \ggg 2) \fzXOR (A \ggg 13) \fzXOR (A \ggg 22)$

$M := (A \fzAND B) \fzXOR (A \fzAND C) \fzXOR (B \fzAND C)$

$T_2 := S_0 \fzADD M$

$H,\; G,\; F,\; E,\; D,\; C,\; B,\; A := G,\; F,\; E,\; D \fzADD T_1,\; C,\; B,\; A,\; T_1 \fzADD T_2$

{\ }

There are modifications that doesn't change the result for binbits, but make it different for fuzbits,
e.g. $P := (E \fzAND F) \fzXOR (\fzNOT(E) \fzAND G)$.

{\ }

To specify the number of rounds (defaults are 64 for MD5, 80 for SHA1, 64 for SHA2, and 24 for SHA3/Keccak),
we add the second argument: SHA1($M$, 4) is SHA1 with 4 rounds.

{\ }

For the ASCII-encoded message $M=$ ``\texttt{The quick brown fox jumps over the lazy dog.}'' the hash
SHA1($M$, 2) is $H = $ \texttt{c30b13efe3eaa95cf28c25be8c25bf585c8dbeee} in hexadecimal bytes.
The 6th bit of $M$ (here the count starts from 1) is $m_6 = 0$, the 32nd bit of $H$ is $h_{32}=1$.
If we change $m_6$ to $1$, then $h_{32}$ is still $1$.
If we change $m_6$ from 0 to 1 \textit{gradually}, $h_{32}$ is given on (a):

{\ }

\cenlin{\footnotesize{\begin{tabular}{ccc}
\includegraphics[height=2.5cm]{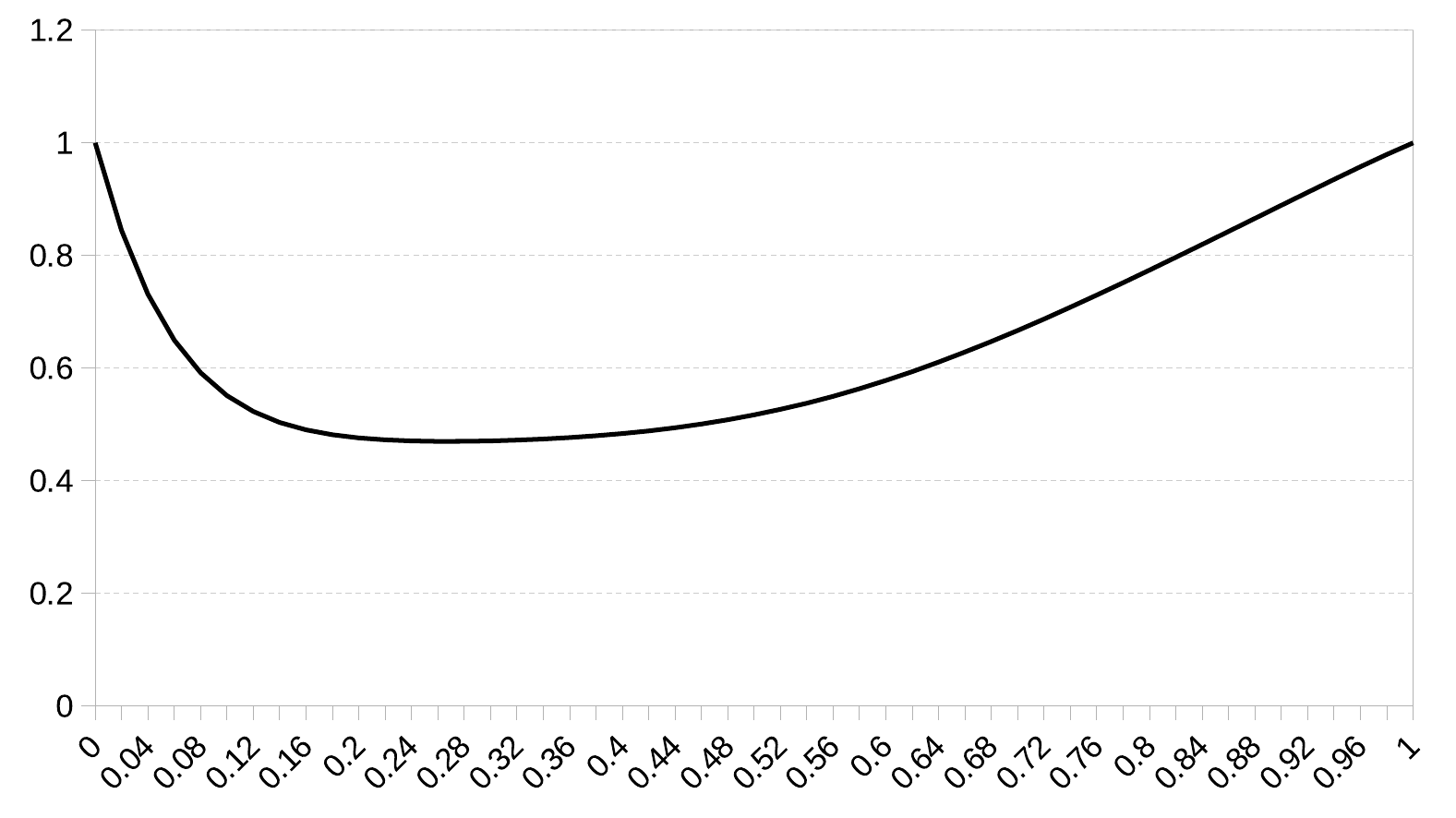}
&
\includegraphics[height=2.5cm]{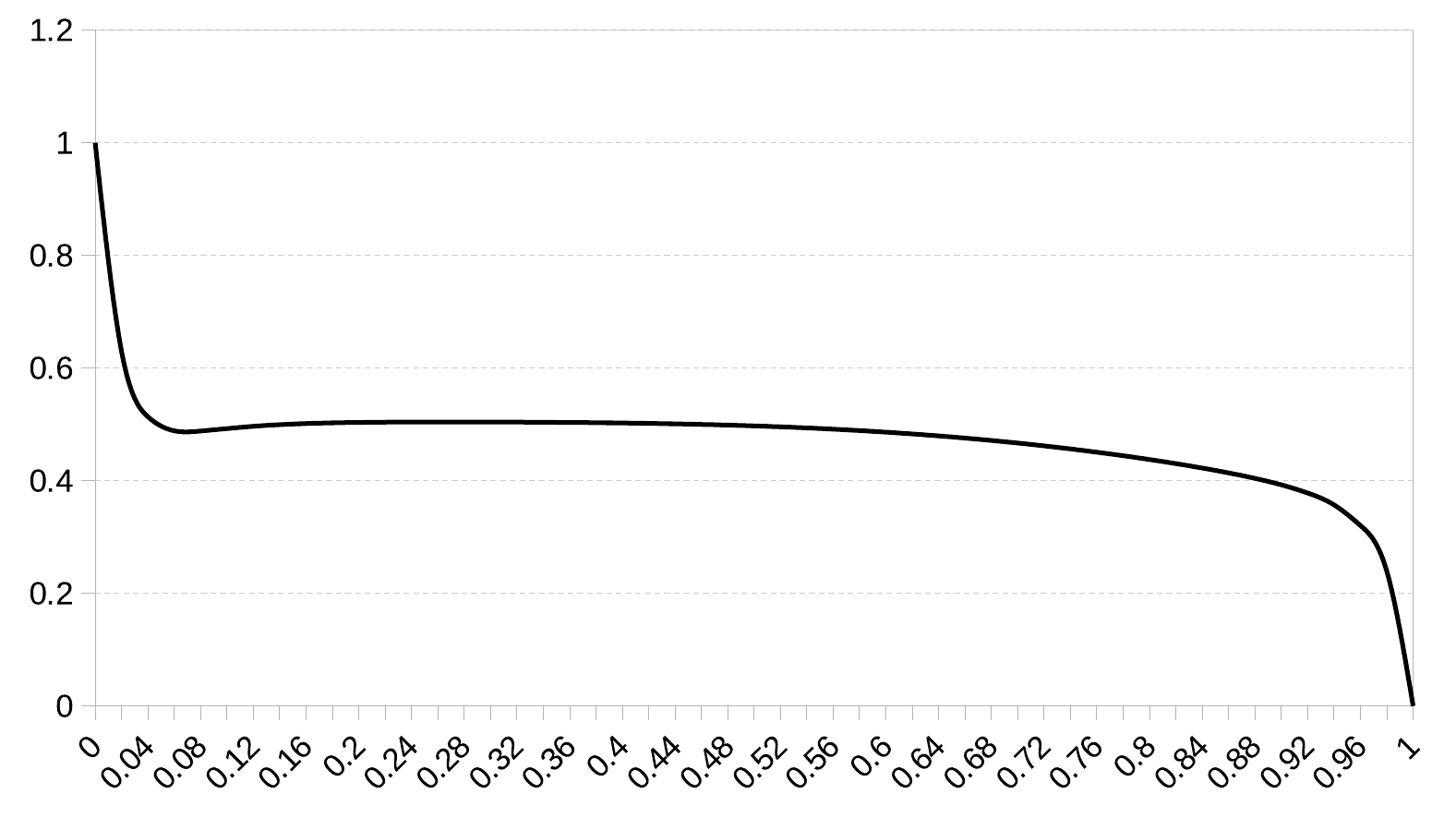}
&
\includegraphics[height=2.5cm]{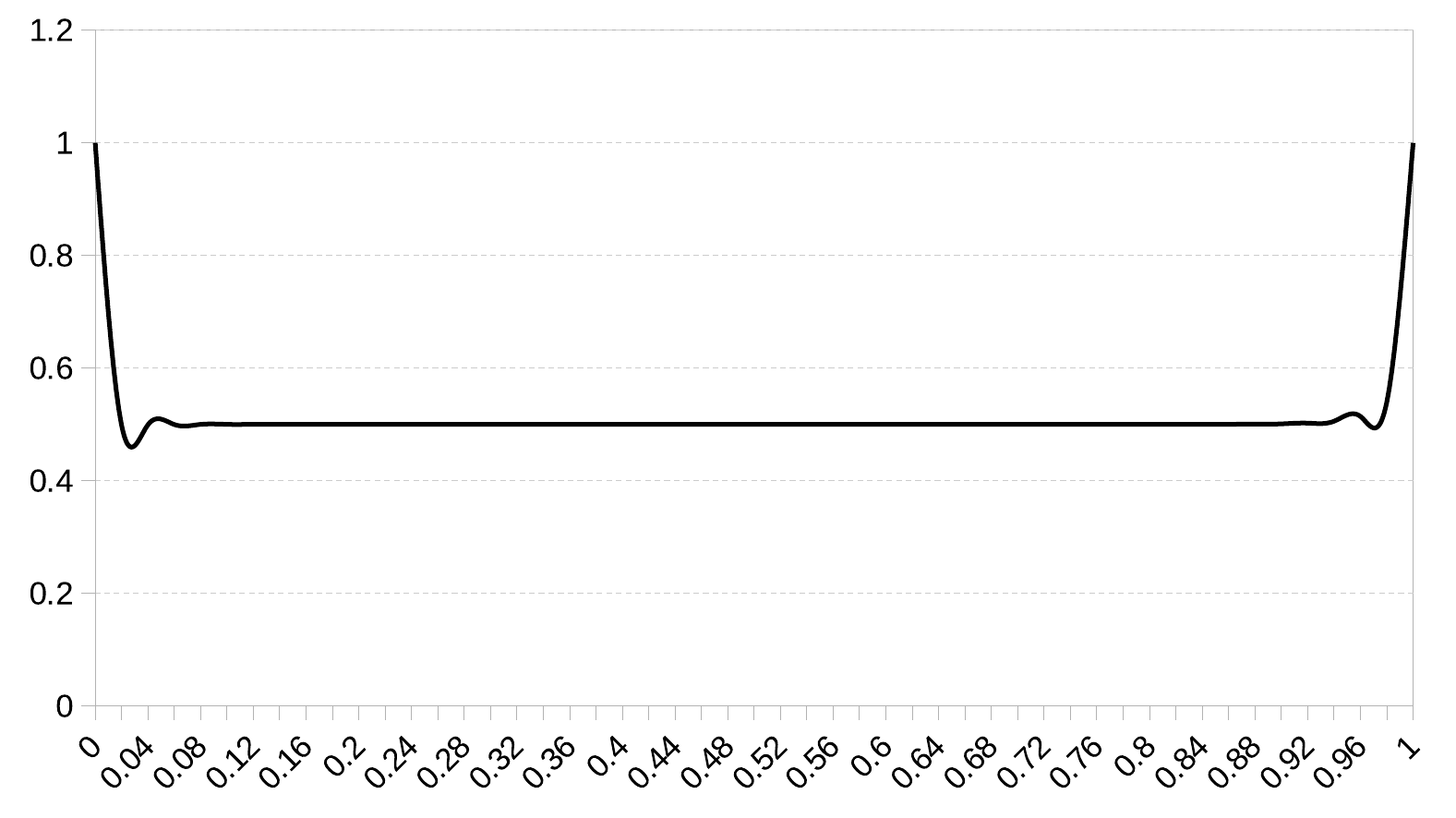}
\\
a (2 rounds) & b (4 rounds) & c (6 rounds)
\end{tabular}}}

{\ }

On (b) we see $h_{32}$ w.r.t. $m_6$ for 4-round SHA1, and on (c) --- for 6-round SHA1.

(c) reveals the flaw of our fuzzy ops on $[0;1]$:
if fuzbits are not binbits, then the result quickly converges to $\frac{1}{2}$ as the number of ops increases.
For 20 rounds, which is only $\frac{1}{5}$ of the full SHA1,
and \texttt{float64} precision, $m_6 = 10^{-12}$ already gives $h_{32} = 0.5$.
In some sense, it corresponds with ``certainty'' meaning of a fuzzy bit:
when we combine uncertaint entities, the certainty diminishes.

In attempt to circumvent this flaw we can modify fuzzy ops as follows:
the result $r$ is transformed smoothly so that $r \in [0;\frac{1}{2})$ moves closer to $0$
and $r \in (\frac{1}{2}; 1]$ moves to 1. One of such transformations is $t(r) = 3r^2 - 2r^3$.
On the other hand, if used without other limitations, it makes the result to change very slowly near 0 and 1,
while near $\frac{1}{2}$ the change is very rapid.

Being given binbit hashes, NN with the loss function based on such fuzzy hashes learns to predict the messages
whose fuzbits are close to binbits as well.

{\ }

\textbf{Circular fuzziness.} Let $M$ be the same message and the CHF be SHA3-256.
When fuzbits and ops are circular, --- $m_i, h_i \in (-1;1]$, --- we have the following dependency of $h_{14}$ on $m_{6}$:

{\ }

\cenlin{\footnotesize{\begin{tabular}{ccc}
\includegraphics[height=2.5cm]{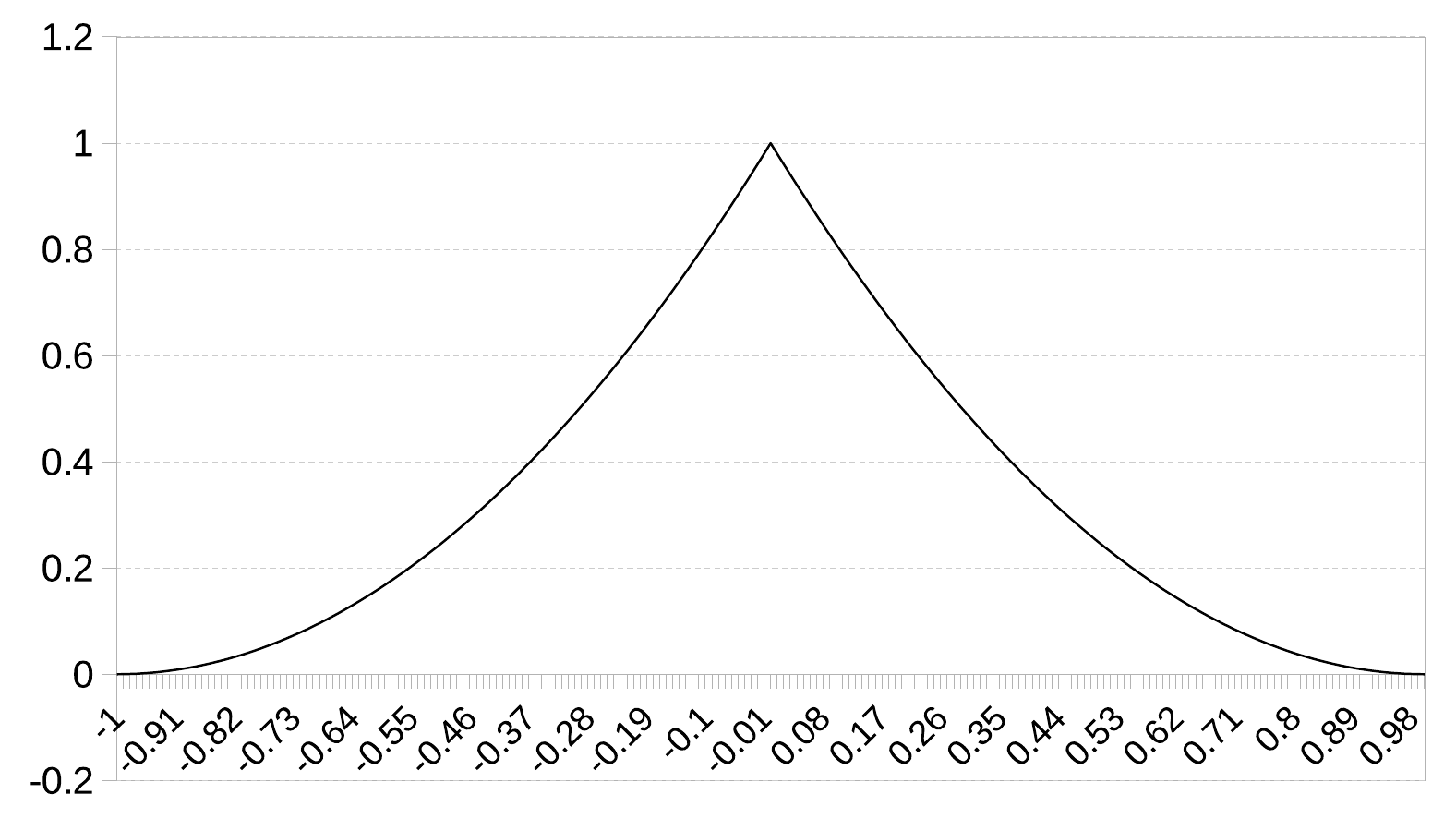}
&
\includegraphics[height=2.5cm]{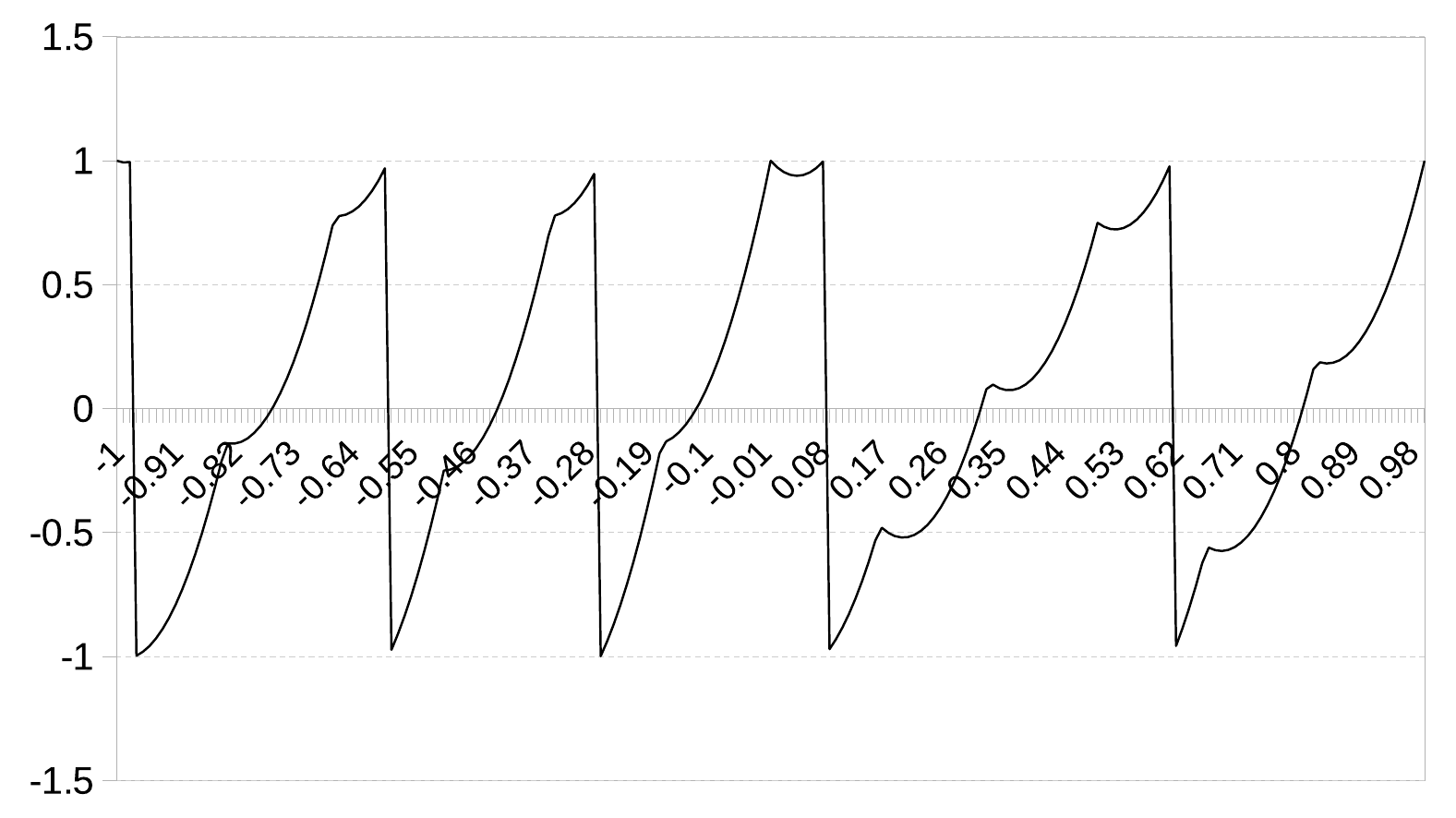}
&
\includegraphics[height=2.5cm]{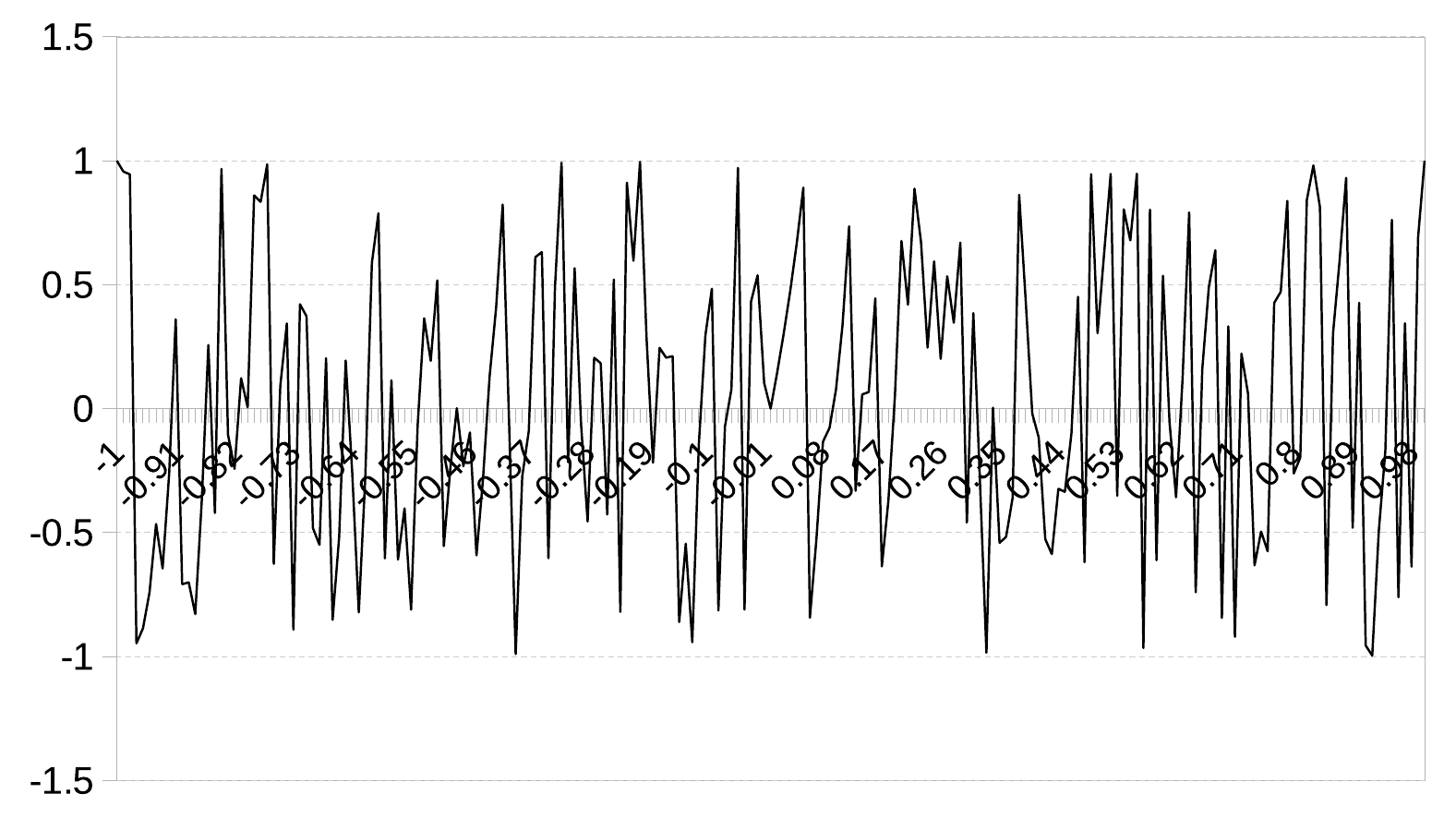}
\\
2 rounds & 3 rounds & 4 rounds
\end{tabular}}}

{\ }

\noindent
indicating another potential flaw --- an ``excessive'' non-linearity.
Of course, from CHF design's point of view this is a merit, not a flaw.

\subsection{Neural networks and training}

\forceparindent
We apply, in general, the traditional fully-connected perceptron: NN is the sequence of the layers,
each layer consists of the cells, the output of each cell is the function of the weighted sum of the outputs
of all cells from the previous layer:

\cenlin{$s_{ij} = a(b_{ij} + \sum\limits_{k=1}^{n_{i-1}} w_{i,j,k} s_{i-1,k})$}

We extend it with the so-called batch normalization (\cite[8.7.1]{goodfellow2017}, \cite{ioffe2015}) after each inner layer;
not only does it accelerate learning, but improves the final results, ---
the average number of wrong hash bits becomes 1.5--2 times less.
Occasionally, regularization techniques (dropout, weight regularization etc.) may be added.

Perhaps more sophisticated (e.g. deeper) architectures will work better, perhaps not.

The hyperparameters of such NN are the number of layers and the number of cells in each layer.
The size of the input layer is the size of the hash, the size of the output layer is the size of the predicted message
(so any NN of this kind predicts the messages of certain fixed length).

To implement this architecture, we use Keras framework (\cite{keras})
with TensorFlow backend (\cite{tensorflow}).

The activation function for all layers except the output one is ELU: $a(x) = \begin{cases}
x, & x\greq 0,\\
e^x - 1, & x \lesq 0.
\end{cases}$

The activation for the output layer is sigmoid $a(x) = (1 + e^{-x})^{-1} \in (0;1)$
or ``hard sigmoid'' $a(x) = \min \{ \max \{ 0; \frac{1}{5} x + \frac{1}{2} \}; 1 \}$, ---
for $[0;1]$- \textit{and} for $(-1;1]$-fuzbits
(using something like $\tanh(x)$ or $\sin(x)$ for the latter, although seeming more convenient,
practically prevents NN from learning).
Hard sigmoid usually accelerates learning, but with it NN may ``stick'' to non-optimal states.

The loss function is calculated as follows:
let $h()$ be the fuzzy CHF and $H = h(M)$ for the randomly generated message $M$. $M$ is unknown to NN,
and $H$ is the input to NN. NN predicts fuzzy message $M'$, for which we obtain the fuzzy hash $H' = h(M')$
and the single loss $L = L(H; H')$. If NN is trained by batches of size $n$ (so there are $n$ input hashes
$H_1$, ..., $H_n$), then the total loss on the batch is $\frac{1}{n} \sum\limits_{i=1}^n L(H_i; H'_i)$.
$L$ is $\| H - H' \|_1$: $L(H; H') = \sum\limits_{j=1}^k | h_j - h'_j |$ (denoted by $L_1$)
or it may be a binary crossentropy
$L(H; H') = -\frac{1}{k} \sum\limits_{j=1}^k \bigl[ h_j \log h'_j + (1 - h_j) \log (1 - h'_j) \bigr]$ (denoted by BCE),
which sometimes slightly improves the accuracy of inversion.

The optimizer is \texttt{Nadam} (Nesterov Adam) with default learning rate 0.002,
it can be replaced by \texttt{Adam} or \texttt{RMSprop} with no significant impact on the results;
see \cite[8.5]{goodfellow2017}, \cite{kingma2014}.

\section{Results}

\noindent
are given in Addendum. Each of them consists of 6 items:

{\ }

\textbf{1.} CHF, number of rounds, message length, and the \textit{mask} --- the bits of hash
that contribute to the loss and miss count (regardless of the mask, input to NN is always a full hash).

\textbf{2.} Hyperparameters of NN (cells in each layer) and training parameters per epoch
(train samples and batch size).

\textbf{3.} Training chart: $X$ is epoch, $Y$ is loss.

{\ }

When the training finishes, we proceed to the testing phase. We randomly generate 1024 binbit messages,
calculate the CHF for them and let the trained NN predict the messages with the same hashes.
Then we bitwisely round predicted messages so that fuzbits become binbits, calculate their usual hashes
and for each pair (true hash; hash of rounded predicted message) count the number of misses --- different binbits of hashes.
After that we generate another 1024 random messages, calculate their hashes and compare with the true ones to count misses;
this shows the advantage of the trained NN over ``zero level'' of randomly guessing the message with a given hash.

{\ }

\textbf{4.} Histograms of \textit{total number of} misses:
$X$ is number of misses, $Y$ is number of hashes with so many misses,
for predicted and for random messages.

\textbf{5.} Histogram of misses \textit{for hash bits} of predicted messages:
$X$ is the index of bit, $Y$ is the frequency of miss at this bit.

\textbf{6.} Average, minumum, and maximum number of misses, for predicted and random messages.

Only the ranges with non-zero values are shown.

{\ }

In this scheme, instead of training NN on the actual values of certain CHF for the messages,
we can input \textit{random} bit strings, including the hashes that are ``impossible''
(e.g. 1-round SHA1 produces the hash with majority of its bits being constant).
As with many other choices, it makes almost no difference: the speed of learning is the same,
and although the final loss is greater due to ``unreachable'' wrong hash bits,
in the end NN inverts test hashes equally well/not well.

We present ``general inversion'' results (see Introduction),
because we have found that in average they are usually close to ``single inversion'' ones.
When the task is to invert a specific hash, it is certainly more preferable to train many NNs, each given this hash,
and collect ``the best'' preimages with minimal number of hash bit misses.

In most cases, the loss function at the end of the training has apparently come close to an asymptotic value,
but sometimes the training was interrupted before that, and the loss had not yet reached its minimum.

The results are stable: if we run NN training with the same hyperparameters, for the same number of epochs etc.,
the final loss and the accuracy for the test hashes will be nearly identical.

\section*{Conclusion and further development}
\addcontentsline{toc}{section}{Conclusion and further development}

\forceparindent
Obviously, as the number of rounds increases,
all four CHFs quickly become too complicated for NNs of considered architectures to invert,
or, in other terms, to approximate the inverse function.

{\ }

\textit{SHA1:} NN is able to invert 1 round with positive probability, for 2 rounds there are at least 10 misses,
then the number of misses becomes greater than the number of matched bits.
The full hash for more than 5--6 rounds with the accordingly long messages is almost uninvertible;
the quick-convergence-to-$\frac{1}{2}$ flaw of (unadjusted) fuzzy ops contributes to this barrier.
Partial matching of a small group of hash bits is possible.

\textit{MD5:} slightly \textsl{harder} than SHA1, but closer to it than to SHA2 from NN-inversion difficulty's point of view.

\textit{SHA2:} similar to SHA1, harder.

\textit{SHA3/Keccak:} Inversion of 1 round for short (what's short depends on rate and capacity)
messages is good, can be exact; of 2 and more is very bad.
Weakened rounds move the ``barrier'' further.
Learning may delay for few initial epochs.
A small group of hash bits can be matched partially.

{\ }

Even for relatively small number of rounds this approach in its presented form is much weaker than e.g. those from
\cite{de2007}, \cite{morawiecki2013}, \cite{nejati2017}.
One of its most important shortcomings is that there is no guarantee to obtain the exact preimage,
however much time NN has spent on training. Another drawback is the fixed length of the preimage returned by given NN,
though it's possible to train many NNs for different message lengths.

On the other hand, a tiny but statistically significant increase of accuracy in hash bits
for NN-predicted messages over that for randomly guessed messages
indicates the possibility to ``trace back'' the hash bits to the message ones, which can be exploited.

{\ }

Few directions of further development, in no particular order,
and not without assumption that something has been done already, somewhere:

$\bullet$ Tune fuzzy ops and NN for this problem by defining ops, selecting activations, loss function,
optimizer and its parameters etc. more carefully.
Few times throughout this paper we've mentioned some alternatives and that switching between them doesn't improve the final accuracy;
however, there may be much better ones that slipped past our attention.
Objection: this is a common ``belief'' when dealing with NNs,
but maybe it can be proven that NNs will always have a lower positive bound of the accuracy.

$\bullet$ If the value of $m$-round CHF provides its full internal state, then we can use NN inversion iteratively
to restore the state before the $m$-th round, then before the $(m-1)$-th one, and so on.
Objection: multiple rounds may depend on the one and the same part of the input message,
like SHA1 after the first 16 rounds; also, the hash may come from a truncated state (SHA3/Keccak).

$\bullet$ Modify NN architecture, add connections between non-adjacent layers, convolutions, recurrency,
add combinators/solvers that produce fuzzy bits within NN and are controlled by outputs of the cells, and so on.
Objection: see 1st item.

$\bullet$ Combine NN with other known preimage-searching methods, like SAT solvers.
Objection: NNs may not provide any advantage over those, wherever into the ``chain'' NNs are inserted
(if so, it would be useful to prove the statements about the absence of such advantage).
Since SAT solvers actually apply machine learning approaches, this one is probably implemented by now.

$\bullet$ Train the ensembles of NNs, then combine their outputs.
Objection: if NNs predict different messages with partial hash matching, there is no ``averaging'' to increase the accuracy.

$\bullet$ Train NN to predict the message for which not only the final hash matches the input one,
but some bits of intermediate states match the respective bits of such states at the calculation of the true hash.
Objection: different states transform to the same hash.

$\bullet$ Generate training samples from the messages that satisfy some constraints, maybe changing these constraints with time,
so that NN learns to invert the dependence of CHF on different parts of the message gradually.
Objection: NN may not converge, it will learn at first to predict the beginning and not the end,
then to predict the end and not the beginning, then again etc.

$\bullet$ Apply different fuzzy ops for different bits of the state, make them depend on the index of the round etc.,
to distinguish previous states leading to the same current state. Objection: inverse function may become more complex, not simpler.

$\bullet$ Make input hashes and messages they're obtained from more fuzzy, so that their bits aren't only 0/1.
Objection: tried on NNs described here, it does not improve training or its results.

$\bullet$ Abandon this approach completely and do not waste time on it\textellipsis

{\color{white} \tiny $\bullet$ ???}

\newpage

\section*{Addendum}
\addcontentsline{toc}{section}{Addendum}

\forceparindent
Before actual results for CHFs, we show that a single $\fzADD$ op as we defined it already presents
certain difficulty to NN.

{\ }

\resultitem
\textbf{``CHF''}: ADD. The input to NN is 32-bit string,
and NN learns to predict such 64-bit string that the sum of its two 32-bit halves
is the input. Mask: full, all ``hash'' bits.

\textbf{NN:} 3 layers, 512 cells each. 1024 samples divided into batches of 64 samples.

\traintestcharts{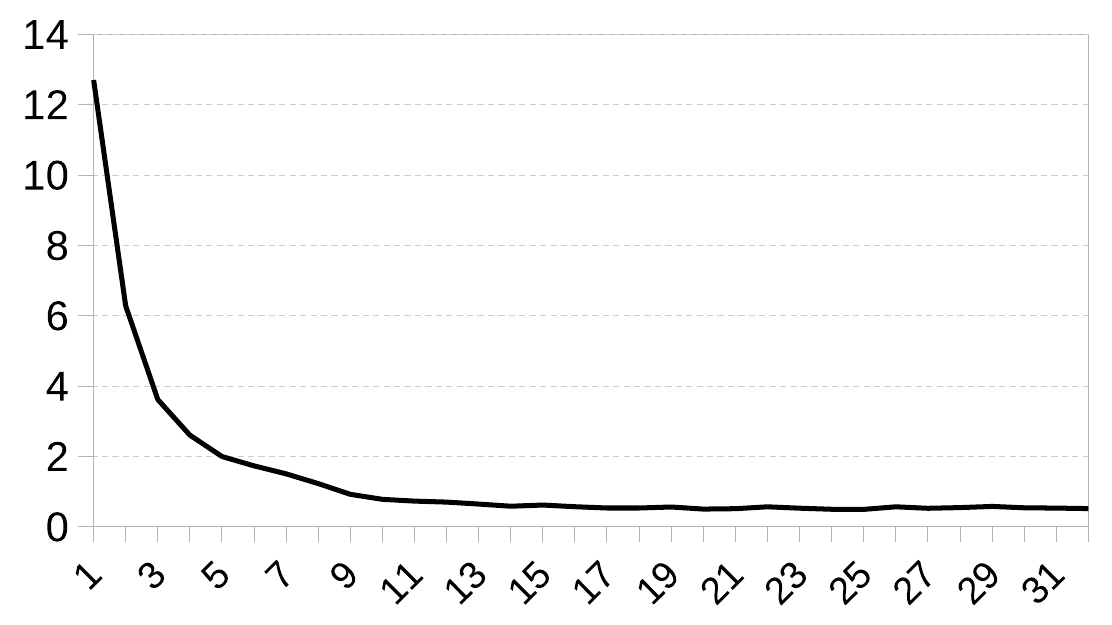}{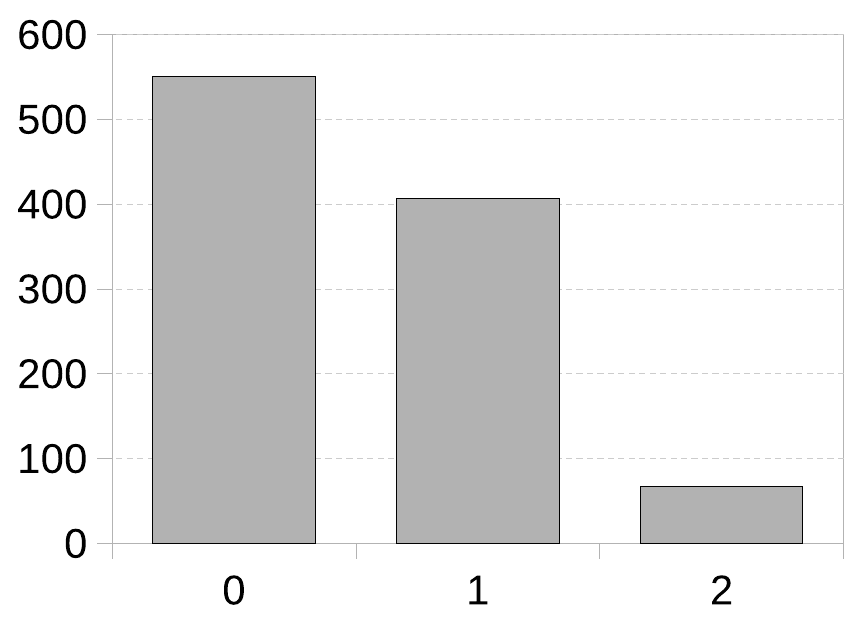}{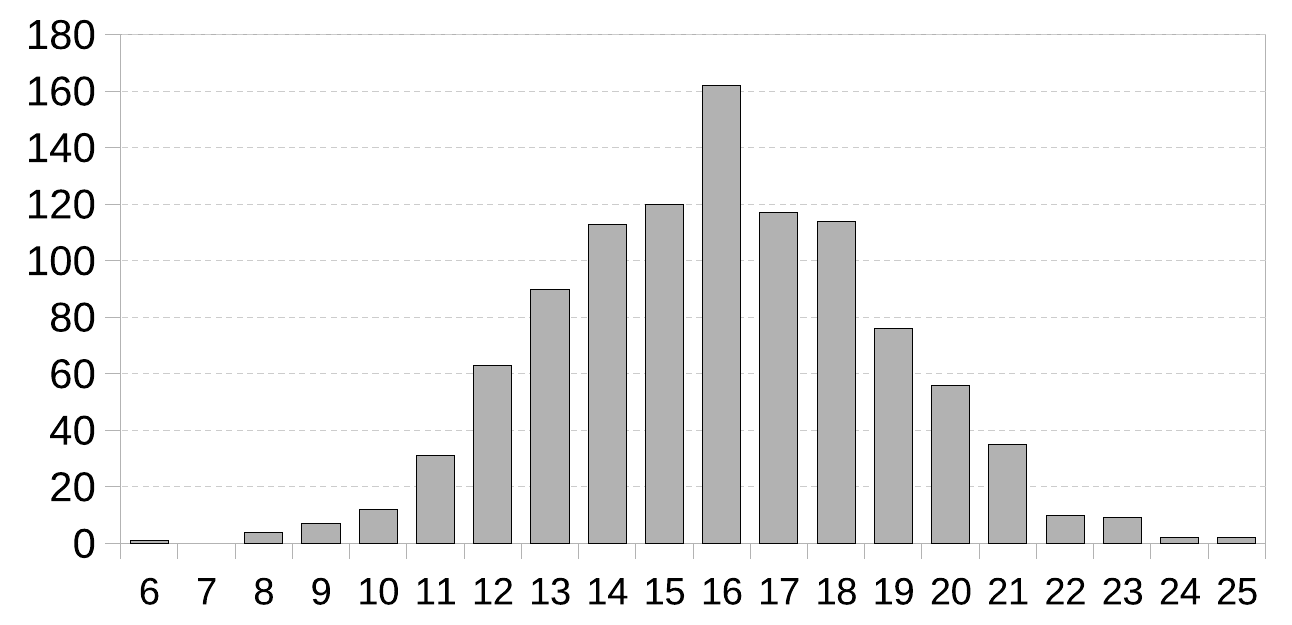}{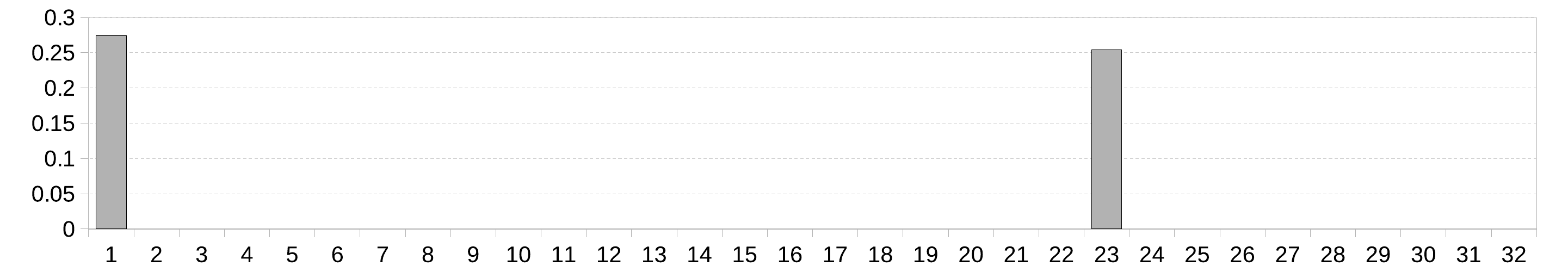}

\textbf{Misses for 1024 predicted messages:} Average = 0.528, Min = 0, Max = 2.

\textbf{Misses for 1024 random messages:} Average = 15.944, Min = 6, Max = 25.

When NN finds two summands with exactly the given sum in only half of all tests,
its ability to invert much more sophisticated CHFs, designed specially to ``cover footprints'', seems questionable.

The histogram of miss frequencies is typical in that it has some bits with ``insurmountable'' uncertainty for NN,
despite of 100\% accuracy for the rest of the bits.

\subsection{SHA1}

\forceparindent
This CHF along with MD5 and SHA2 follows Merkle-Damg\r{a}rd construction (\cite[5.2]{katz2015}) and uses $\fzADD$ op,
which appears to be the most difficult for our NNs to invert.

{\ }

\resultitem
\textbf{CHF:} SHA1, 1 round, message length is 32 bits (1st round doesn't reach farther). Mask: full, all hash bits.

\textbf{NN:} 1024 cells in 1 layer. 1024 samples divided into batches of 64 samples.

\traintestcharts{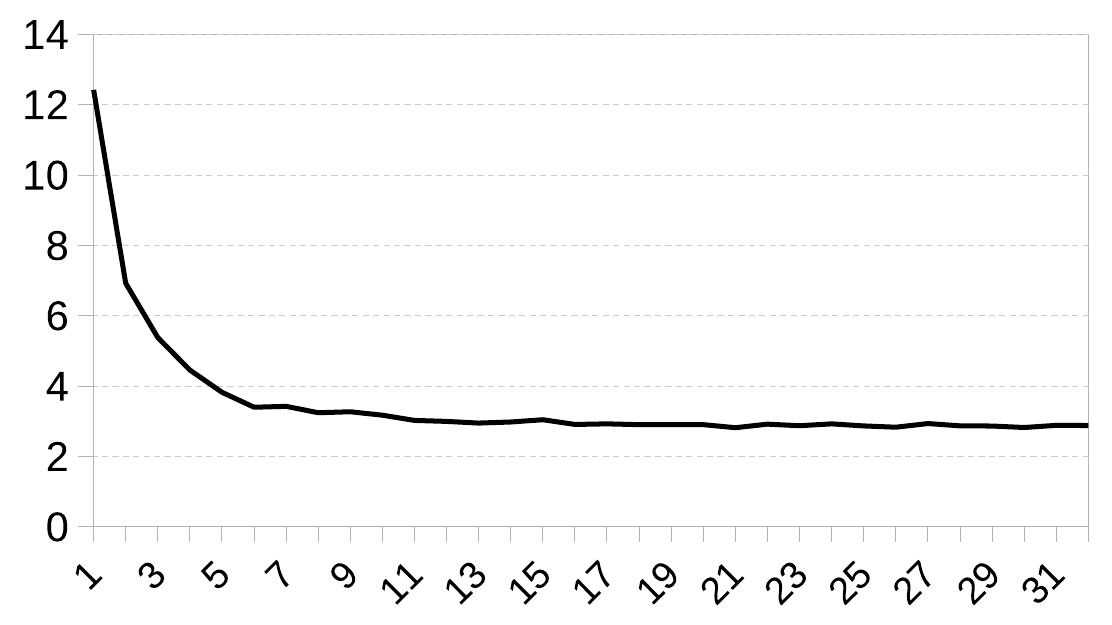}{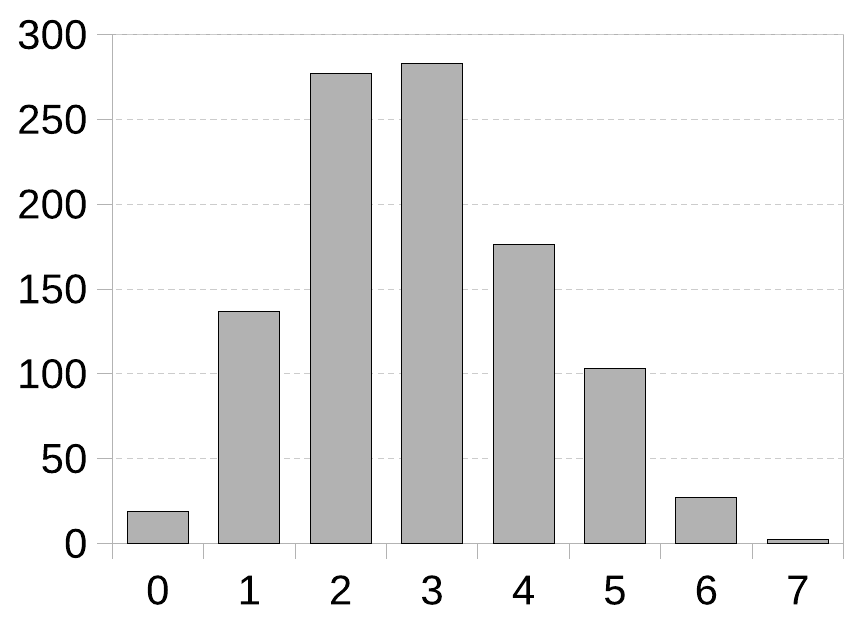}{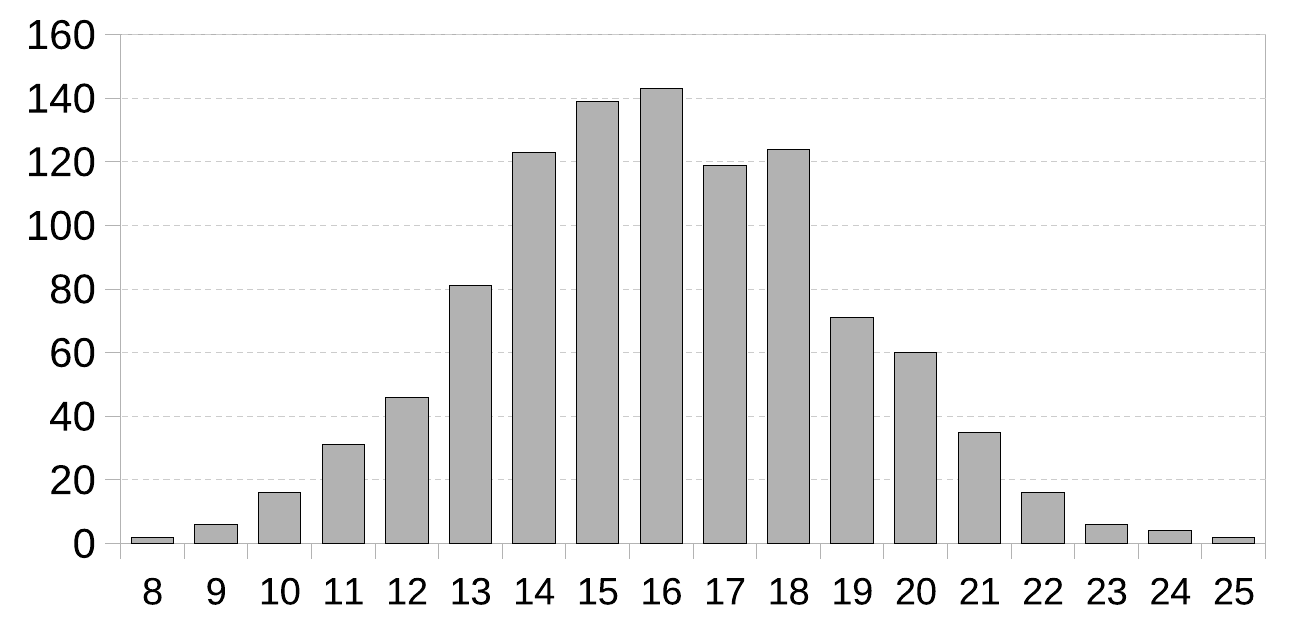}{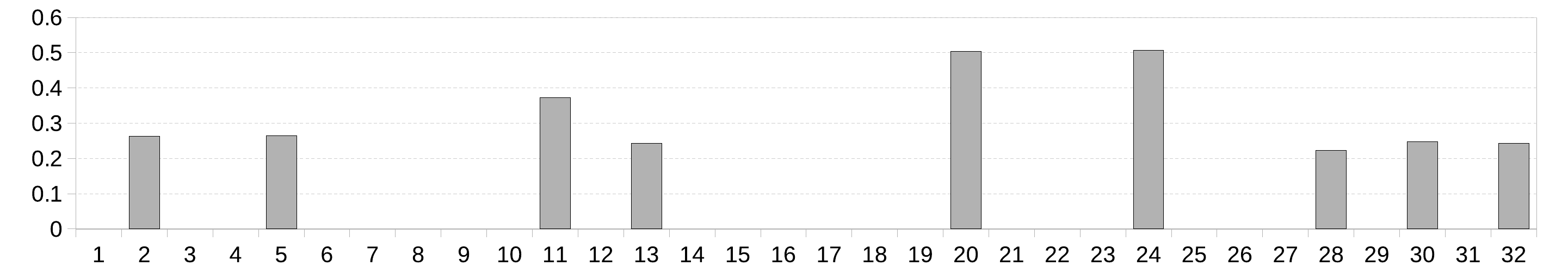}

\textbf{Misses for 1024 predicted messages:} Average = 2.866, Min = 0, Max = 7.

\textbf{Misses for 1024 random messages:} Average = 16.061, Min = 8, Max = 25.

This NN predicts messages such that certain bits of 1-round SHA1 are always right.
The frequencies of miss for the rest of the bits look like they're close to something divisible by $\frac{1}{8}$ or $\frac{1}{4}$.

Increasing the number of layers (towards ``deep'' learning) and cells per layer does not improve these results much,
same if we cut input to only the first 32 bits of hash that can change after 1st round.
Another training of the 4-layer NN ends with similar characteristics and distributions,
including miss frequencies per bit:

{\ }

\cenlin{\footnotesize{\begin{tabular}{c}
\includegraphics[height=2.5cm]{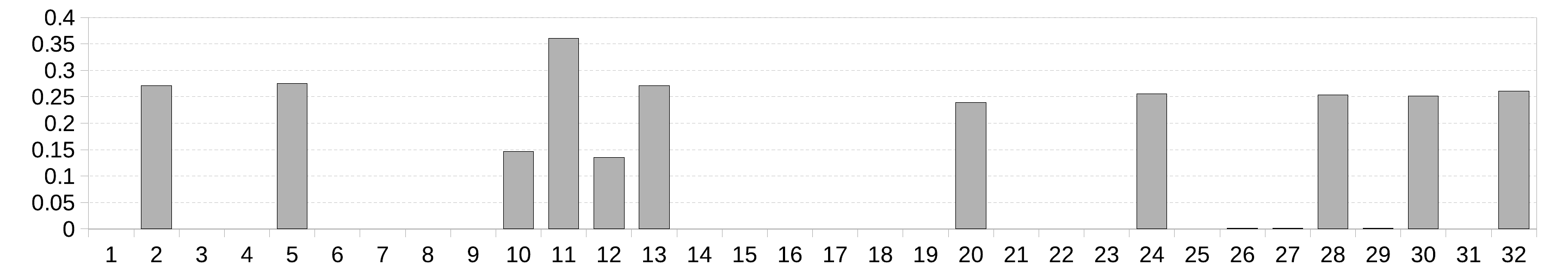}
\\
Miss frequencies per bit (2)
\end{tabular}}}

{\ }

For a comparison we once provide these frequencies for randomly generated messages:

{\ }

\cenlin{\footnotesize{\begin{tabular}{c}
\includegraphics[height=2.5cm]{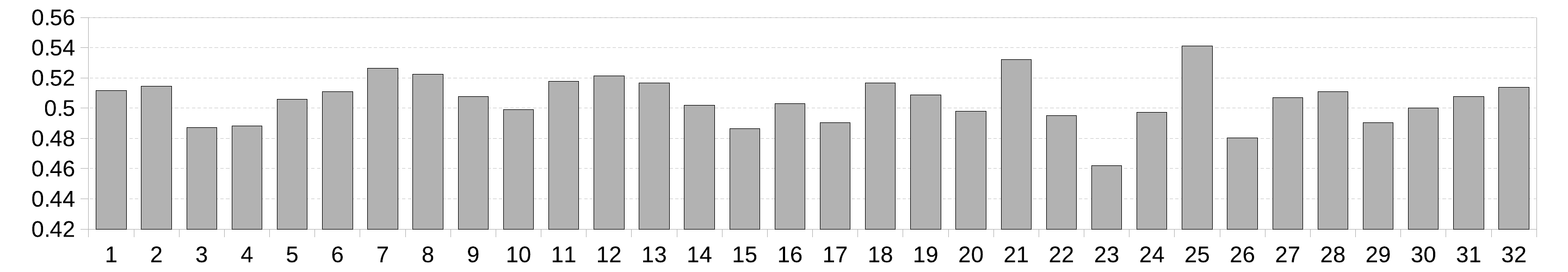}
\\
Miss frequencies per bit for Random
\end{tabular}}}

{\ }

\resultitem\label{resSha1rounds2}
As we turn to 2-round SHA1, the accuracy expectably deteriorates.

\textbf{CHF:} SHA1, 2 rounds, 64-bit message.

\textbf{NN:} 2 layers, 1024 cells each. 1024 samples, 64 in batch.

\traintestcharts{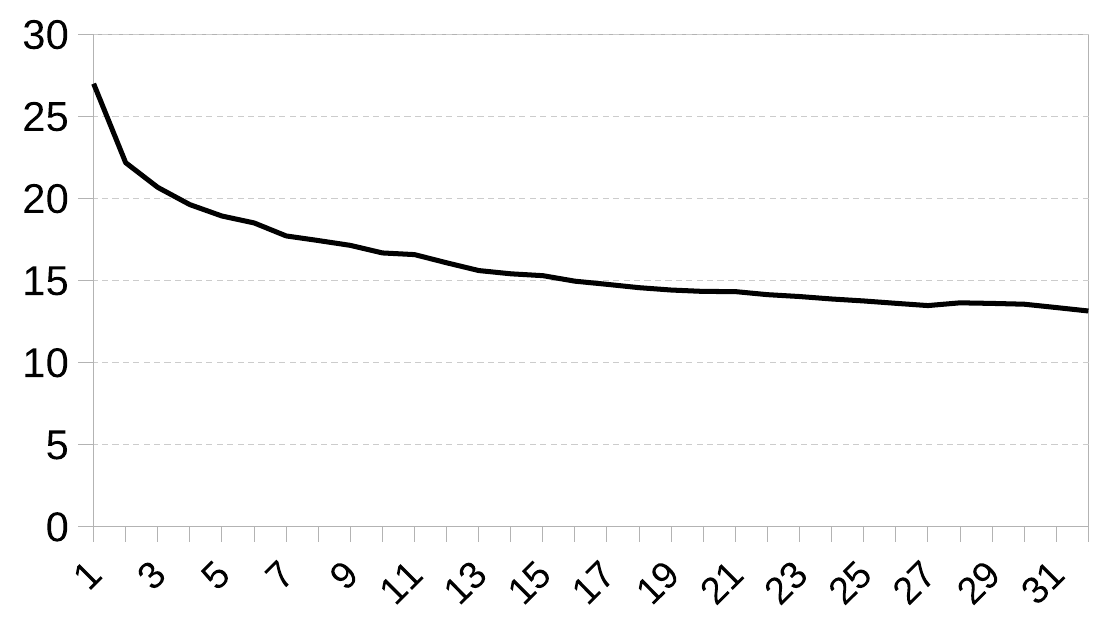}{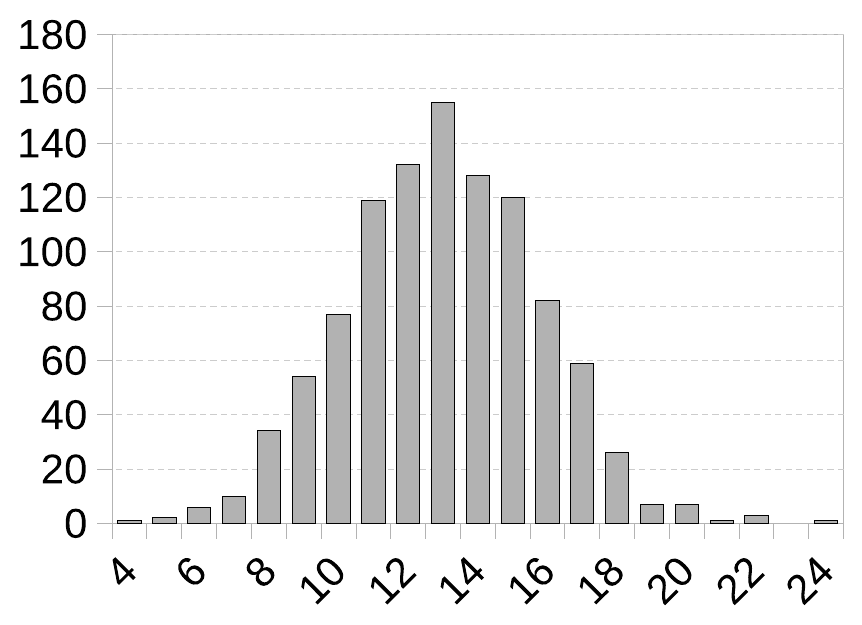}{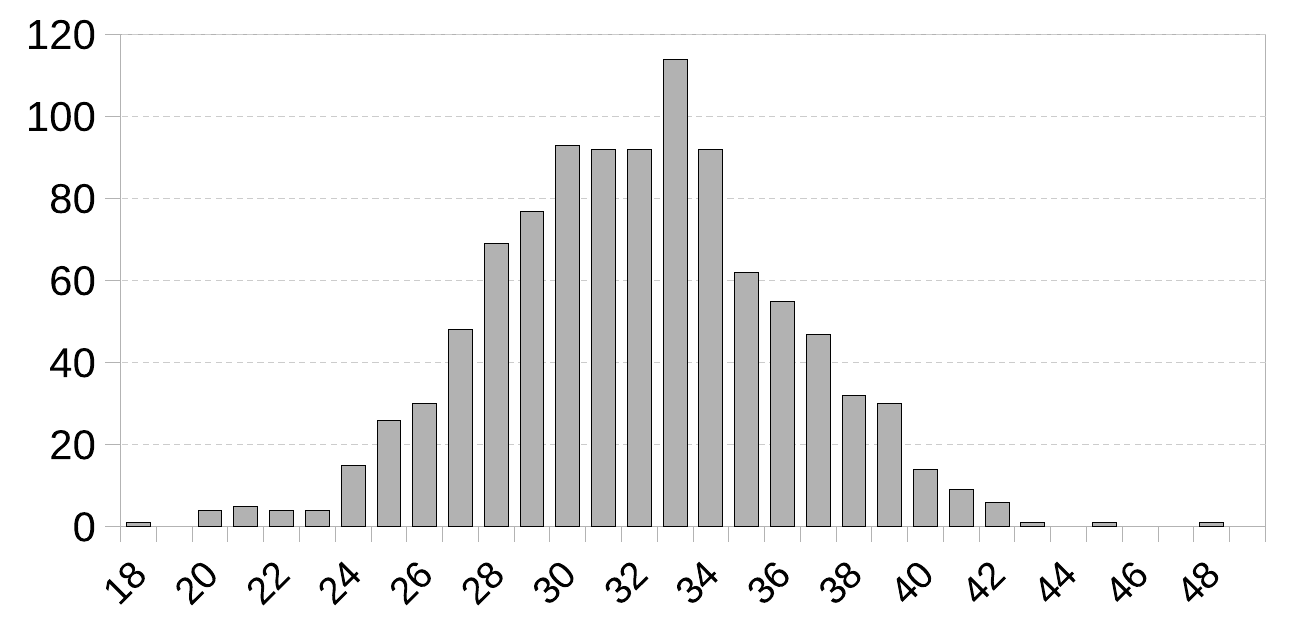}{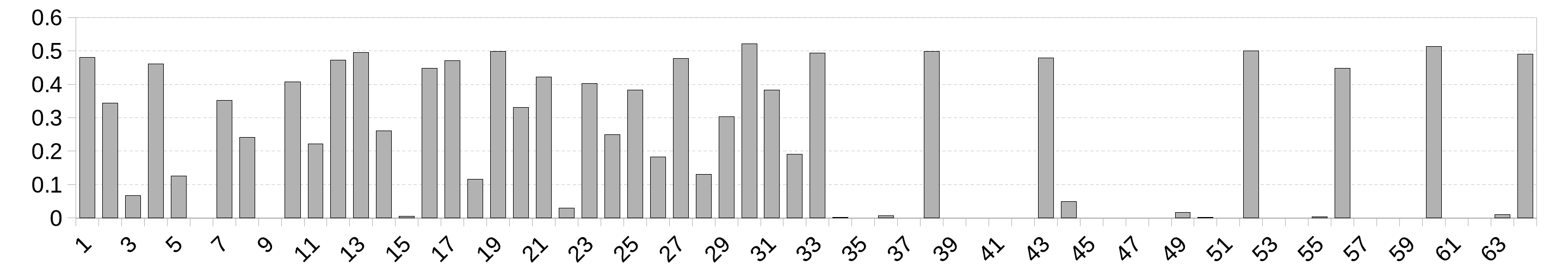}

\textbf{Misses, predicted:} Average = 13.003, Min = 4, Max = 24.

\textbf{Misses, random:} Average = 31.936, Min = 18, Max = 48.

The first 32 bits of hash have become more ``non-linear'', only few of them have ``found the way back to message'' in some sense.
The second half resembles the previous case.

Adding push-from-$\frac{1}{2}$ transformation to $\fzXOR$ op, --- that is, $a \fzXOR' b = 3r^2 - 2r^3$, where $r = a \fzXOR b$, ---
slightly speeds up NN learning. However, if we push $\fzADD$ in a similar way, the learning worsens.

{\ }

\resultitem
\textbf{CHF:} SHA1, 2 rounds, 32-bit message. Mask: full.

\textbf{NN:} 2 layers, 1024 cells each. 1024 samples, 64 in batch.

\traintestcharts{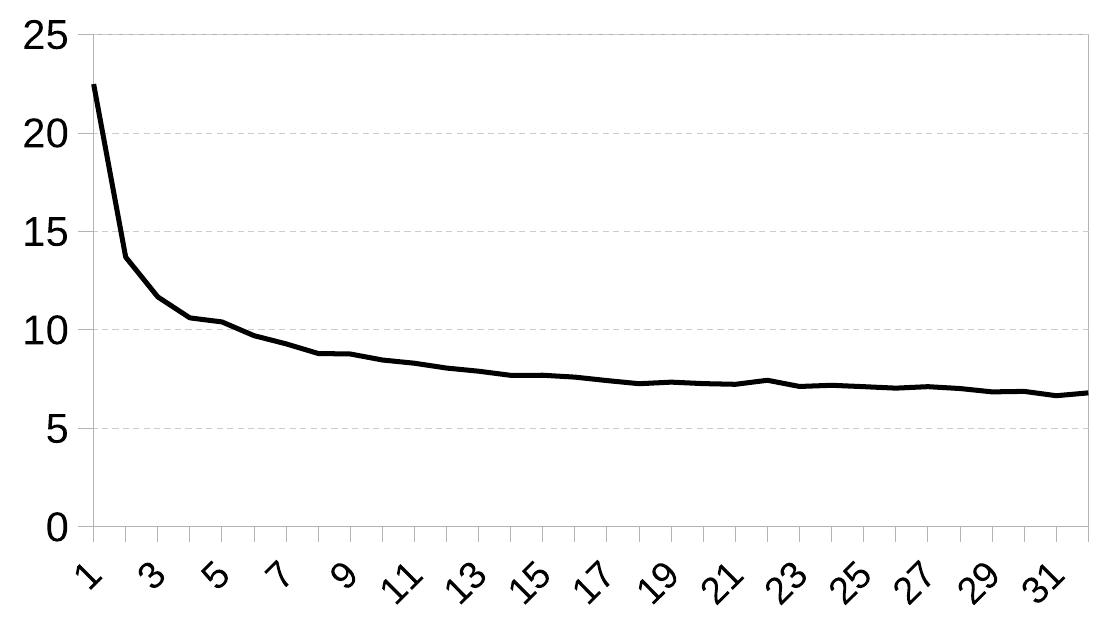}{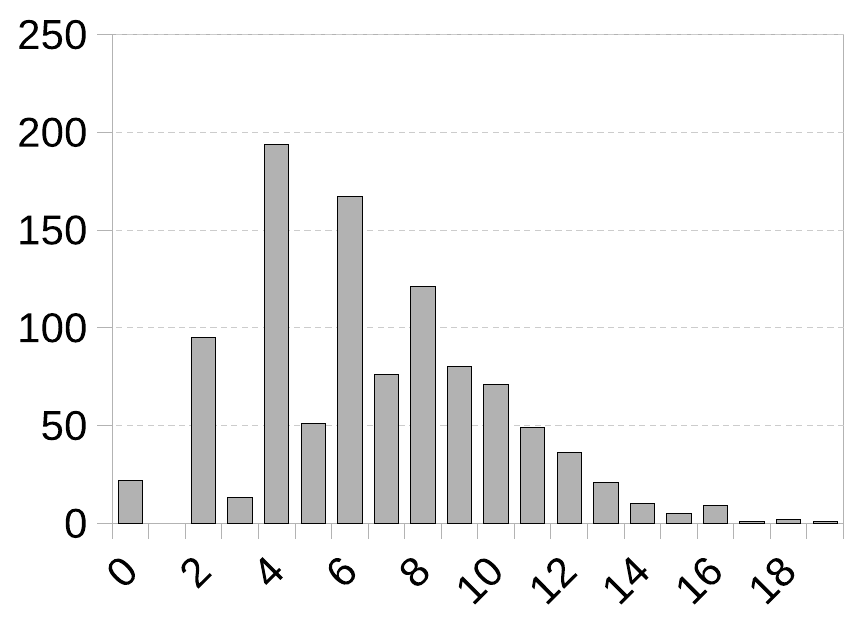}{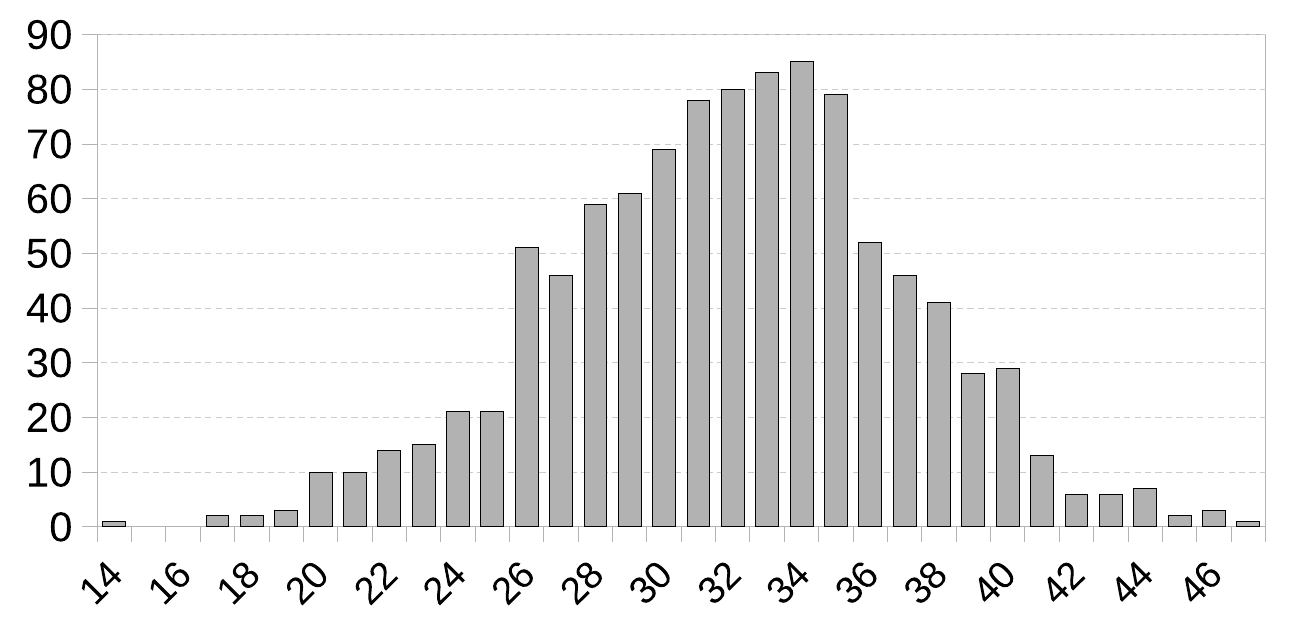}{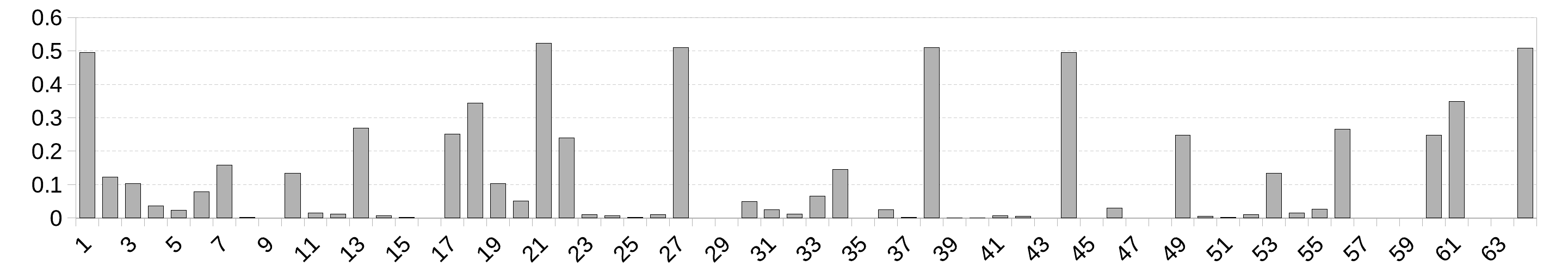}

\textbf{Misses, predicted:} Average = 6.706, Min = 0, Max = 19.

\textbf{Misses, random:} Average = 31.908, Min = 14, Max = 47.

Evidently, the inversion is simpler for shorter messages.

{\ }

\resultitem
5 rounds (every hash bit is able to change) are enough to make the limitations clear.

\textbf{CHF:} SHA1, 5 rounds, 160-bit message. Mask: full.

\textbf{NN:} 4 layers, 512 cells each. 1024 samples, 64 in batch.

\traintestcharts{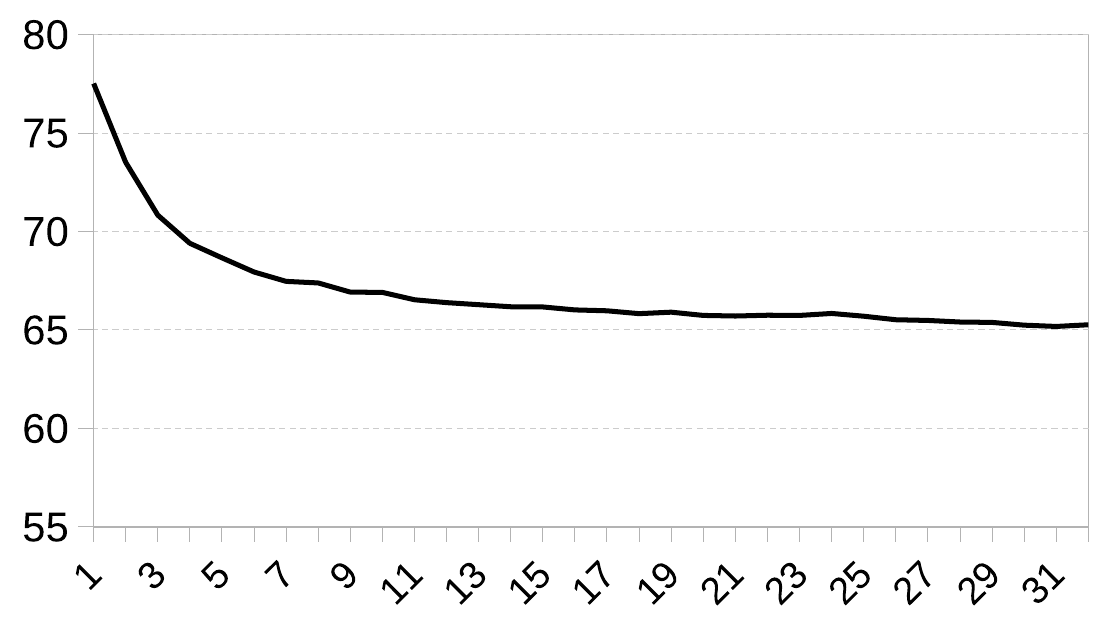}{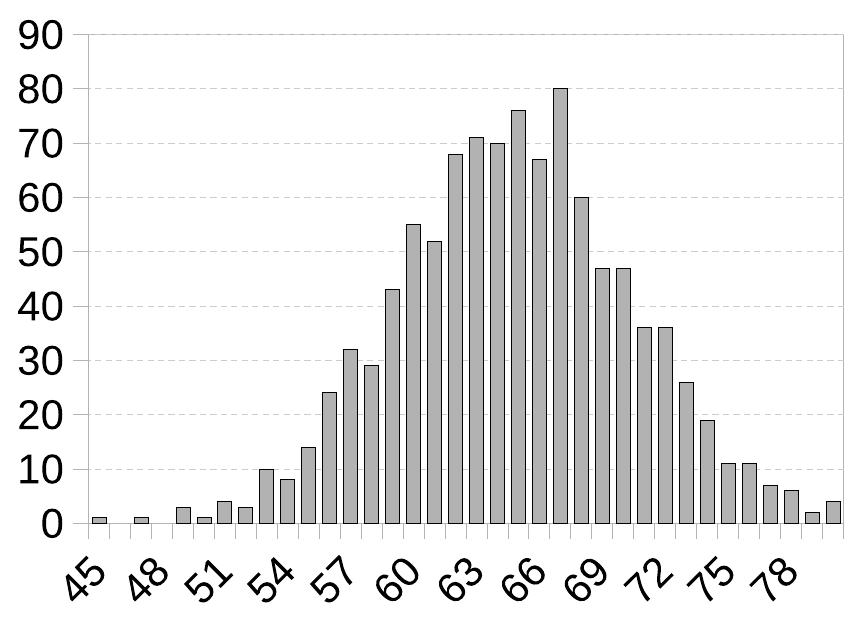}{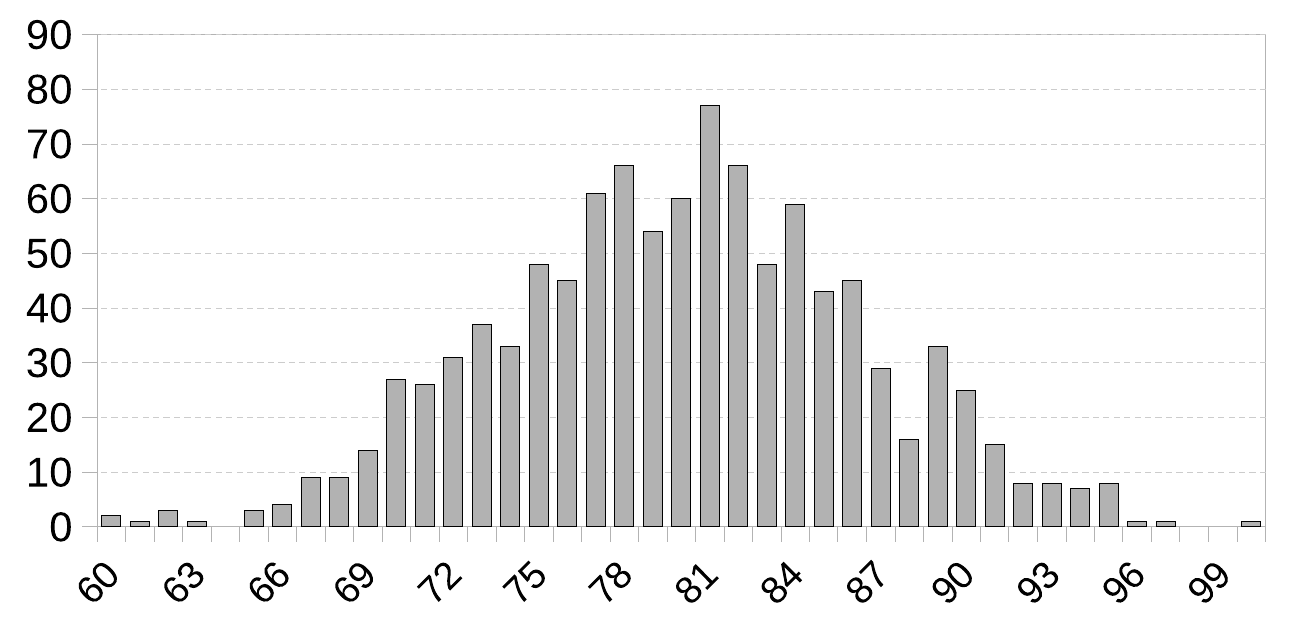}{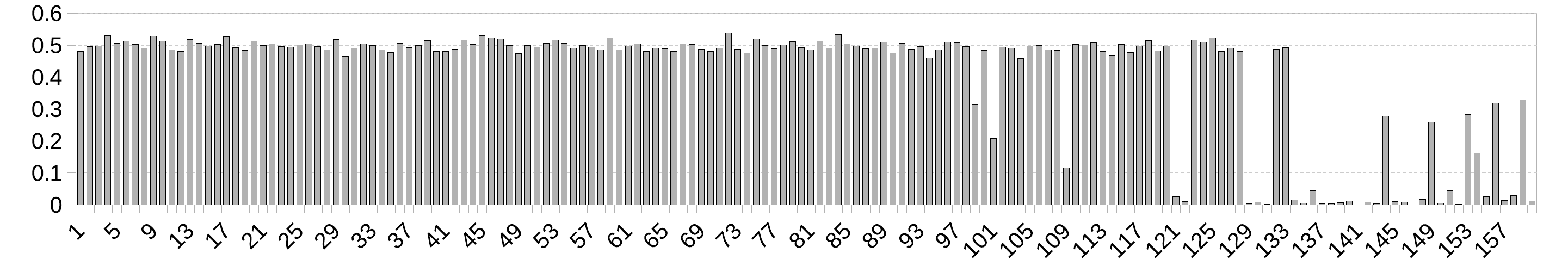}

\textbf{Misses, predicted:} Average = 64.787, Min = 45, Max = 80.

\textbf{Misses, random:} Mean = 79.980, Min = 60, Max = 100.

First 96 bits (3 double words out of 5) of the hash are too random-like for NN, it cannot produce the message to match them.

{\ }

\resultitem
As before, it's easier for shorter messages.

\textbf{CHF:} SHA1, 5 rounds, 32-bit message. Mask: full.

\textbf{NN:} 2 layers, 1024 cells each. 1024 samples, 64 in batch.

\traintestcharts{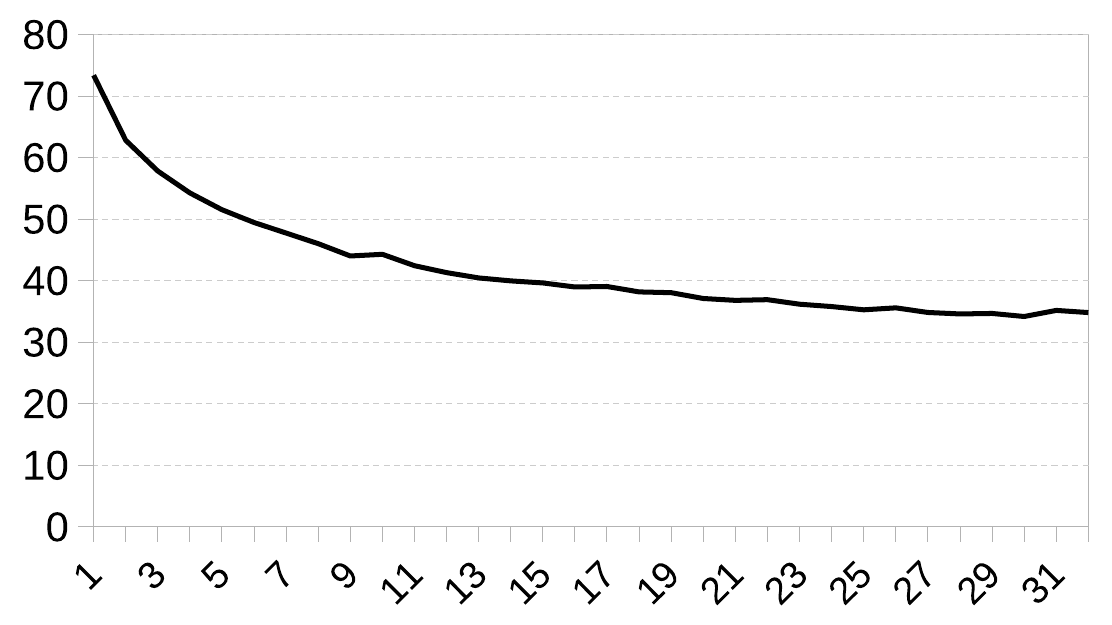}{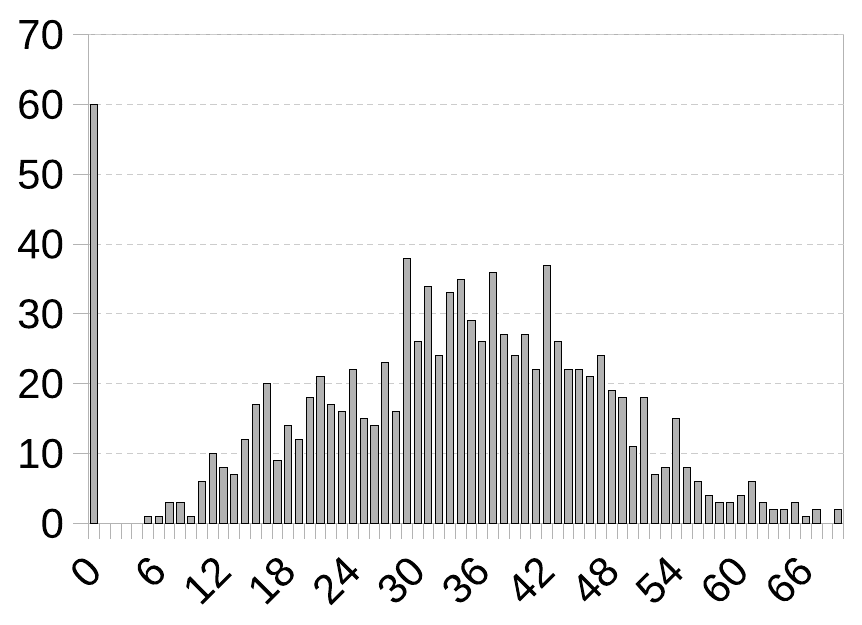}{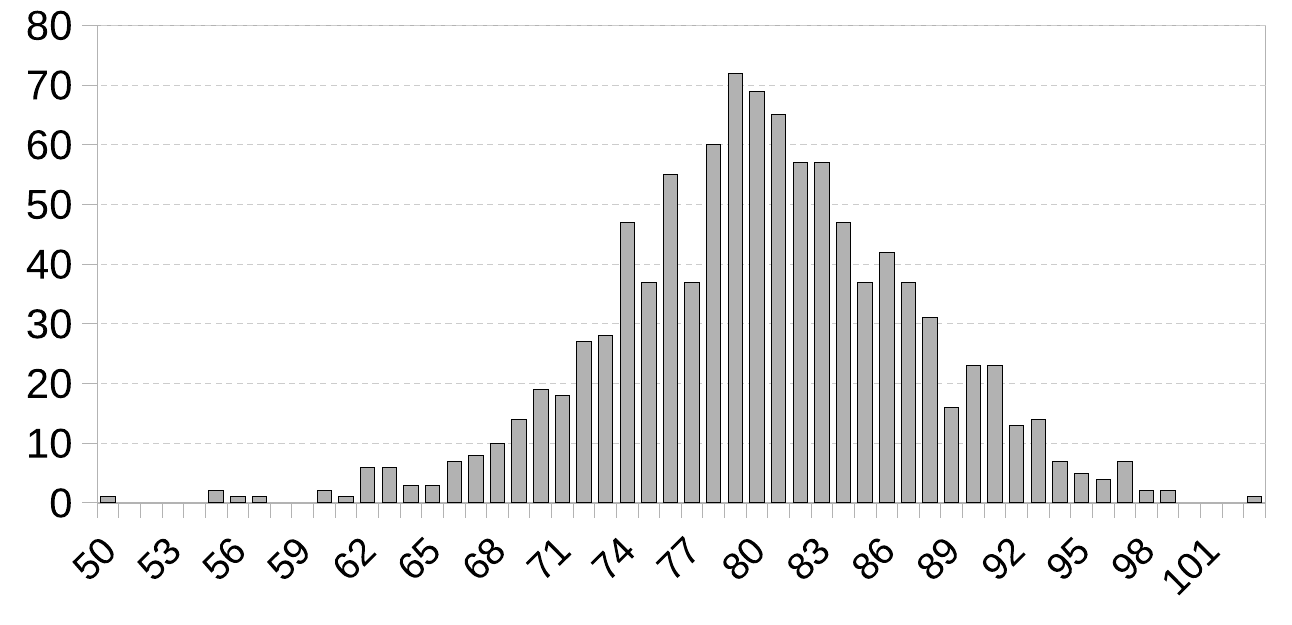}{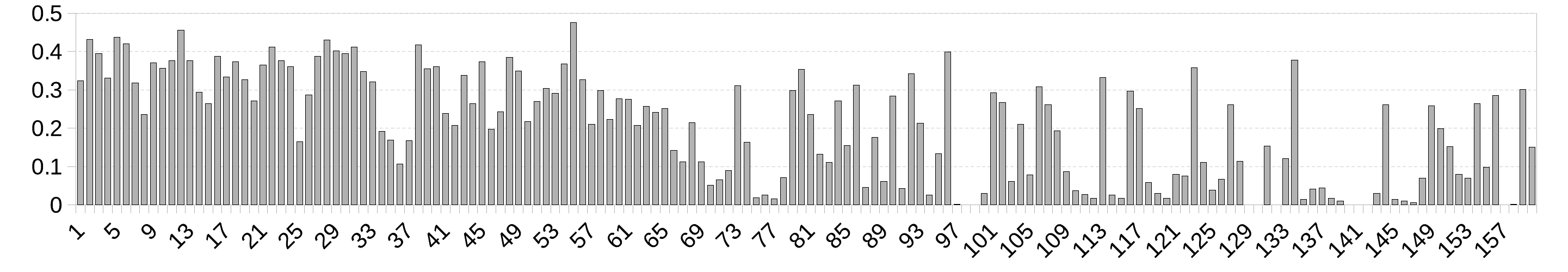}

\textbf{Misses, predicted:} Average = 32.622, Min = 0, Max = 69.

\textbf{Misses, random:} Average = 80.177, Min = 50, Max = 103.

Remarkably, NN exactly inverted 60 hashes out of 1024,
which is at least 1.5 times more than for any other single number of misses.

{\ }

\hrule

{\ }

7--8 rounds for the messages longer than 160 bits are impassable:
the average number of misses is 80 or so, same as if NN were guessing the messages randomly.

{\ }

\resultitem
Next result shows that $\fzADD$ op is the main obstacle for NN.
Here we deal with weakened SHA1 rounds, where all $\fzADD$s are replaced by $\fzXOR$s, which reduces the diffusion greatly.

\textbf{CHF:} SHA1 with $\fzADD$ replaced by $\fzXOR$, 2 rounds, 64-bit message. Mask: full.

\textbf{NN:} 2 layers, 1024 cells each. 1024 samples, 64 in batch.

\traintestcharts{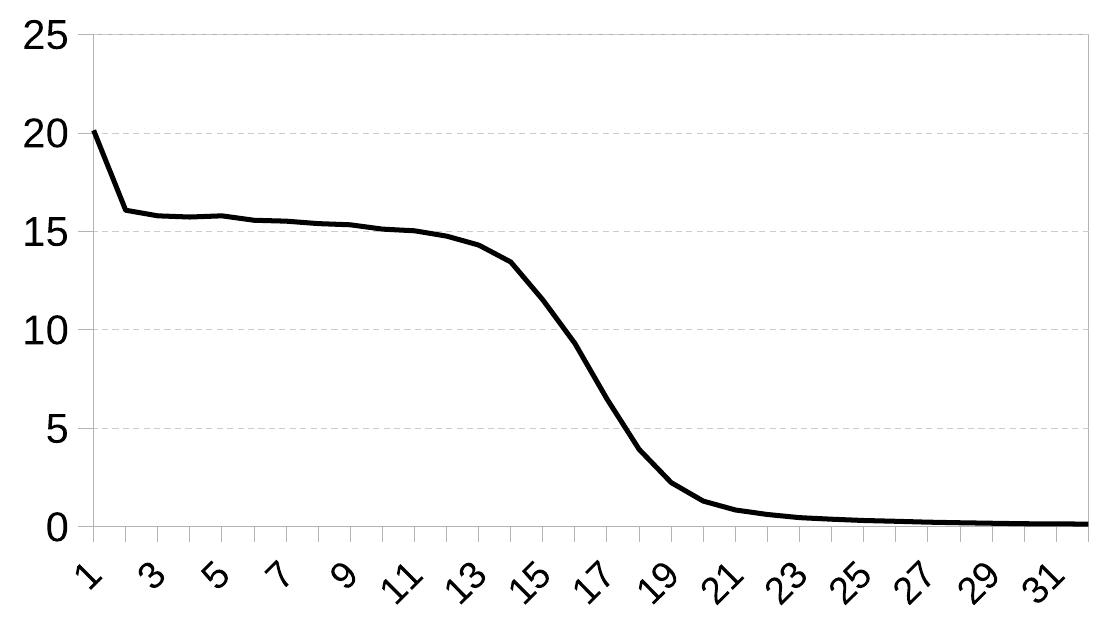}{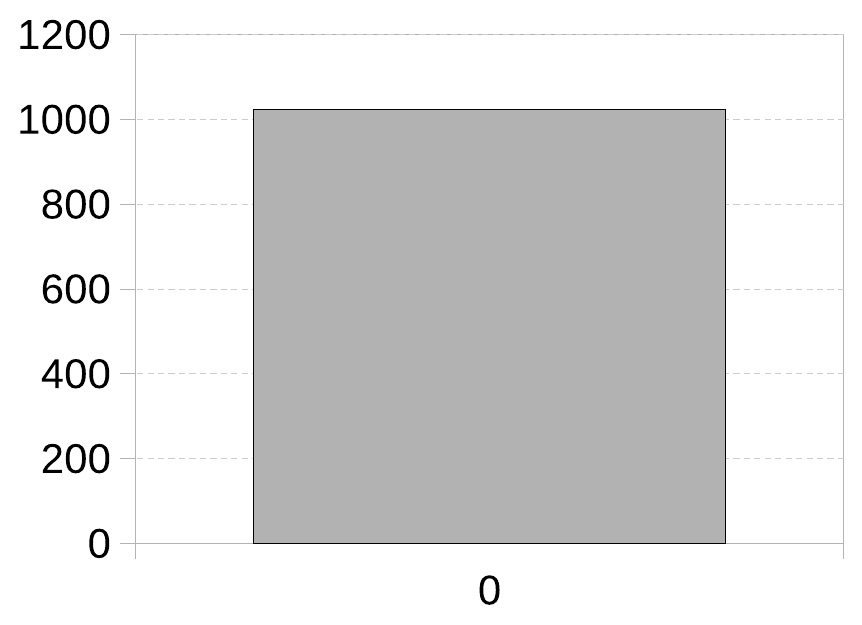}{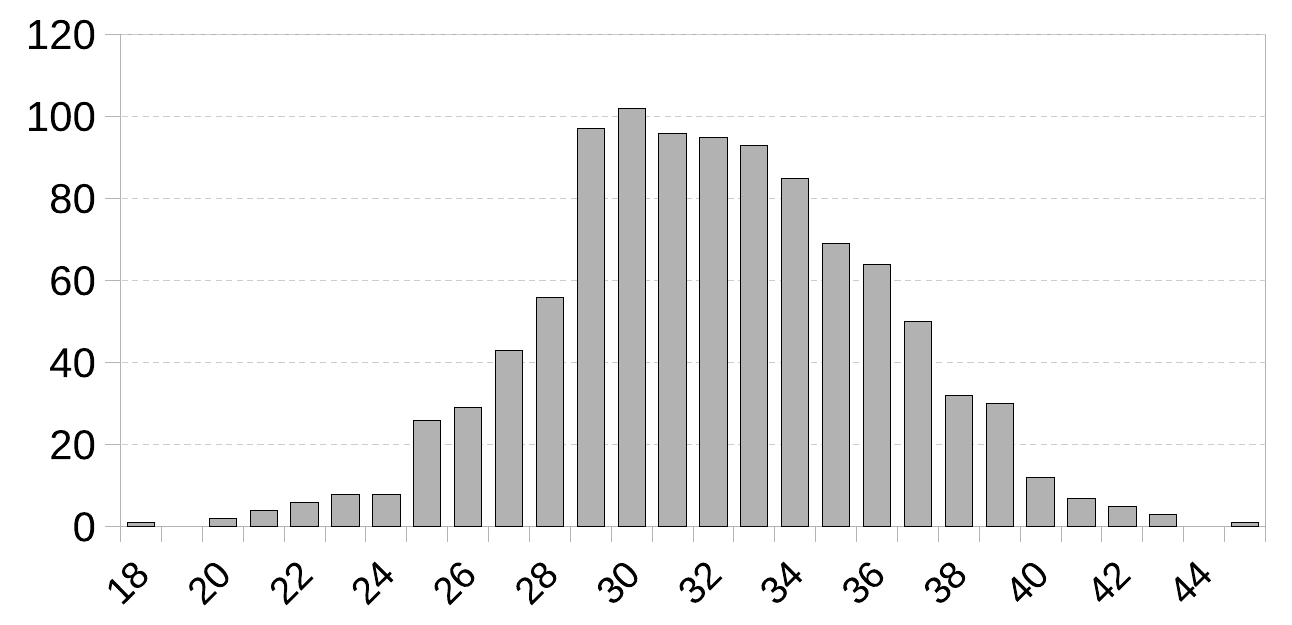}{}

(All miss frequencies are 0, there were no wrong bits.)

\textbf{Misses, predicted:} Average = Min = Max = 0.0.

\textbf{Misses, random:} Average = 31.97, Min = 18, Max = 45.

NN has learnt to invert the hash with 100\% accuracy; compare it with Res. \ref{resSha1rounds2}.

{\ }

\hrule

{\ }

Also, the difficulty depends on the constants $H_i$ and $K_i$ from the SHA1 rounds.
When we set the majority of their bits to 0, or made their bit structure more regular (like \texttt{0x13131313}),
the accuracy of inversion increased (with $\fzADD$ op).

{\ }

\resultitem
When it's too hard to match full hash, NN may be able to match a small group of bits.

\textbf{CHF:} SHA1, 3 rounds, 96-bit message. Mask: bits 0--7.

\textbf{NN:} 3 layers, 512 cells each. 2048 samples, 128 in batch.

\traintestcharts{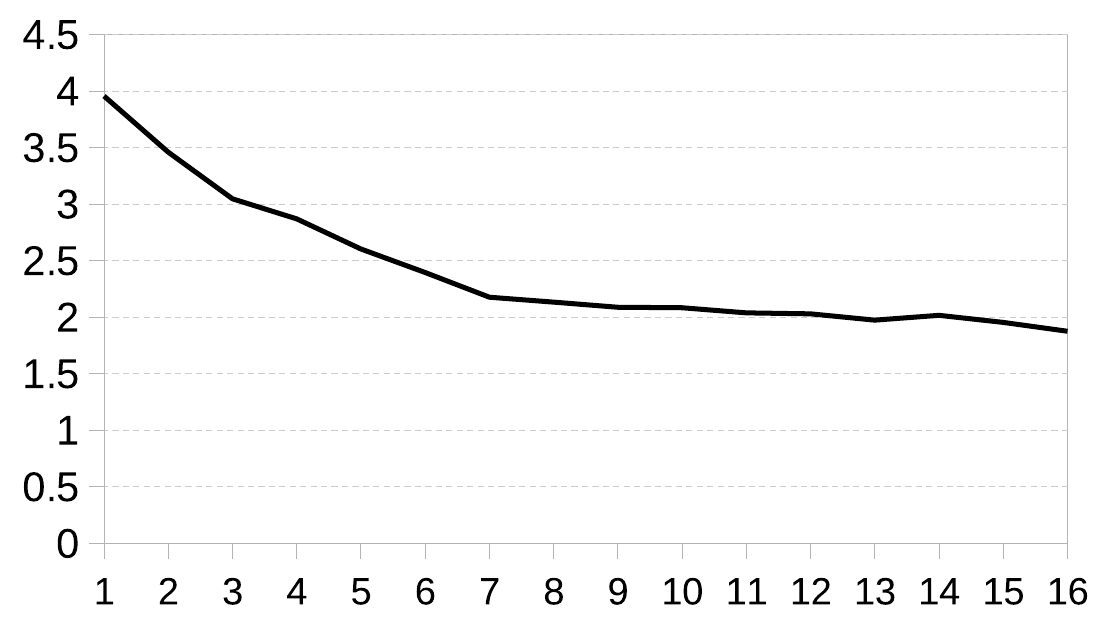}{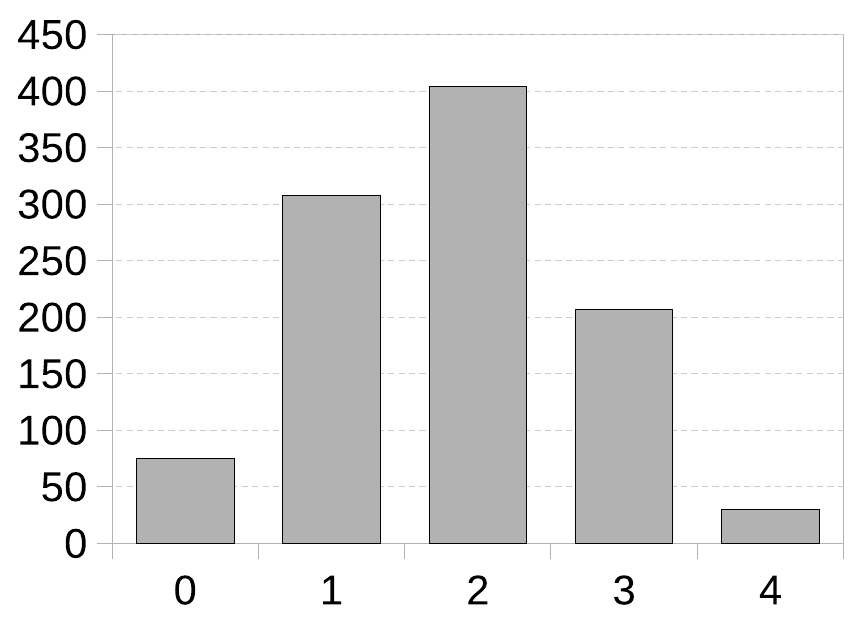}{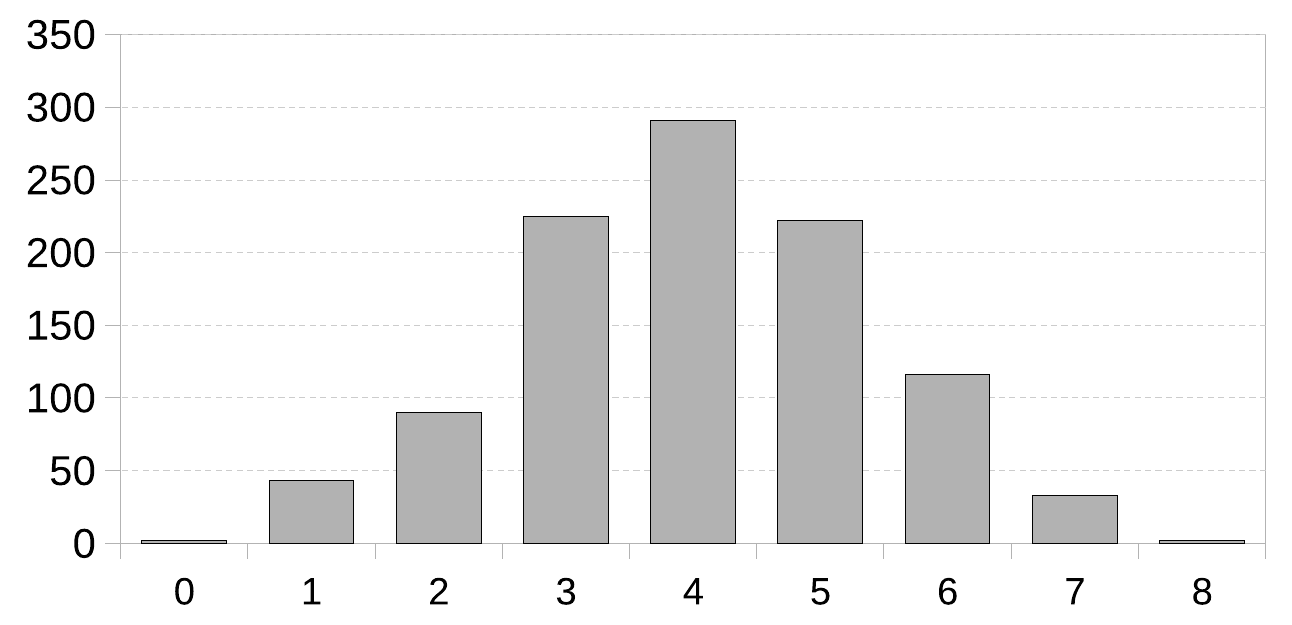}{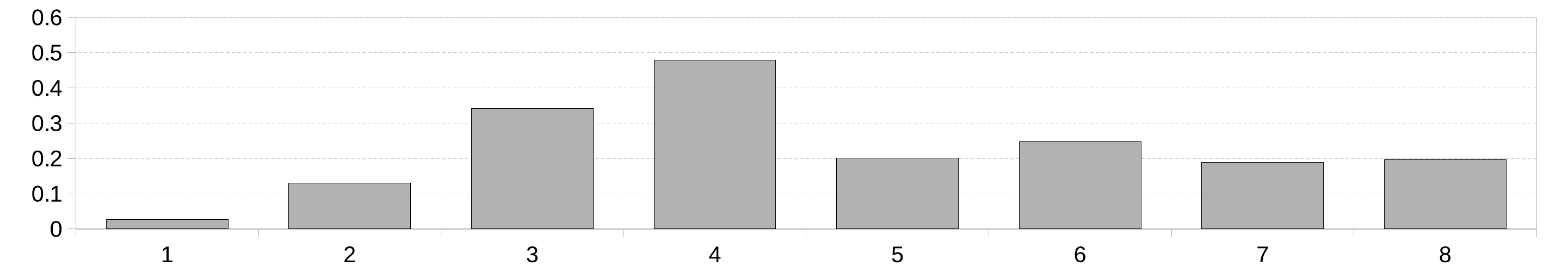}

\textbf{Misses, predicted:} Average = 1.813, Min = 0, Max = 4.

\textbf{Misses, random:} Average = 4.019, Min = 0, Max = 8.

{\ }

\resultitem
Now we try to invert the state transformation during 1st round. The input is 160 bits (five 32-bit words) of the state
after 1st round, NN should predict 160 bits of the state + 32-bit message word before 1st round;
also, there is no addition of initial constant values.

\textbf{CHF:} SHA1, 1st round state transformation, 160+32-bit state+message. Mask: full.

\textbf{NN:} 2 layers, 1024 cells each. 1024 samples, 64 in batch.

\traintestcharts{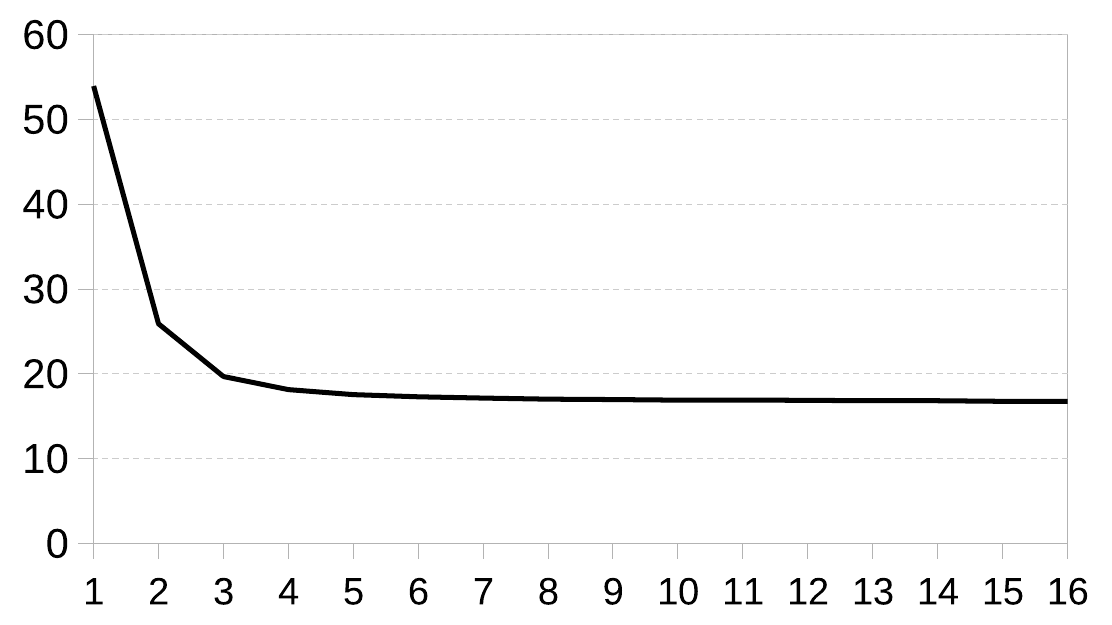}{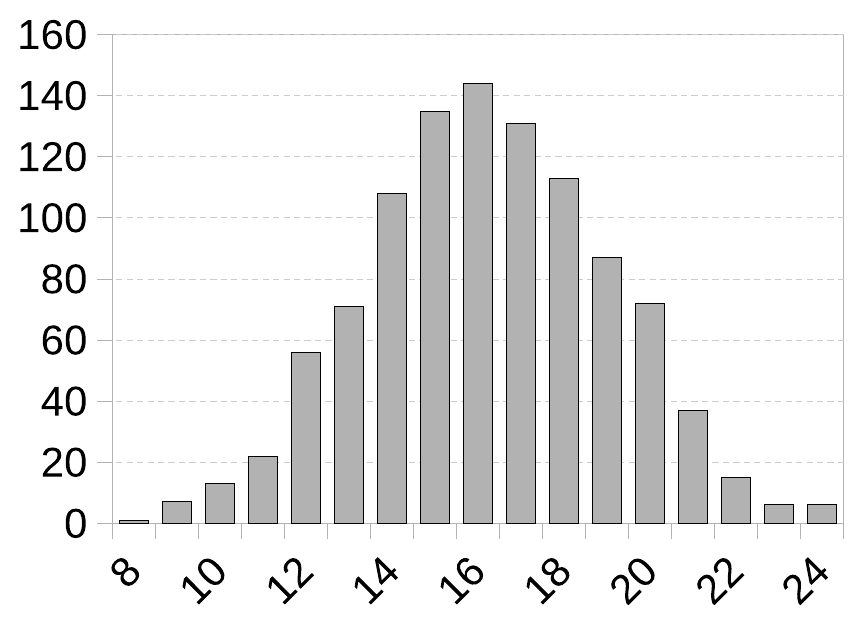}{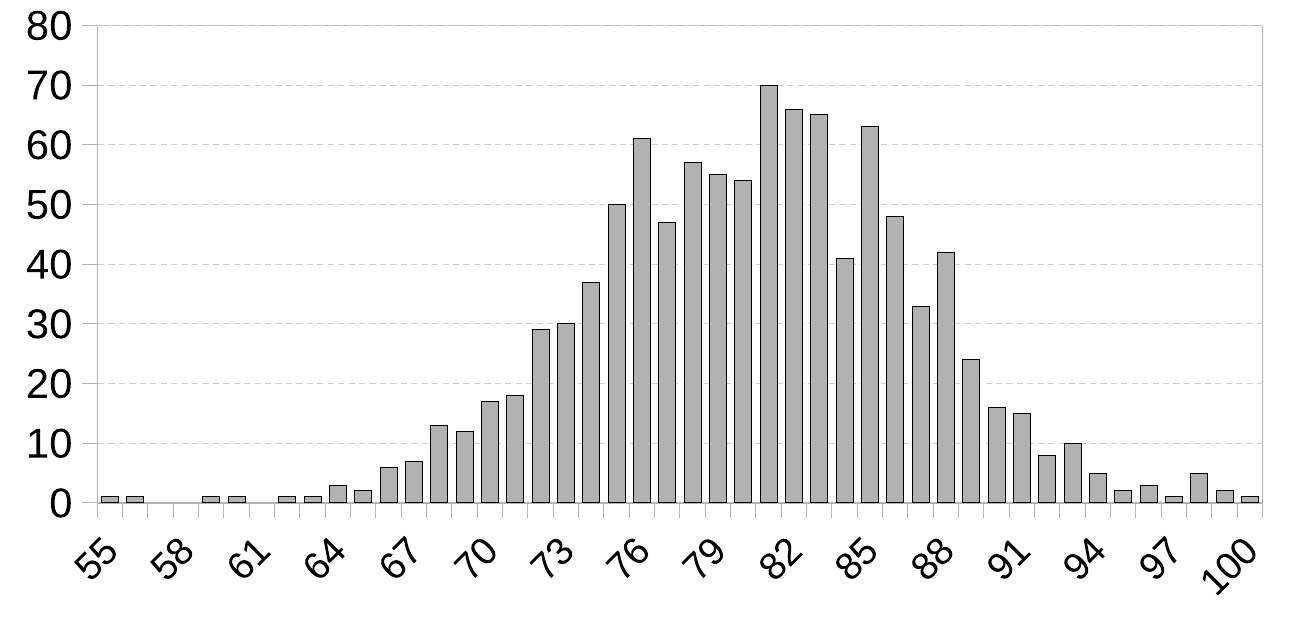}{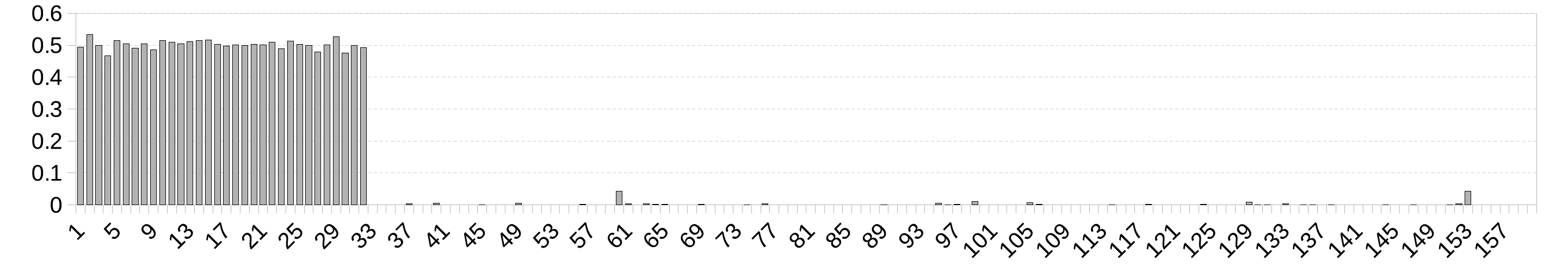}

\textbf{Misses, predicted:} Average = 16.232, Min = 8, Max = 24.

\textbf{Misses, random:} Average = 80.337, Min = 55, Max = 100.

First 32-bit word of the state, which is where the most changes happen, is too random for NN to match.

{\ }

\resultitem
However, for at least one byte of the state the partial inversion is possible.

\textbf{CHF:} SHA1, 1st round state transform, 160+32-bit state+message. Mask: bits 0--7.

\textbf{NN:} 3 layers, 512 cells each. 2048 samples, 128 in batch.

\traintestcharts{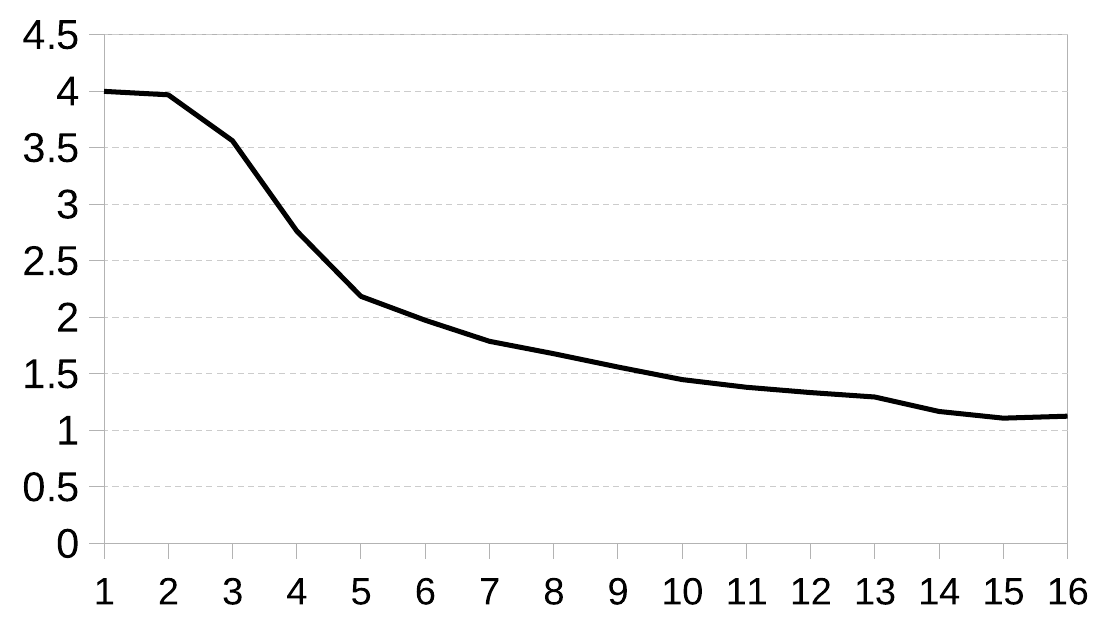}{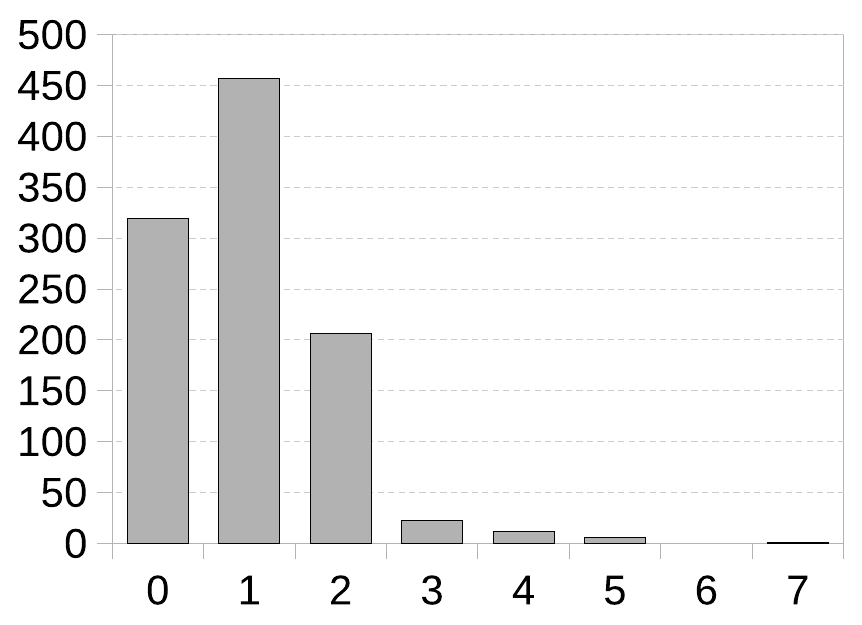}{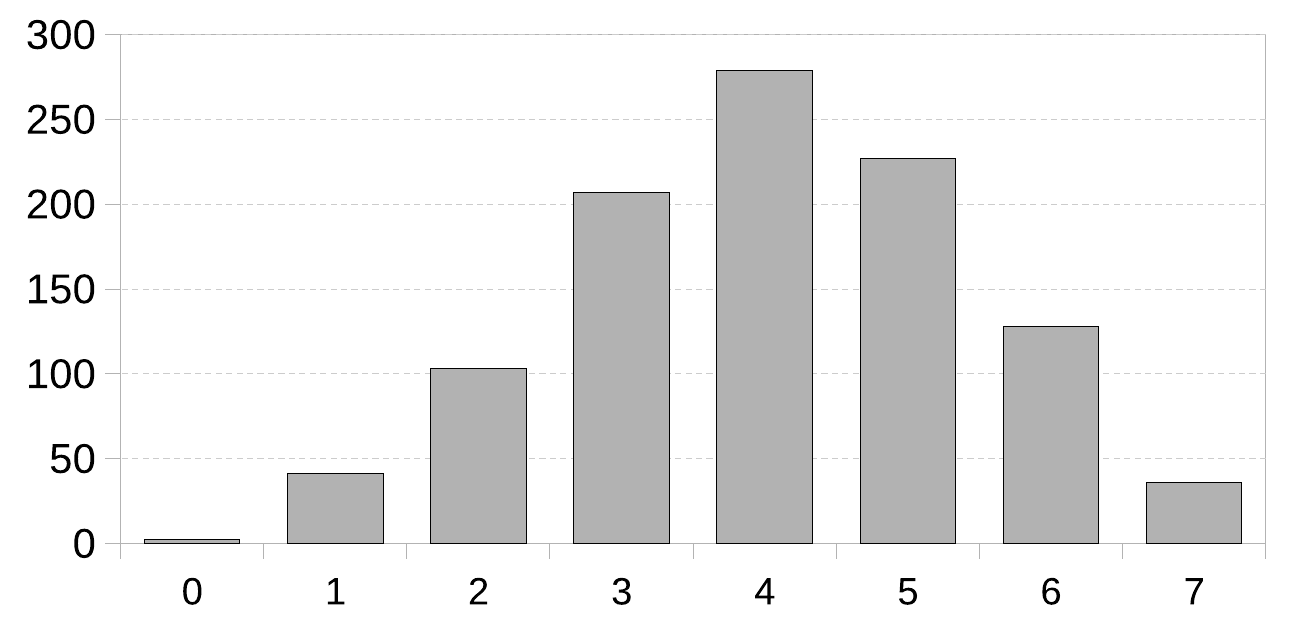}{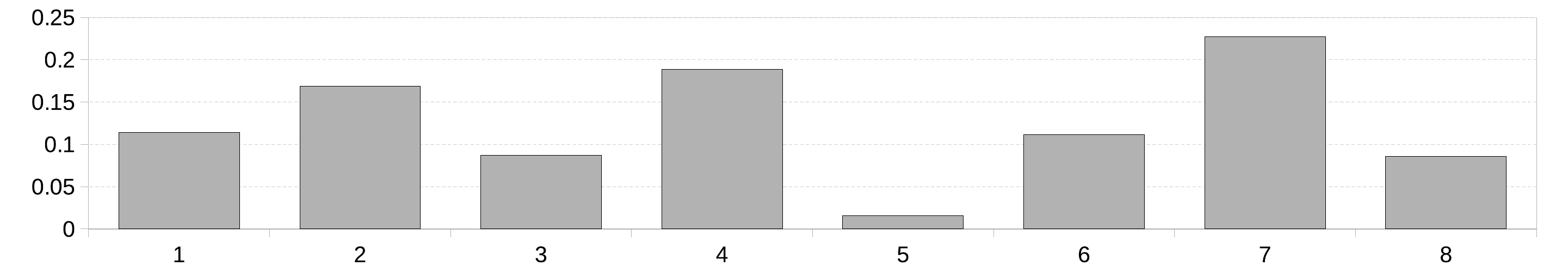}

\textbf{Misses, predicted:} Average = 0.999, Min = 0, Max = 7.

\textbf{Misses, random:} Average = 4.0498, Min = 0, Max = 8.

\subsection{MD5}

\resultitem
Surprisingly, the 1st round of MD5 is tougher for NN than that of SHA1.

\textbf{CHF:} MD5, 1 round, 32-bit message. Mask: full.

\textbf{NN:} 1024 cells in 1 layer. 1024 samples, 64 in batch.

\traintestcharts{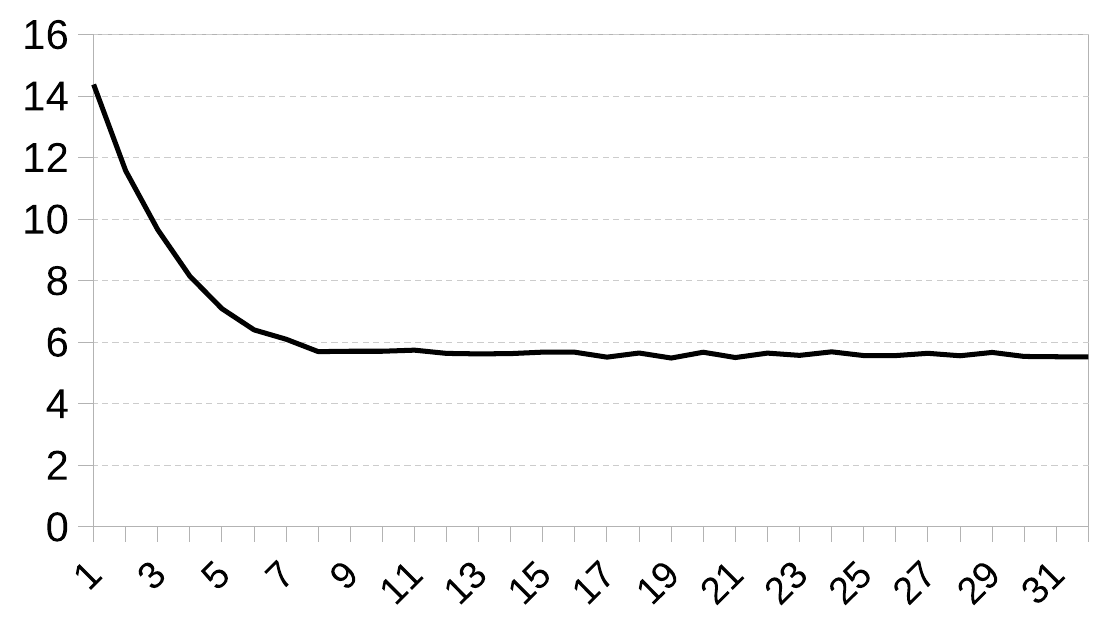}{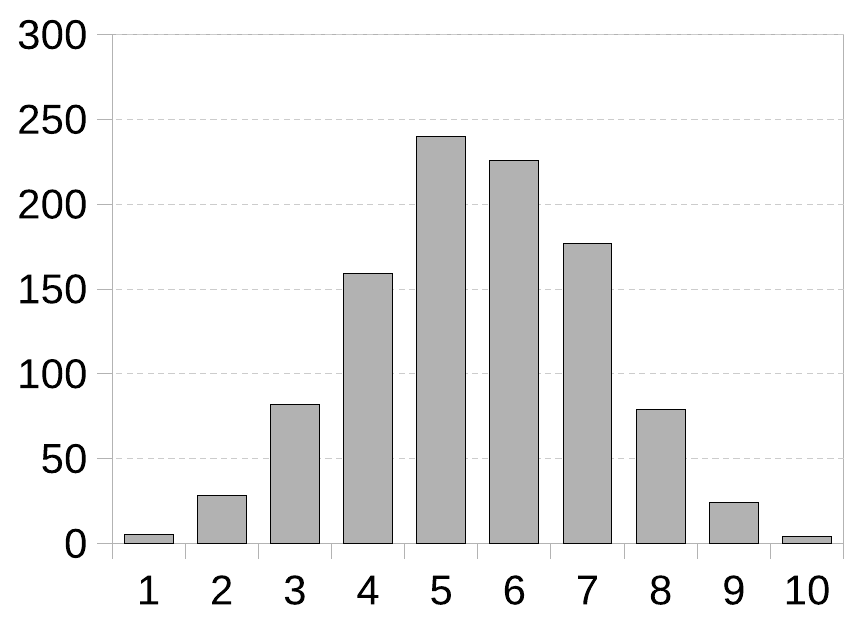}{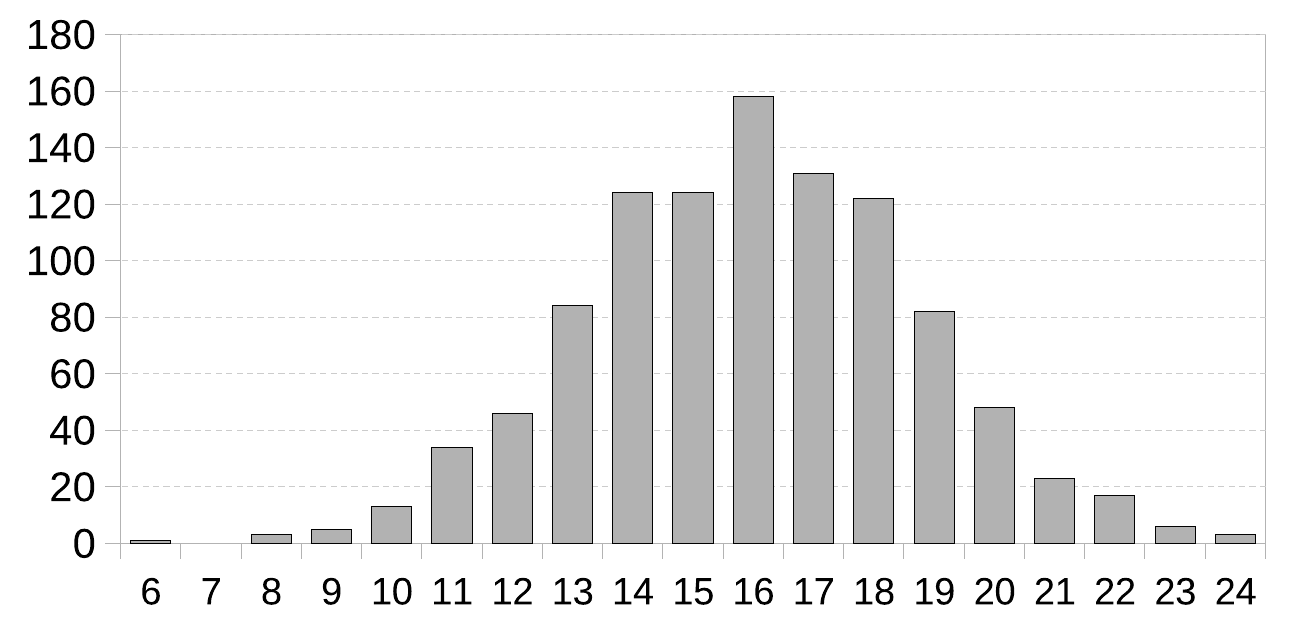}{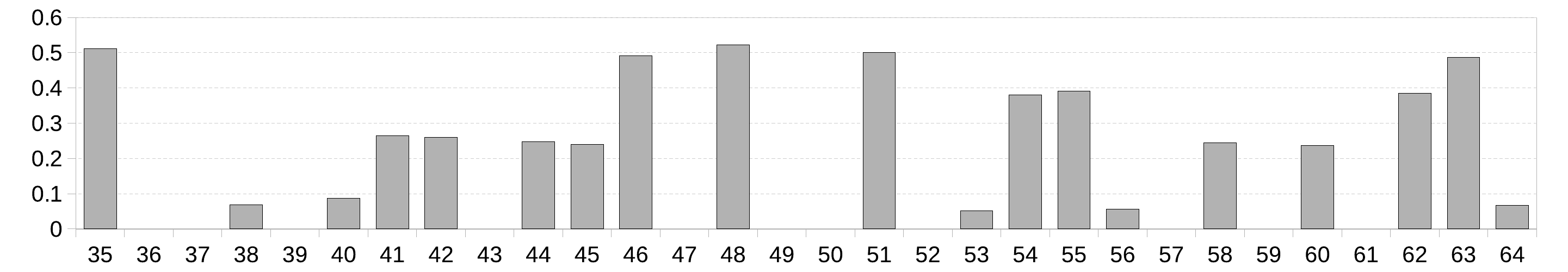}

\textbf{Misses, predicted:} Average = 5.494, Min = 1, Max = 10.

\textbf{Misses, random:} Average = 15.972, Min = 6 , Max = 24.

{\ }

\resultitem
\textbf{CHF:} MD5, 2 rounds, 64-bit message. Mask: full.

\textbf{NN:} 1024 cells in 2 layers. 1024 samples, 64 in batch.

\traintestcharts{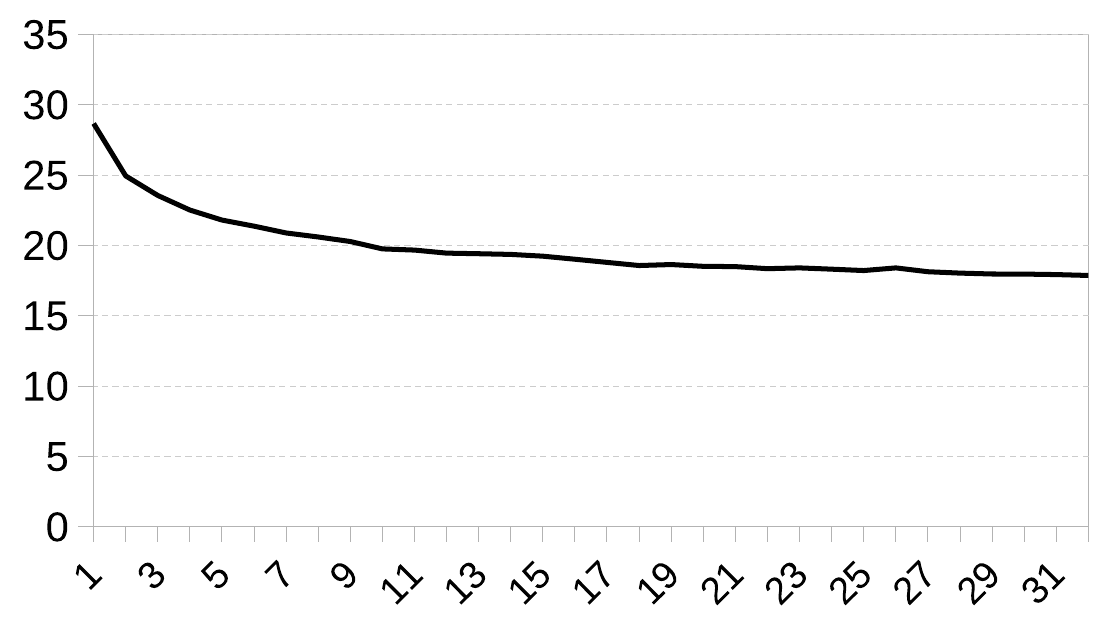}{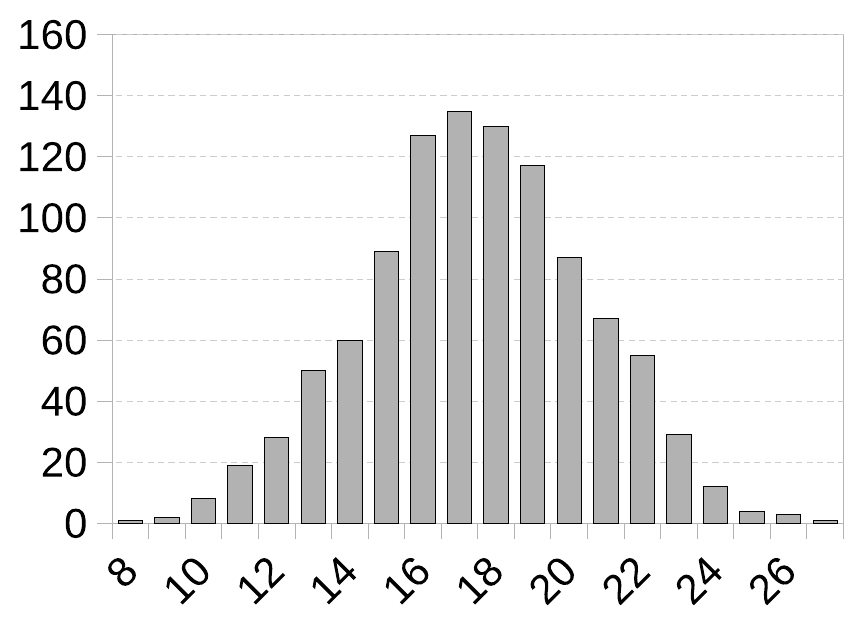}{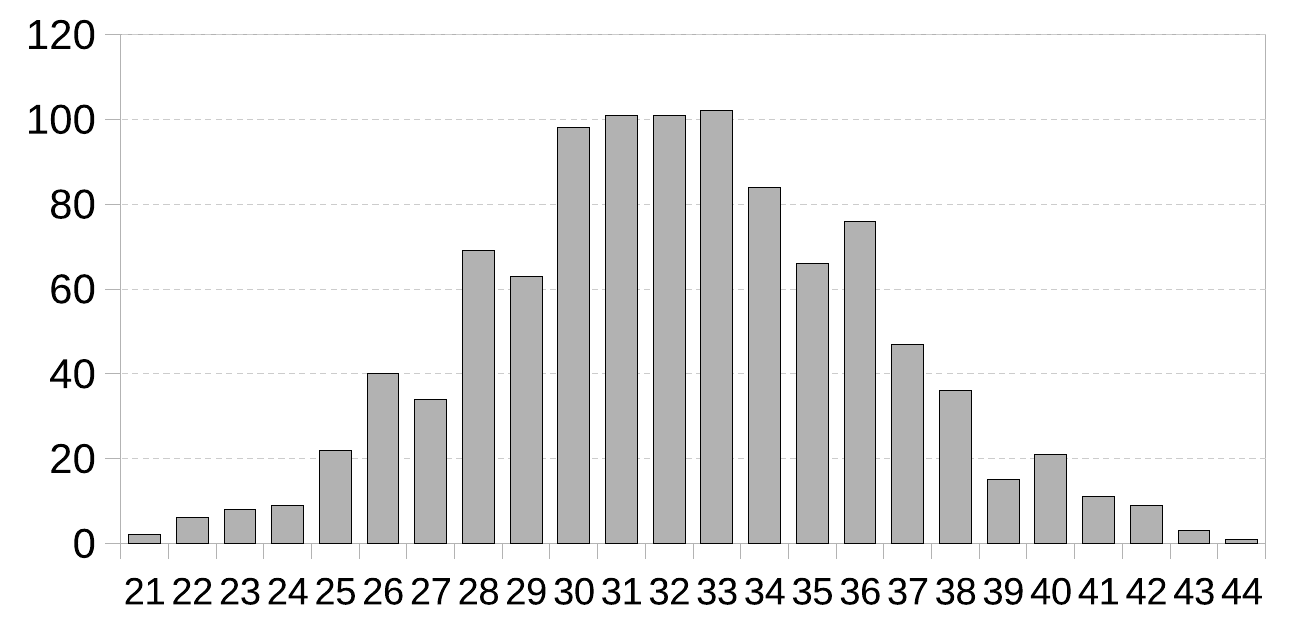}{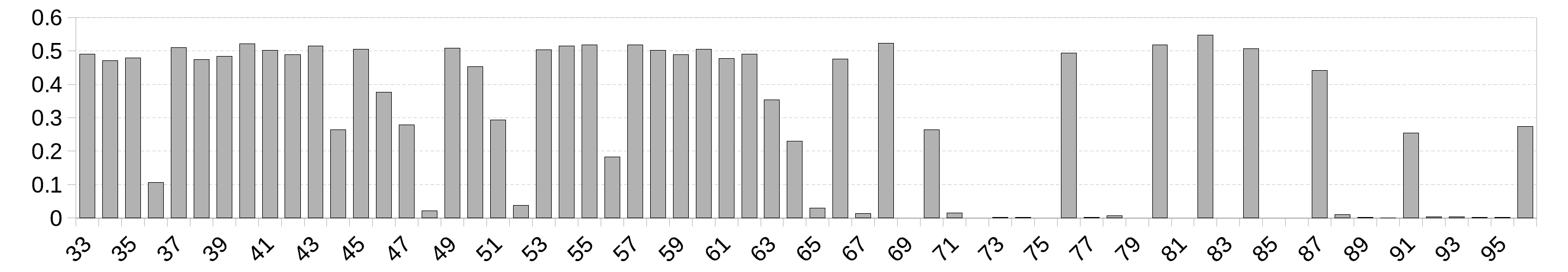}

\textbf{Misses, predicted:} Average = 17.464, Min = 8, Max = 27.

\textbf{Misses, random:} Average = 32.174, Min = 21, Max = 44.

{\ }

\resultitem
\textbf{CHF:} MD5, 4 rounds, 128-bit message. Mask: full.

\textbf{NN:} 1024 cells in 2 layers. 1024 samples, 64 in batch.

\traintestcharts{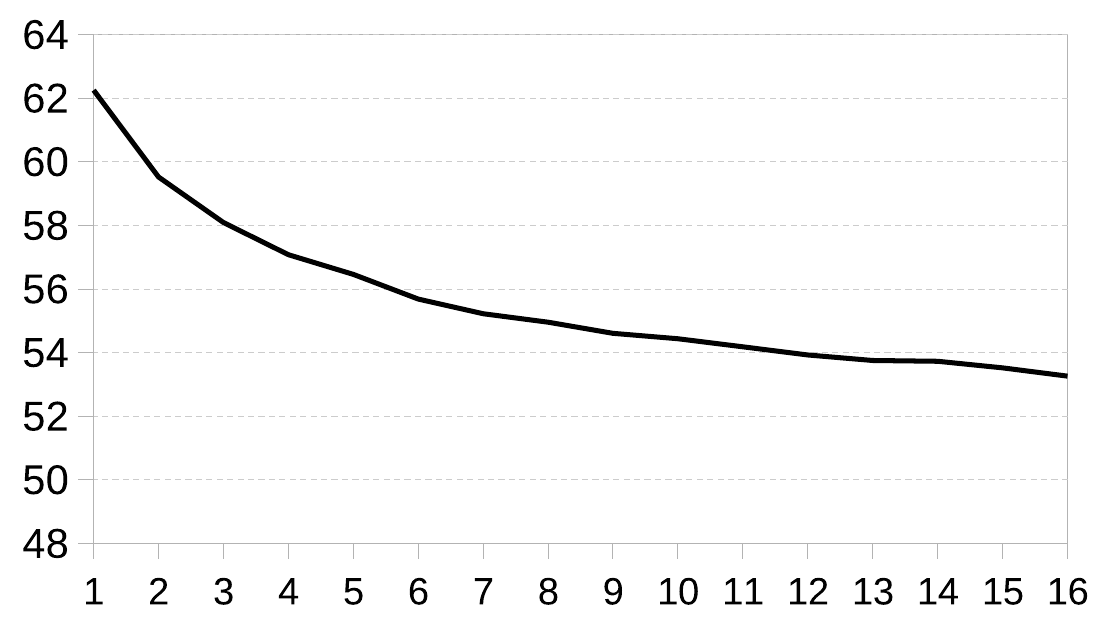}{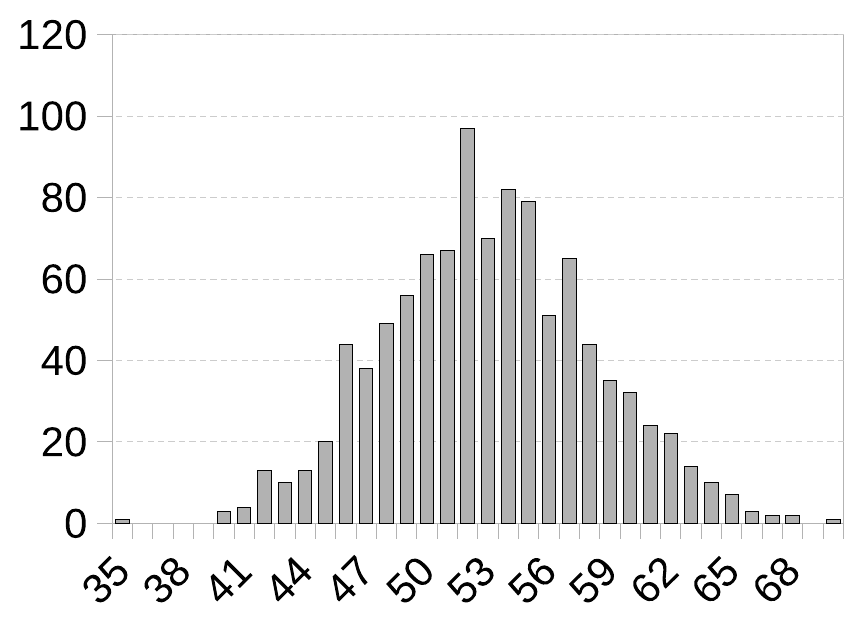}{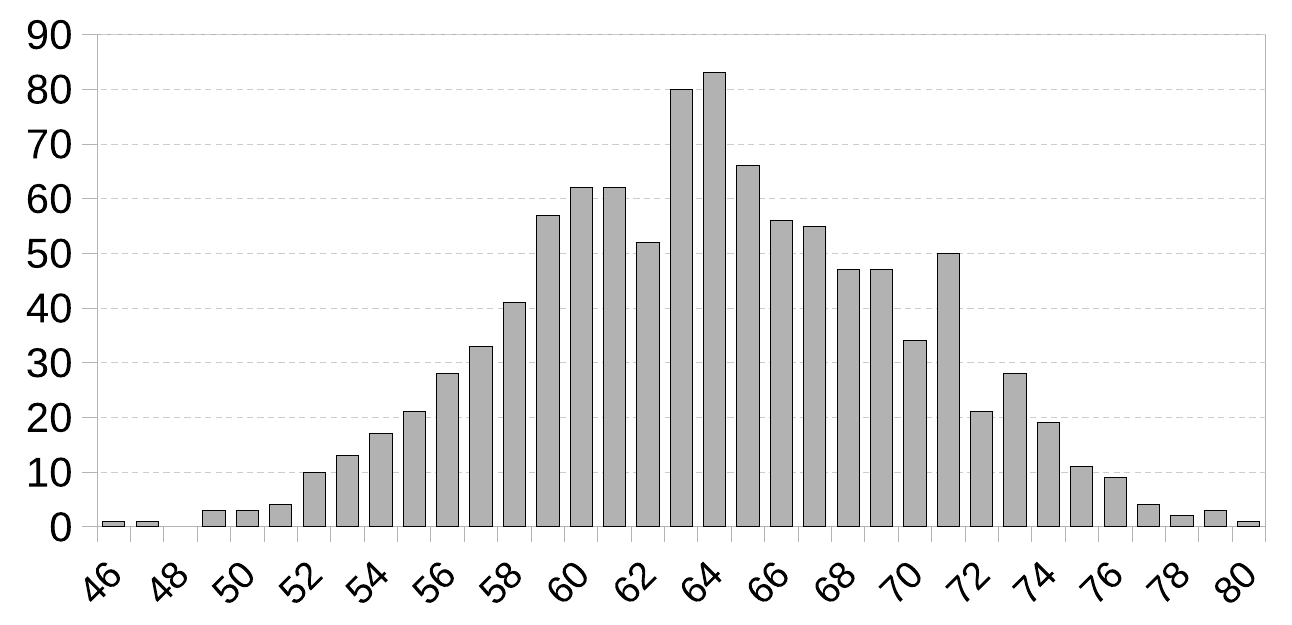}{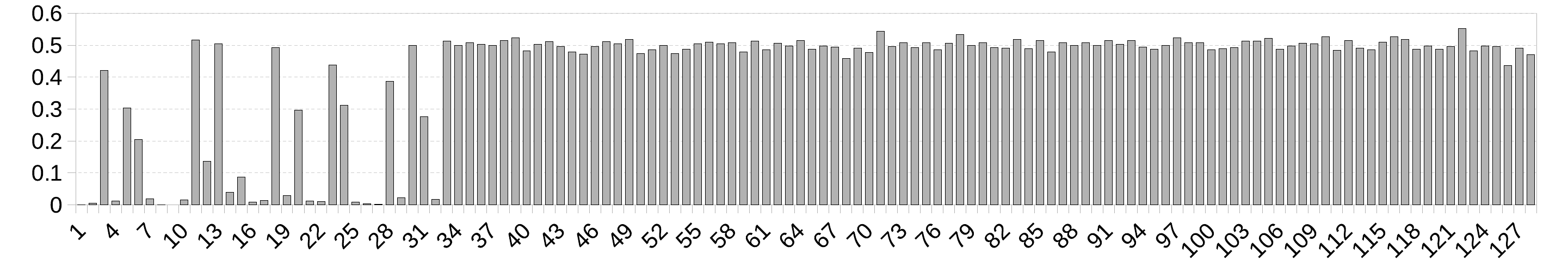}

\textbf{Misses, predicted:} Average = 53.084, Min = 35, Max = 70.

\textbf{Misses, random:} Average = 63.902, Min = 46, Max = 80.

\subsection{SHA2}

\resultitem
\textbf{CHF:} SHA2-256, 1 round, 32-bit message. Mask: full.

\textbf{NN:} 2 layers, 1024 cells each. 1024 samples, 64 in batch.

\traintestcharts{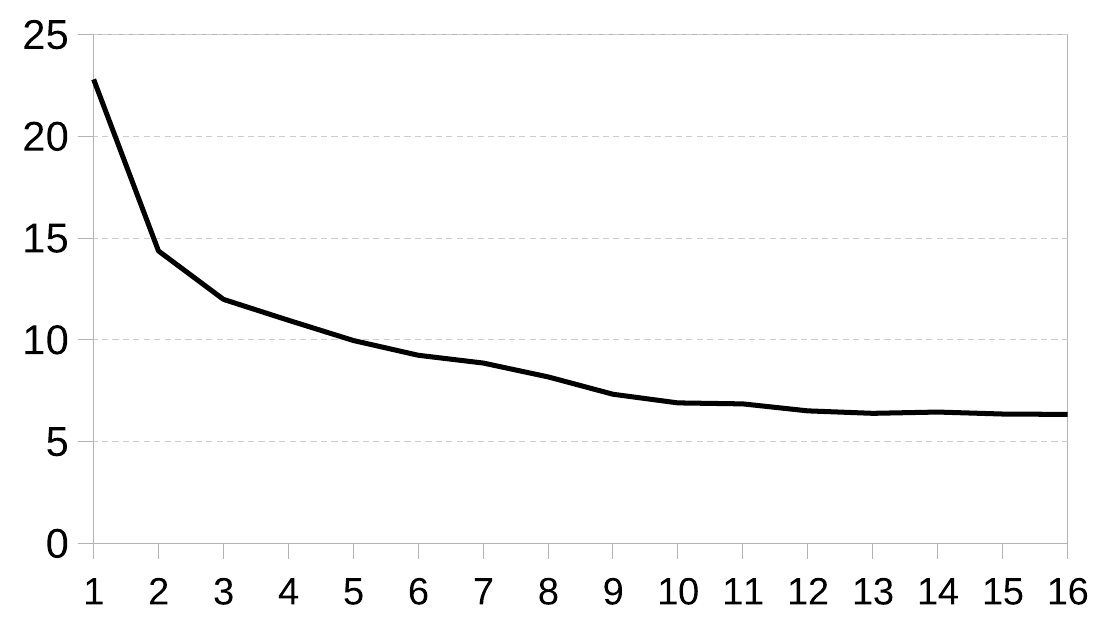}{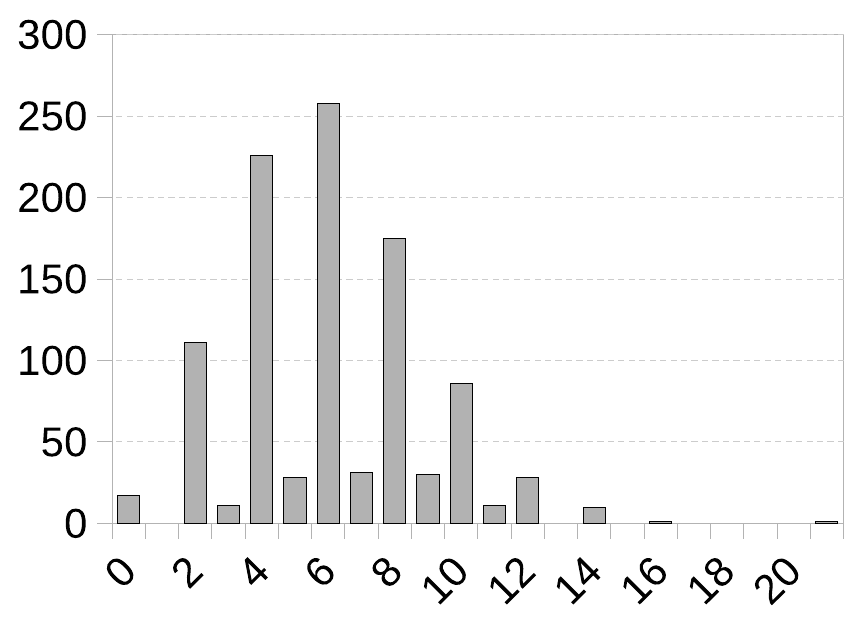}{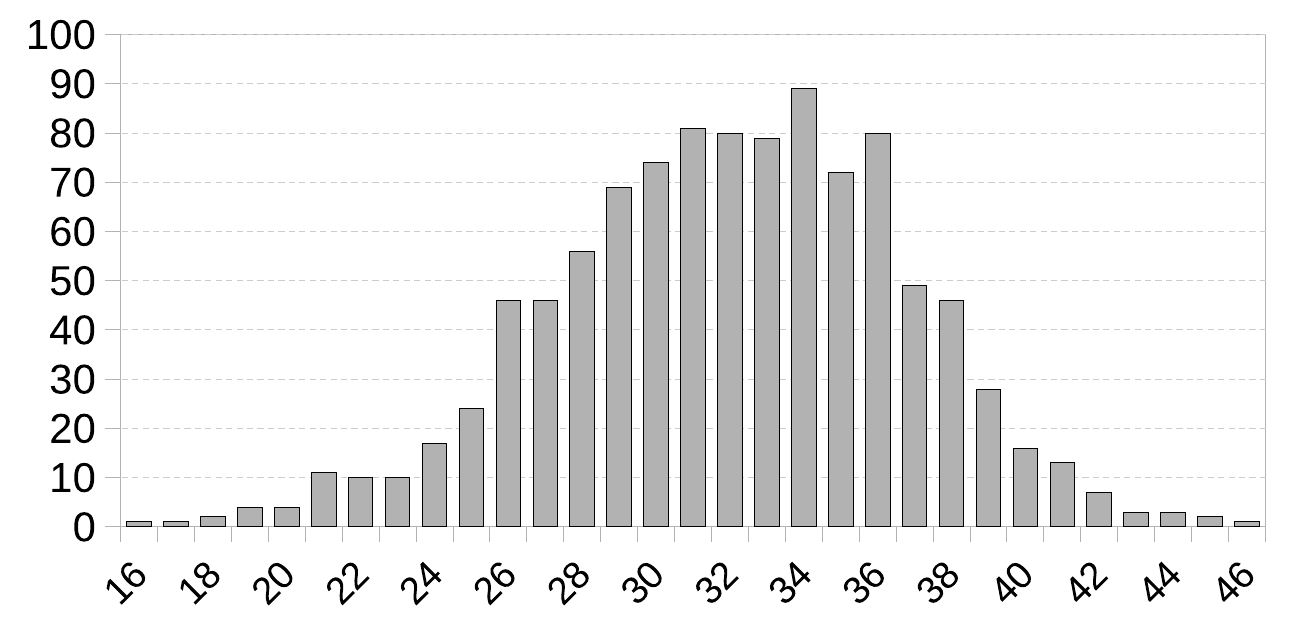}{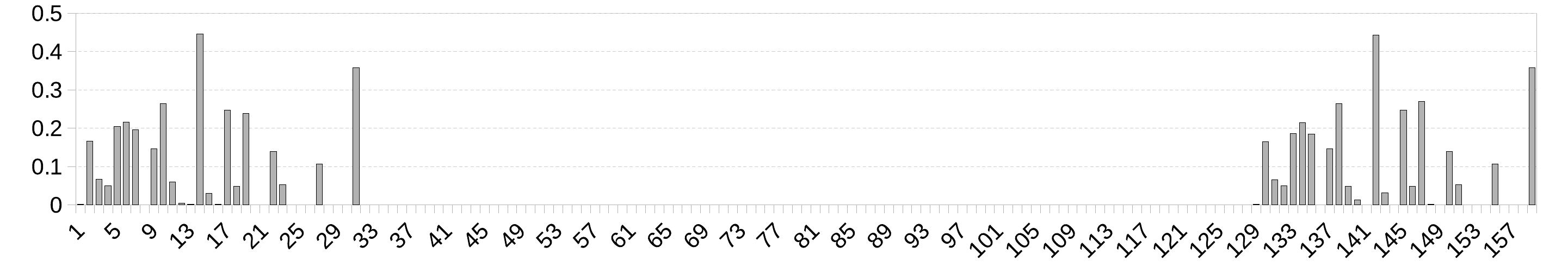}

\textbf{Misses, predicted:} Average = 6.082, Min = 0, Max = 21.

\textbf{Misses, random:} Average = 31.994, Min = 16, Max = 46.

Note that the number of misses is more probably even than odd.

{\ }

\resultitem
\textbf{CHF:} SHA2-256, 2 rounds, 64-bit message. Mask: full.

\textbf{NN:} 2 layers, 1024 cells each. 1024 samples, 64 in batch.

\traintestcharts{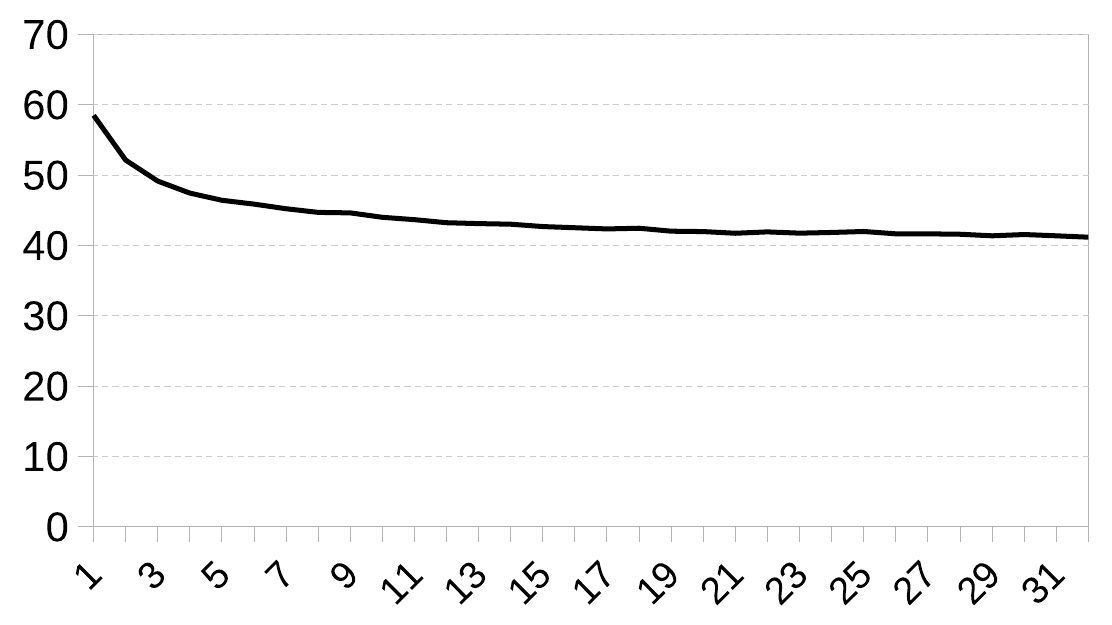}{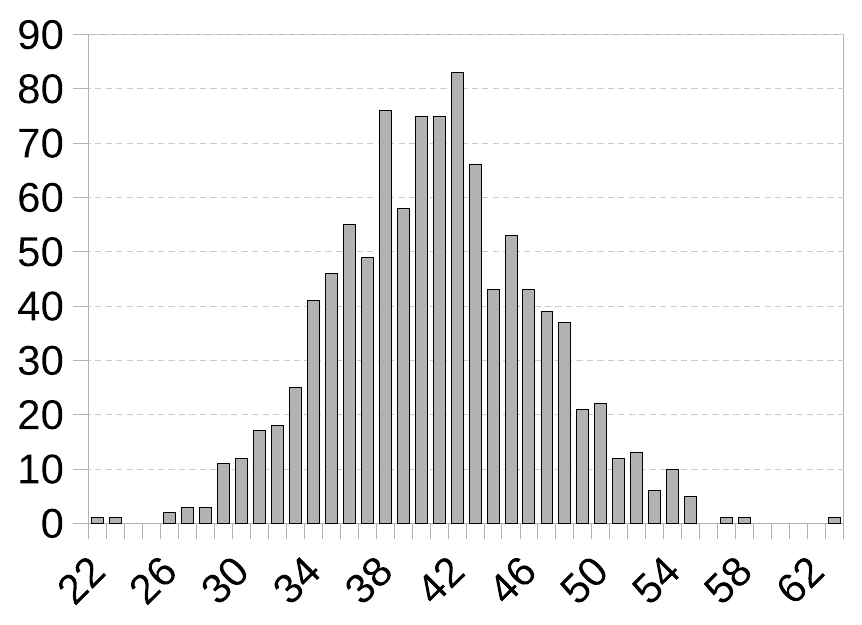}{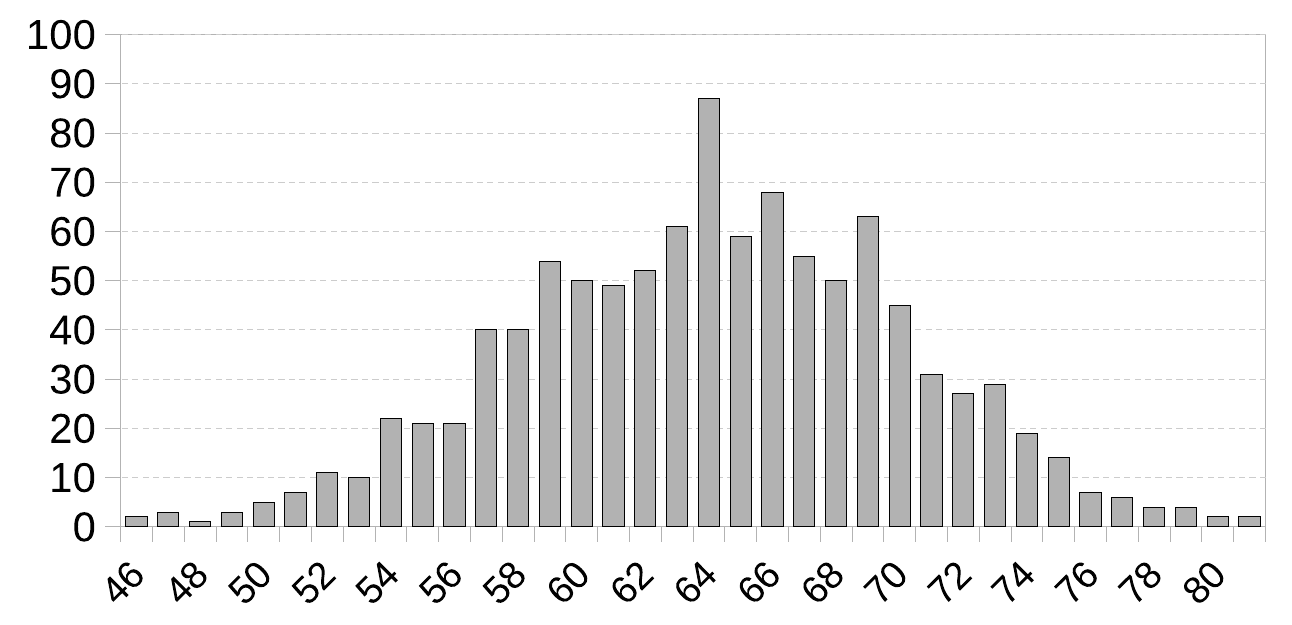}{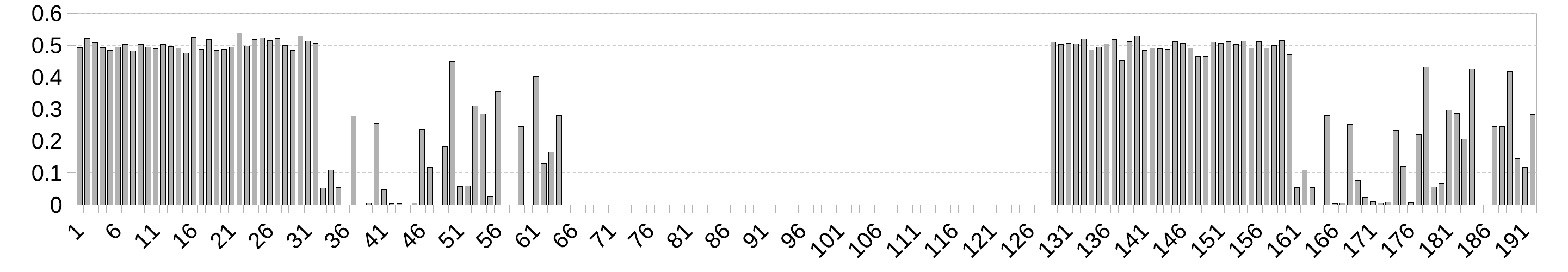}

\textbf{Misses, predicted:} Average = 40.827, Min = 22, Max = 63.

\textbf{Misses, random:} Average = 64.095, Min = 46, Max = 81.

{\ }

\resultitem
\textbf{CHF:} SHA2-256, 4 rounds, 128-bit message. Mask: full.

\textbf{NN:} 4 layers, 512 cells each. 1024 samples, 64 in batch.

\traintestcharts{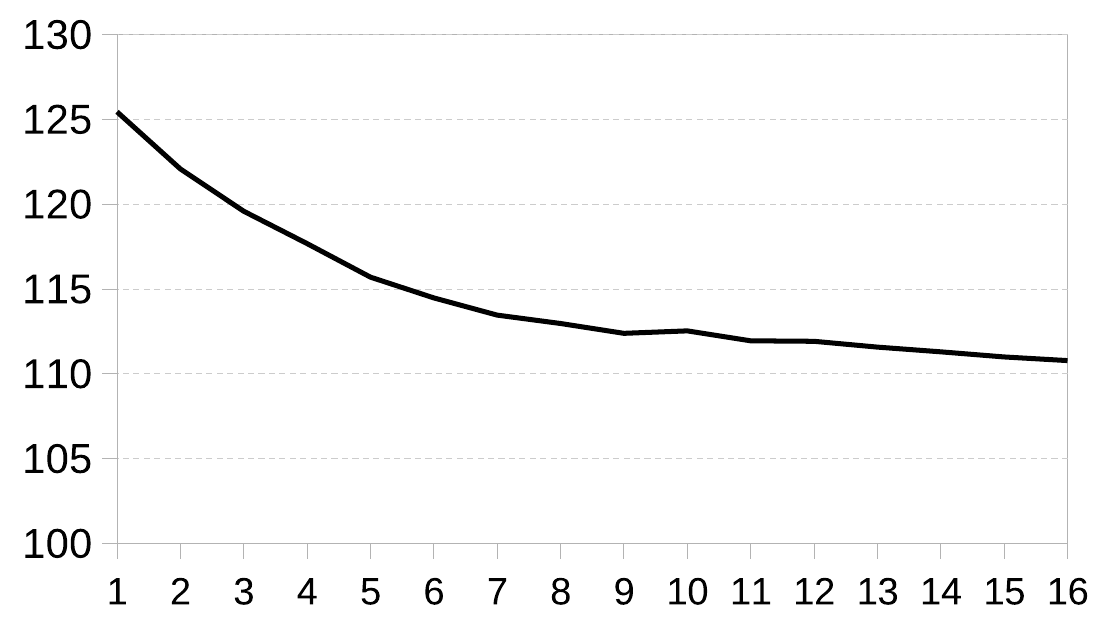}{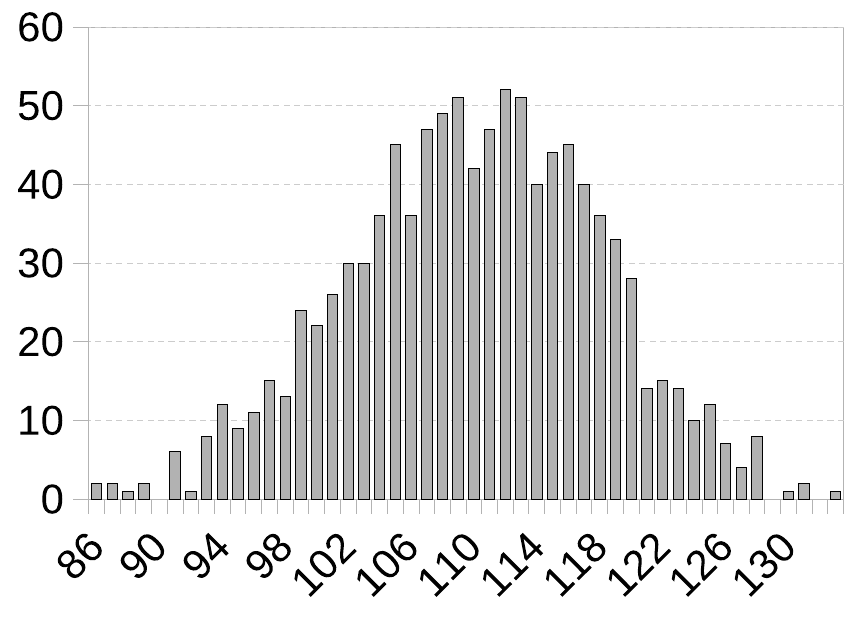}{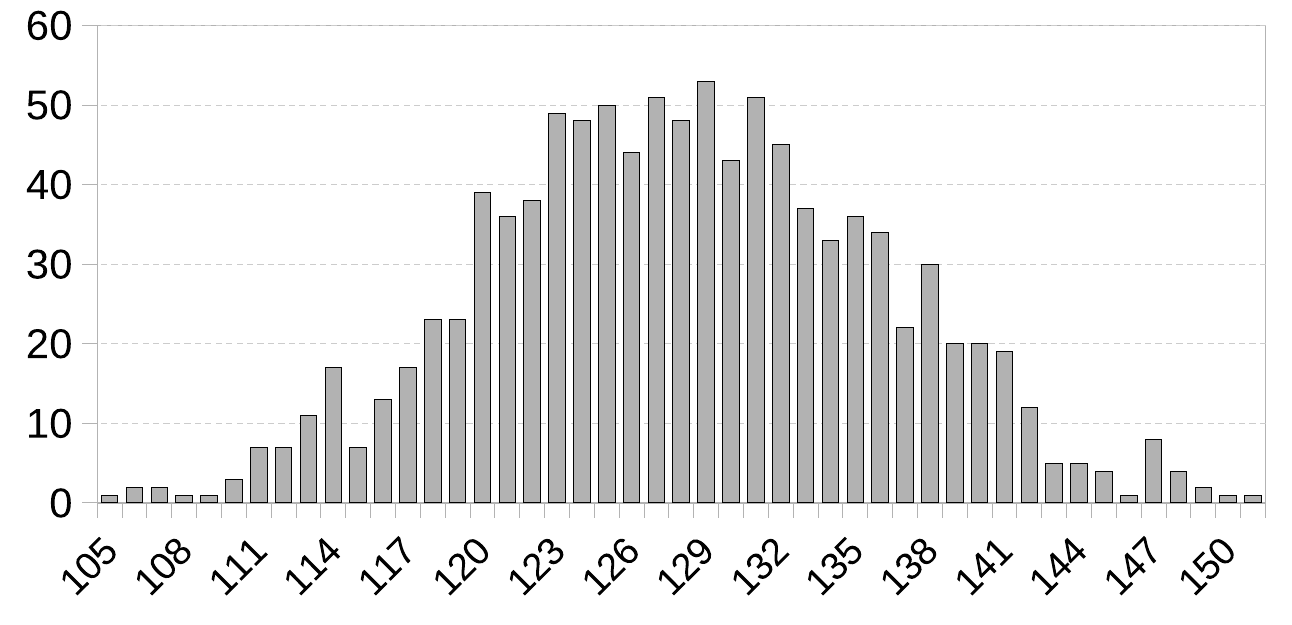}{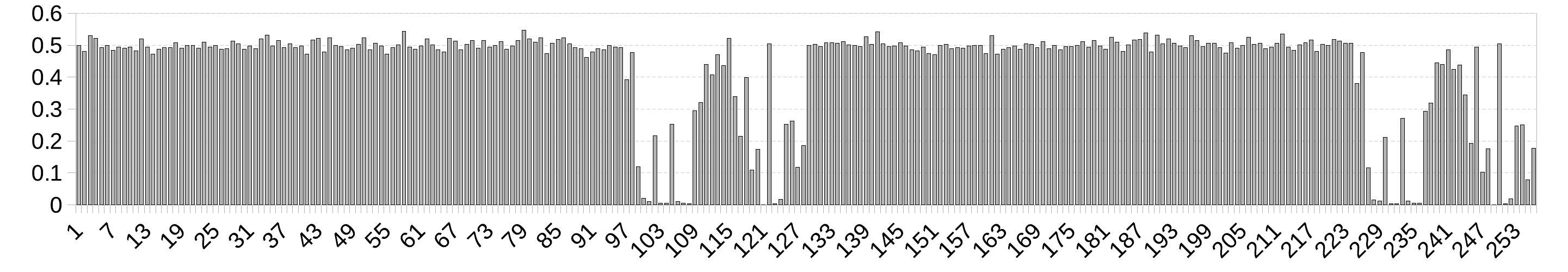}

\textbf{Misses, predicted:} Average = 110.042, Min = 86, Max = 133.

\textbf{Misses, random:} Average = 128.085, Min = 105, Max = 151.

\subsection{SHA3/Keccak}

\forceparindent
There is no $\fzADD$ op in this CHF, only $\fzNOT$, $\fzAND$, $\fzXOR$.
It is based on a sponge construction (\cite{bertoni2011.csf}).

The state size is 1600 bits, by default the hash size is 256 bits
(note that Keccak's authors at \cite{keccak.crunchy.contest}
mention, regarding preimage and collision solutions,
that ``\textsl{smaller versions are harder to break \textellipsis\ they offer much less degrees of freedom}'').

$[0;1]$-fuzbits and $(-1;1]$-fuzbits produce the results on a par with each other,
so we present only the latter-based trainings.

{\ }

\resultitem
When the message is short enough, the inversion of a single (1st) round looks promising.

\textbf{CHF:} Keccak-1600-256, 1 round, 64-bit message. Mask: full.

\textbf{NN:} 256 cells in 1 layer. 1024 samples, 64 in batch.

\traintestcharts{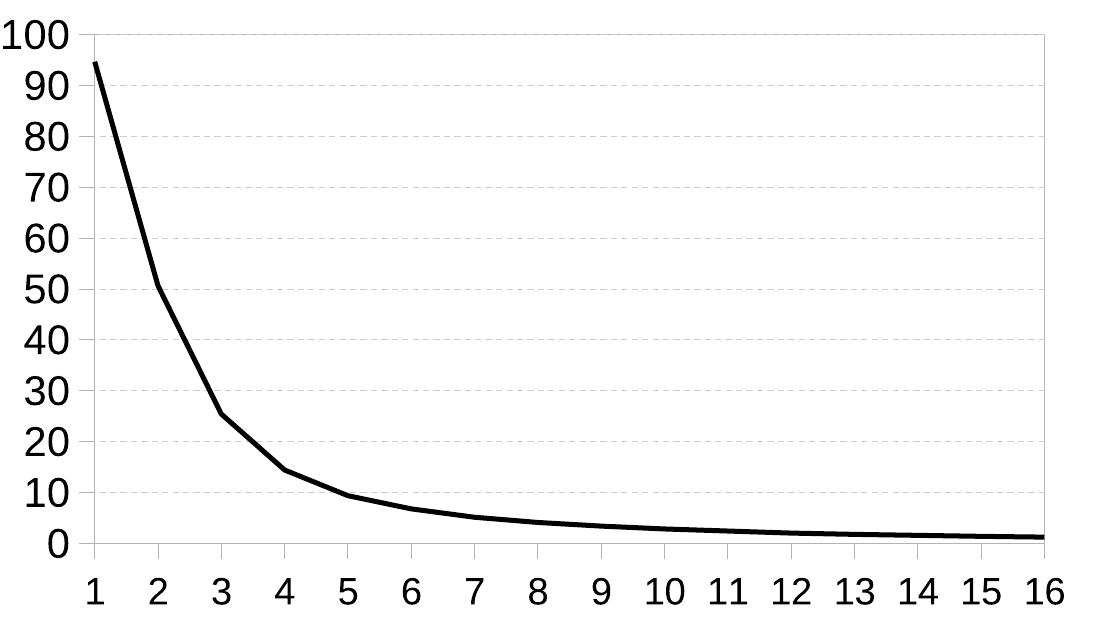}{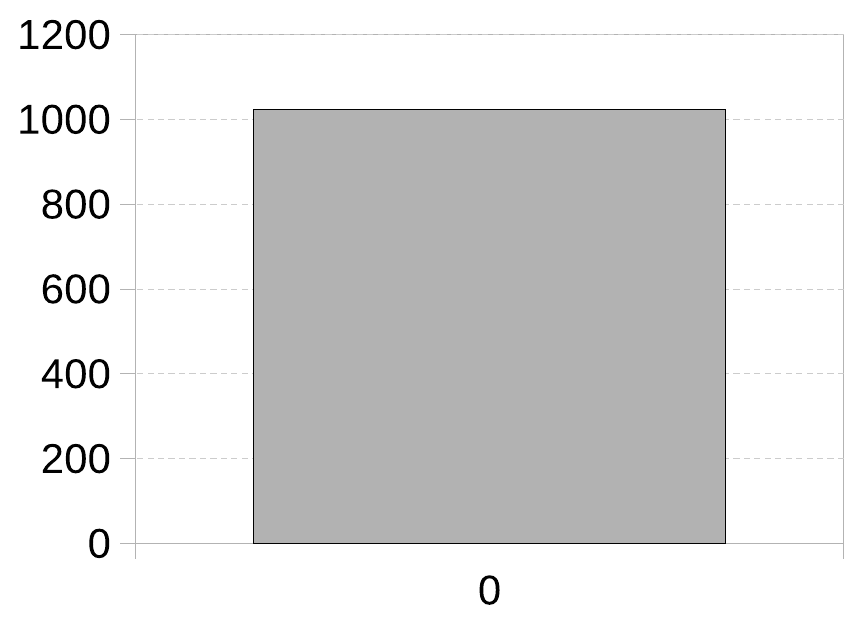}{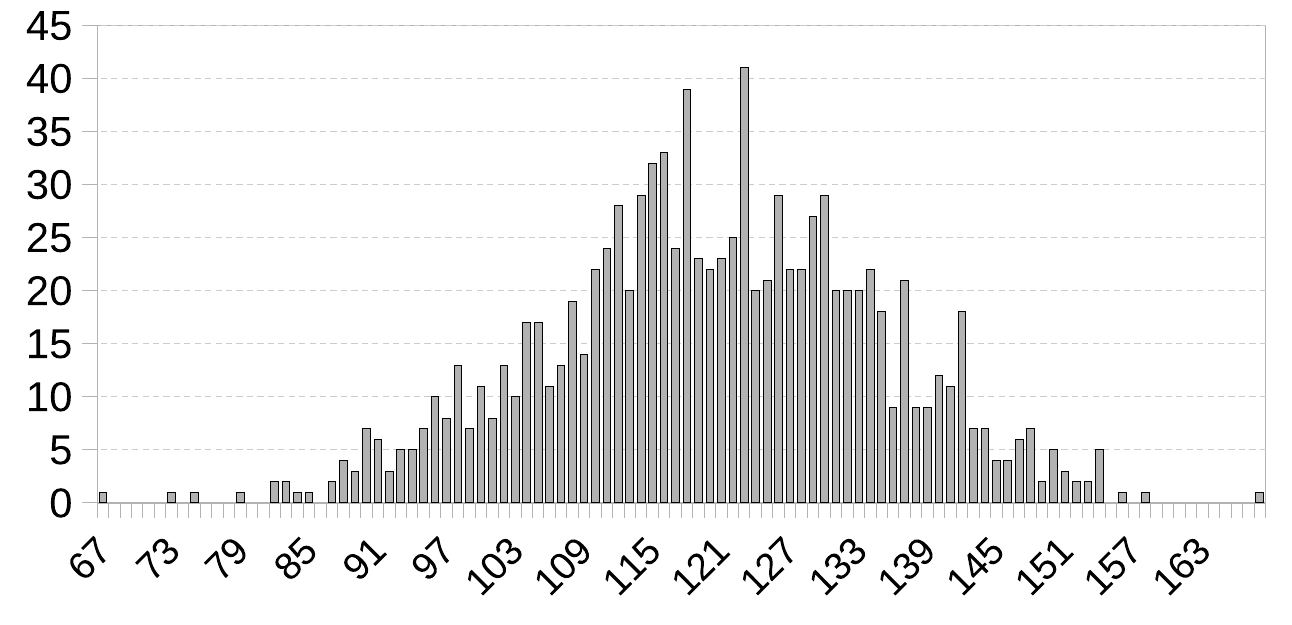}{}

\textbf{Misses, predicted:} Average = Min = Max = 0.

\textbf{Misses, random:} Average = 120.122, Min = 67, Max = 168.

For the similar NN, same message length, and 1 round of Keccak-1600-\textit{512} the results of prediction are identical
(no wrong bits for 1024 test hashes), the misses for random messages have Average = 186.04, Min = 128, Max = 232.

{\ }

\resultitem
However, the messages longer than 64-bit \textit{lane} (see \cite[2.1, 3.1.1]{fips202}) reveal the obstacle.

\textbf{CHF:} Keccak-1600-256, 1 round, 96-bit message. Mask: full.

\textbf{NN:} 2 layers, 1024 cells each. 1024 samples, 64 in batch.

\traintestcharts{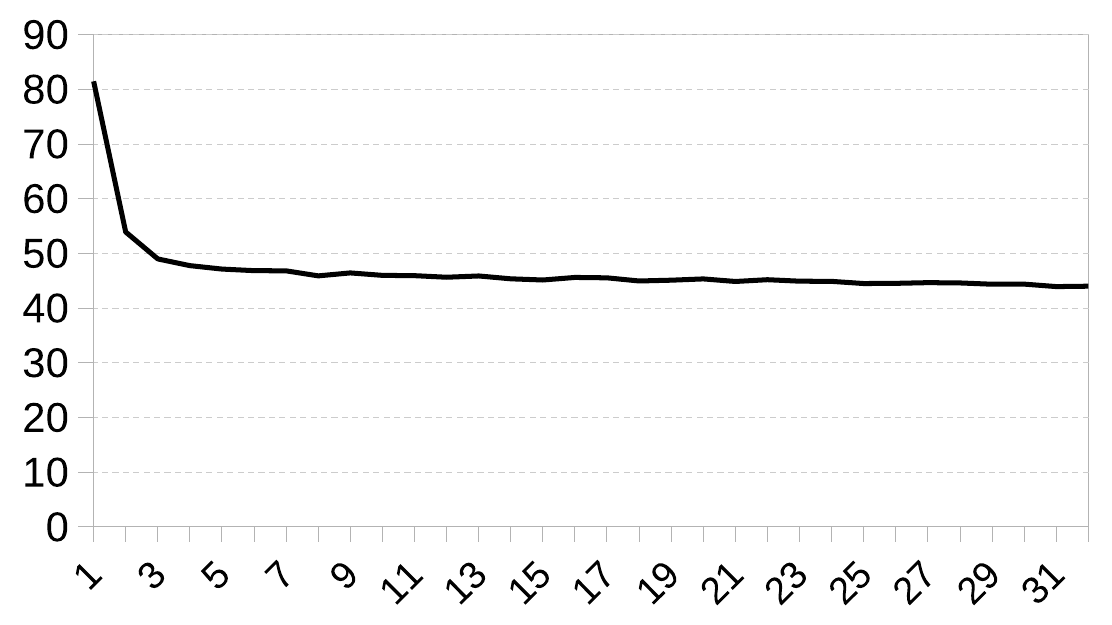}{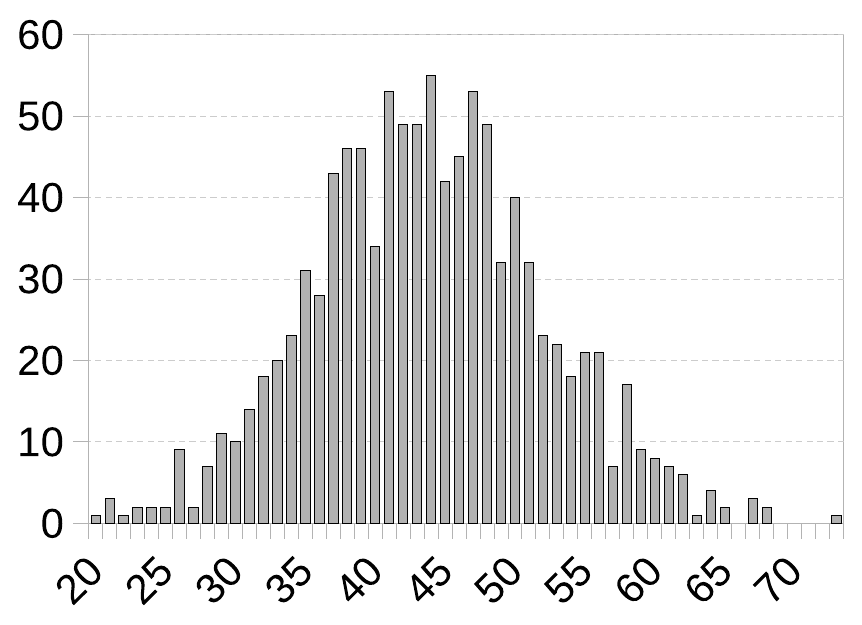}{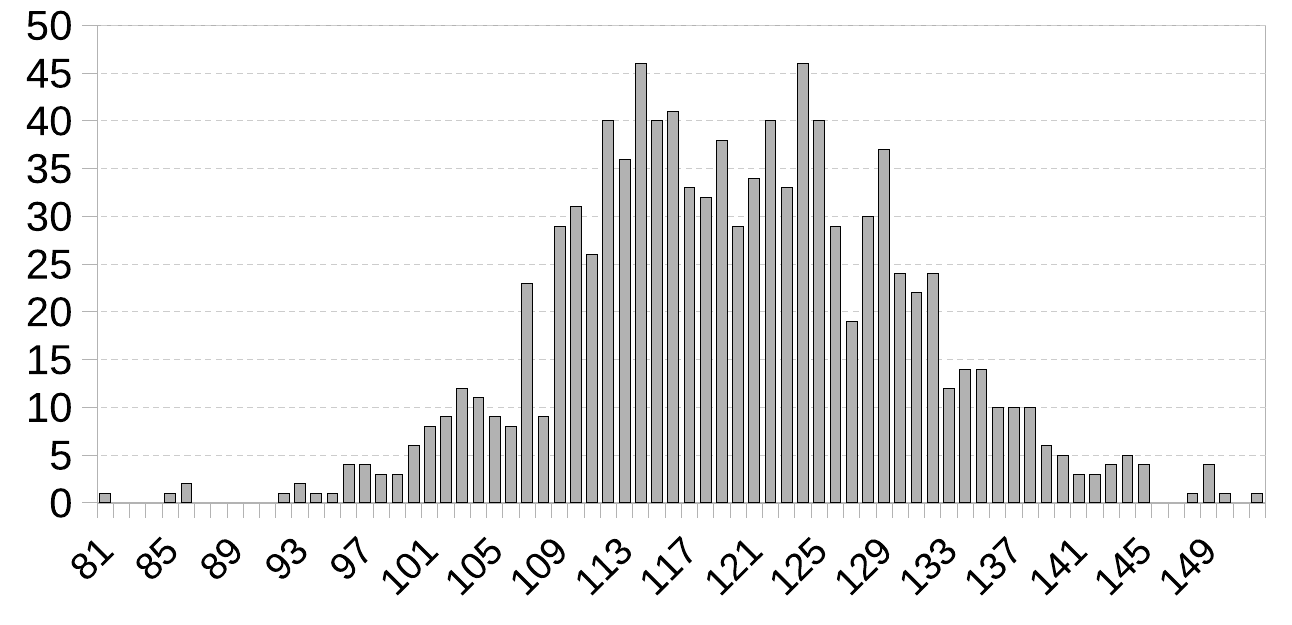}{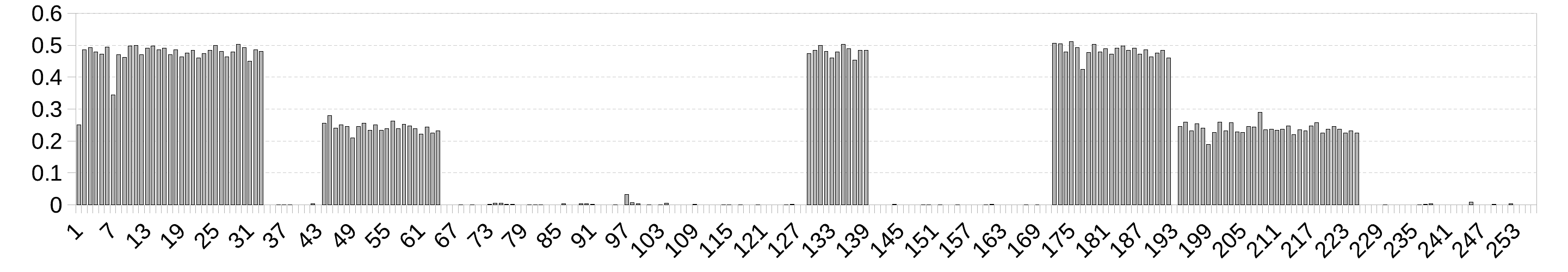}

\textbf{Misses, predicted:} Average = 43.818, Min = 20, Max = 73.

\textbf{Misses, random:} Average = 119.848, Min = 81, Max = 152.

Some hash bits almost always match, others seem random or constant, thus match with probability $\frac{1}{2}$,
and there's the third group, which is ``between''.
This foreshadows the problem increasing rapidly as the number of rounds grow,
for the uncertainty will ``propagate'' to other bits.

{\ }

\resultitem\label{resKec128delay}
That obstacle in the form of lower accuracy bound depends not only on message length,
but on rate and capacity of Keccak transformation too.

\textbf{CHF:} Keccak-1600-128, 1 round, 256-bit message (4 lanes). Mask: full.

\textbf{NN:} 2 layers, 1024 cells each. 1024 samples, 64 in batch.

\traintestcharts{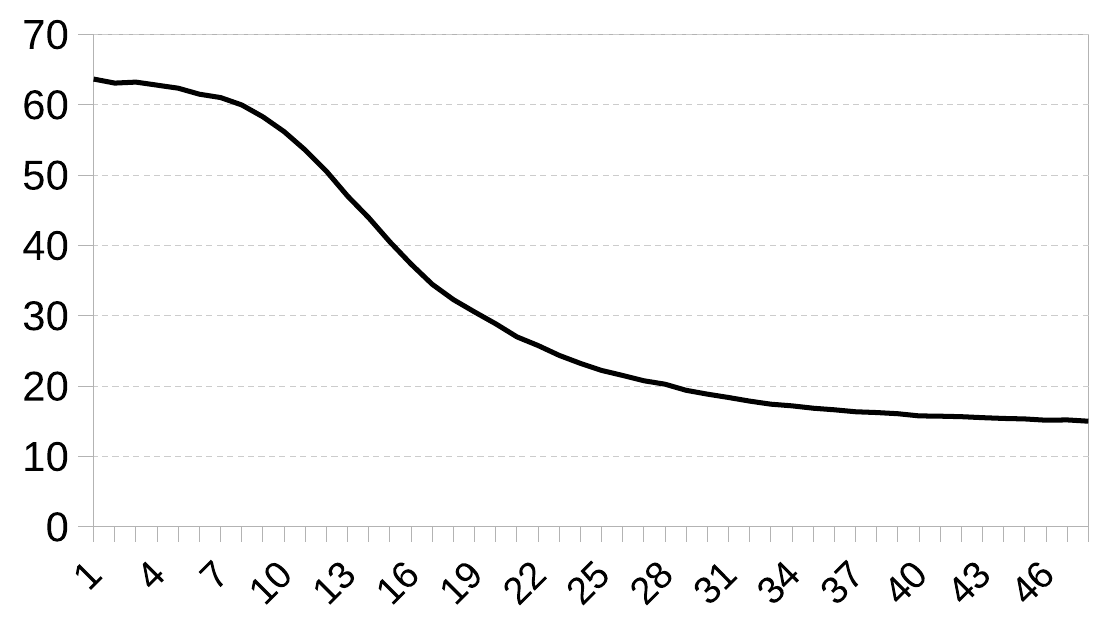}{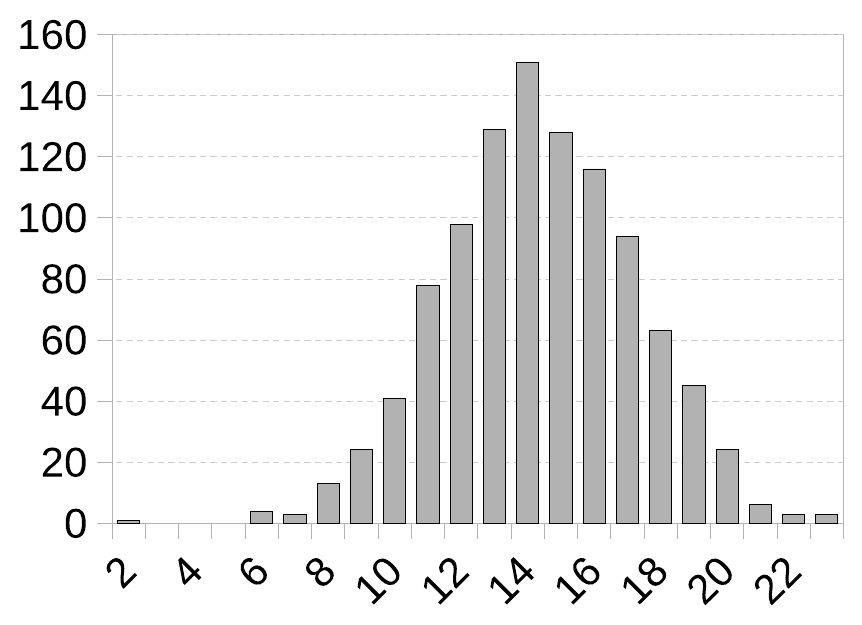}{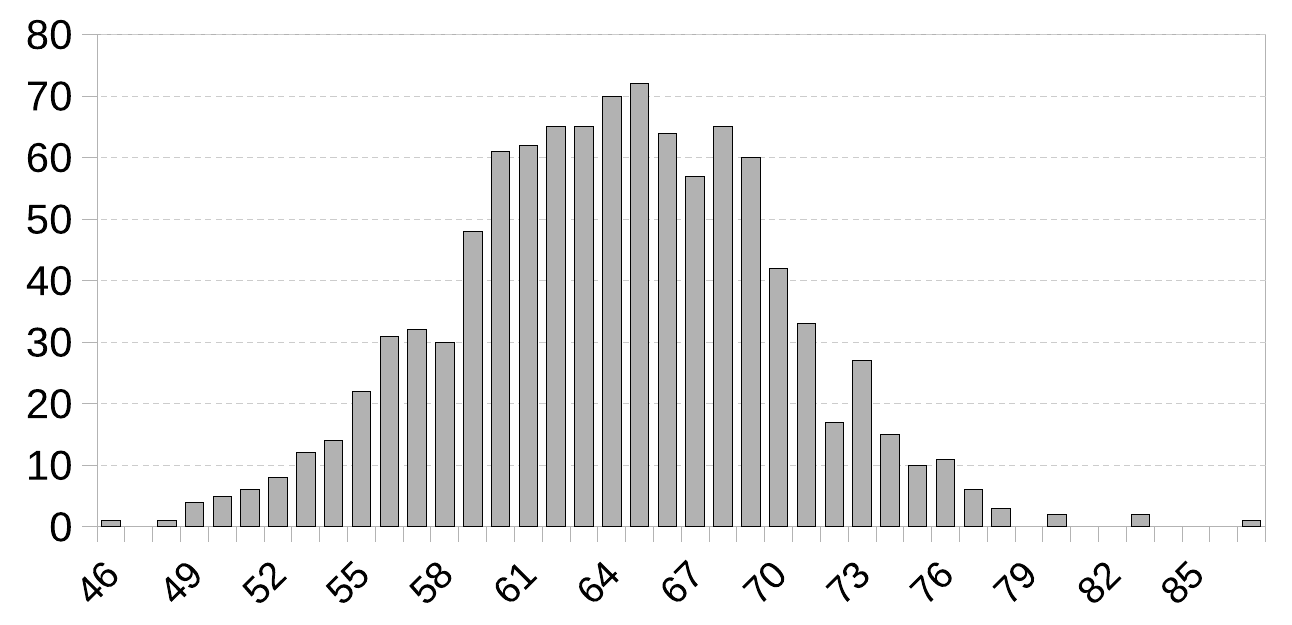}{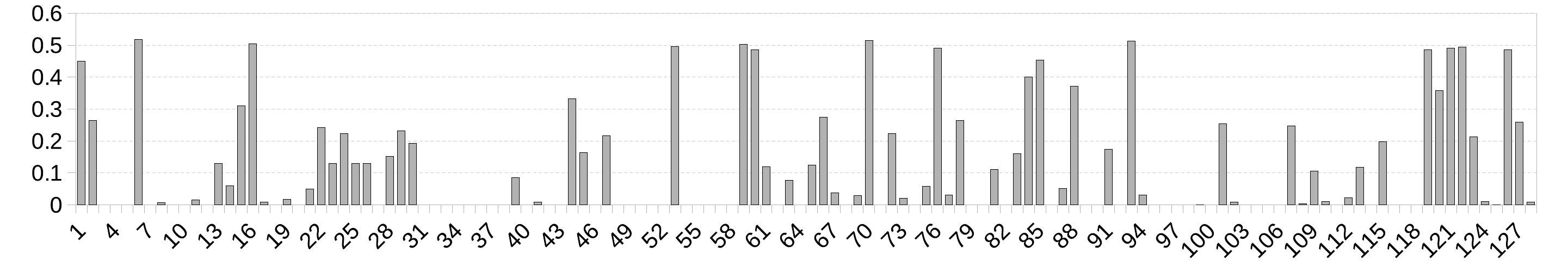}

\textbf{Misses, predicted:} Average = 14.361328125, Min = 2, Max = 23. 

\textbf{Misses, random:} Average = 64.0712890625, Min = 46, Max = 87.

Notice that the increasing of accuracy accelerates after 7 or so initial epochs,
in contrast to the behaviour that we've observed till now.

If the output activation is ``hard'' sigmoid instead of usual one, the learning will slow down.

{\ }

\resultitem\label{resKec256nolearn}
One round more, and we lose everything with these NN and training set.

\textbf{CHF:} Keccak-1600-256, 2 rounds, 64-bit message. Mask: full.

\textbf{NN:} 2 layers, 1024 cells each. 1024 samples, 64 in batch.

\traintestcharts{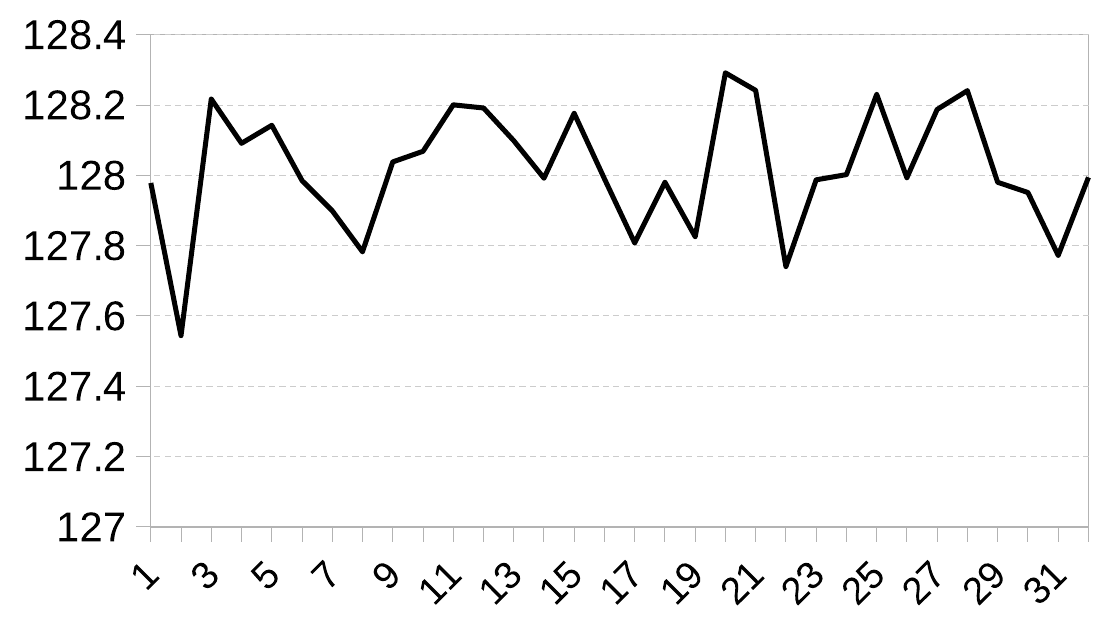}{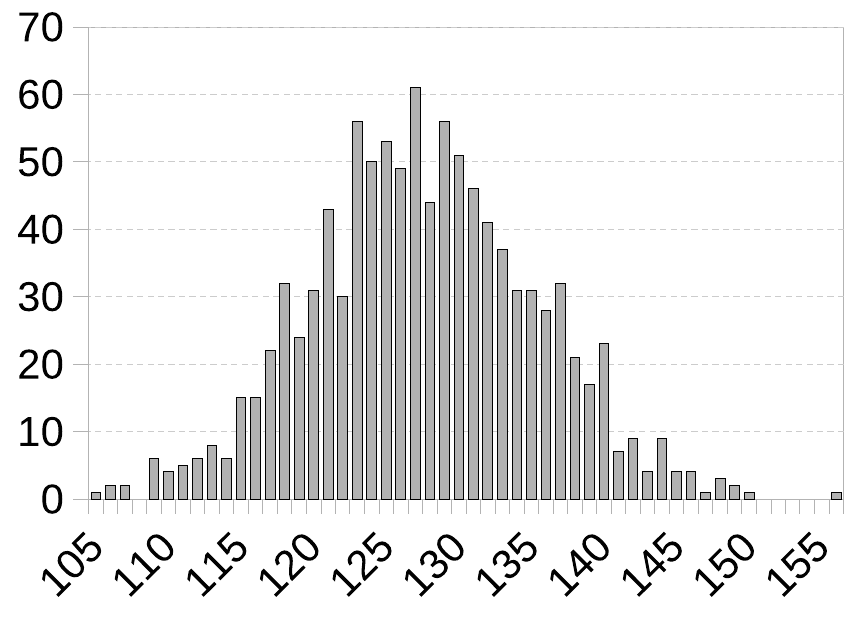}{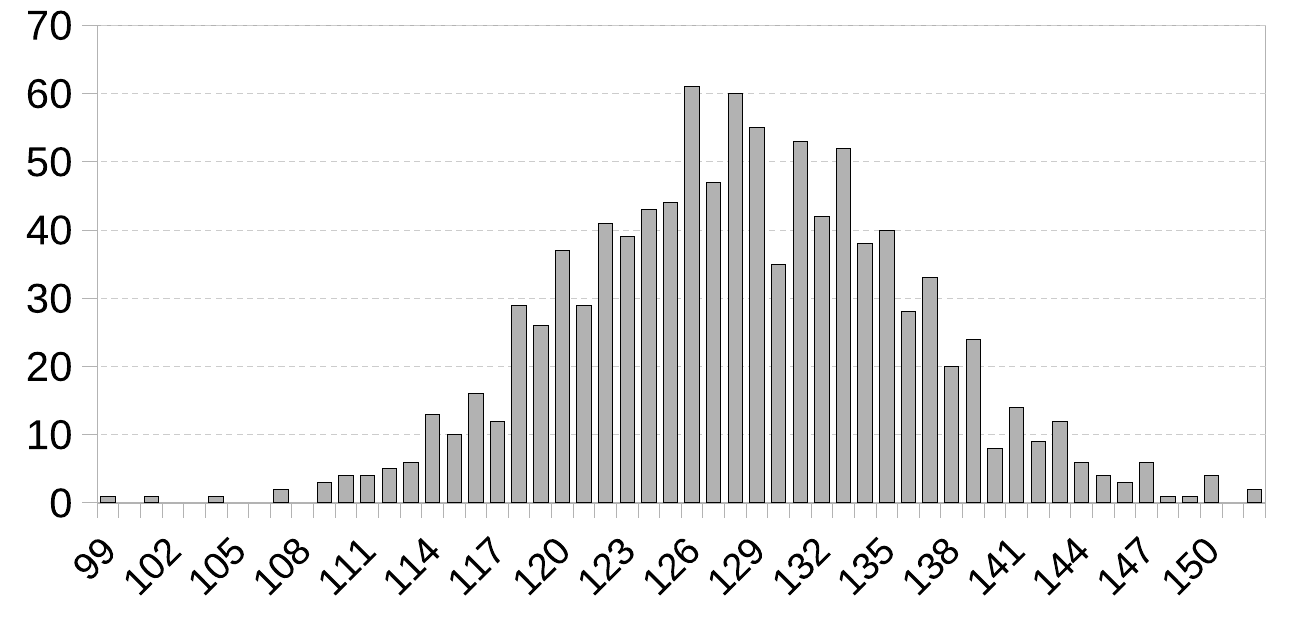}{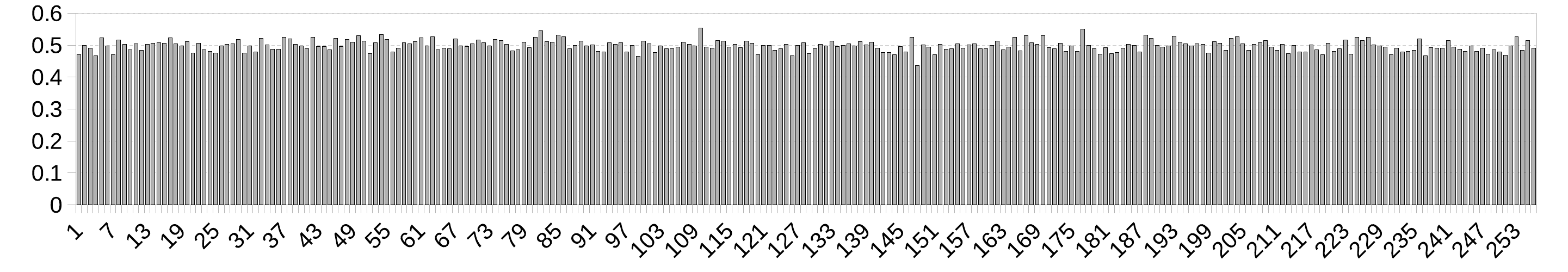}

\textbf{Misses, predicted:} Average = 127.614, Min = 105, Max = 156.

\textbf{Misses, random:} Average = 128.23, Min = 99, Max = 152.

{\ }

\resultitem\label{resKec128lanedelay}
Once again smaller hash size (hence capacity) and bigger training set help a little.

\textbf{CHF:} Keccak-1600-128, 2 rounds, 64-bit message. Mask: full.

\textbf{NN:} 3 layers, 512 cells each. 8192 samples, 512 in batch.

\traintestcharts{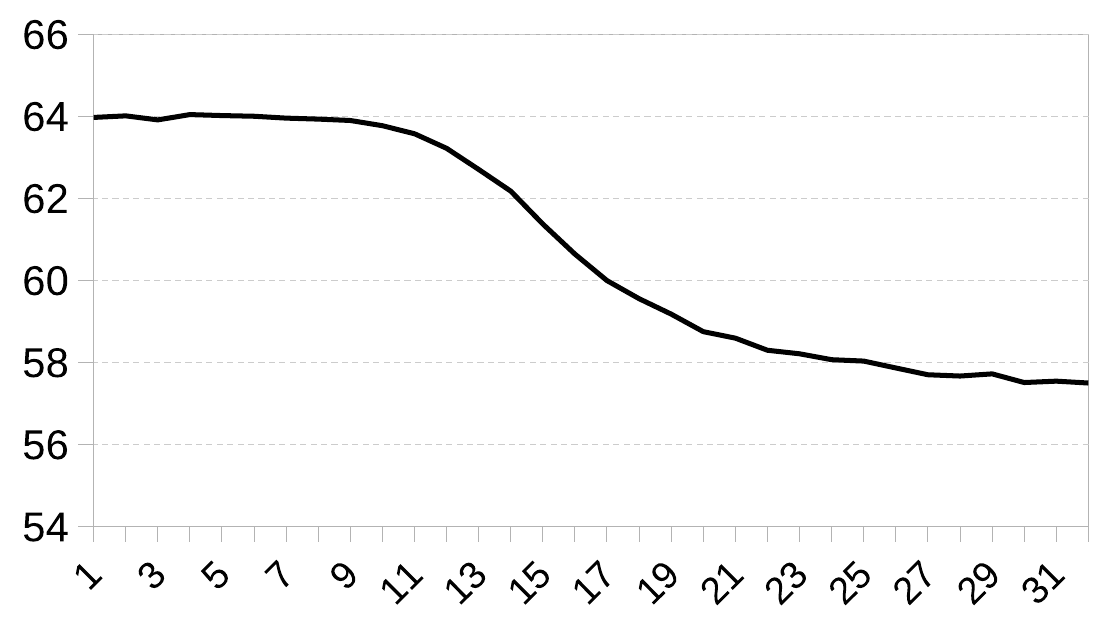}{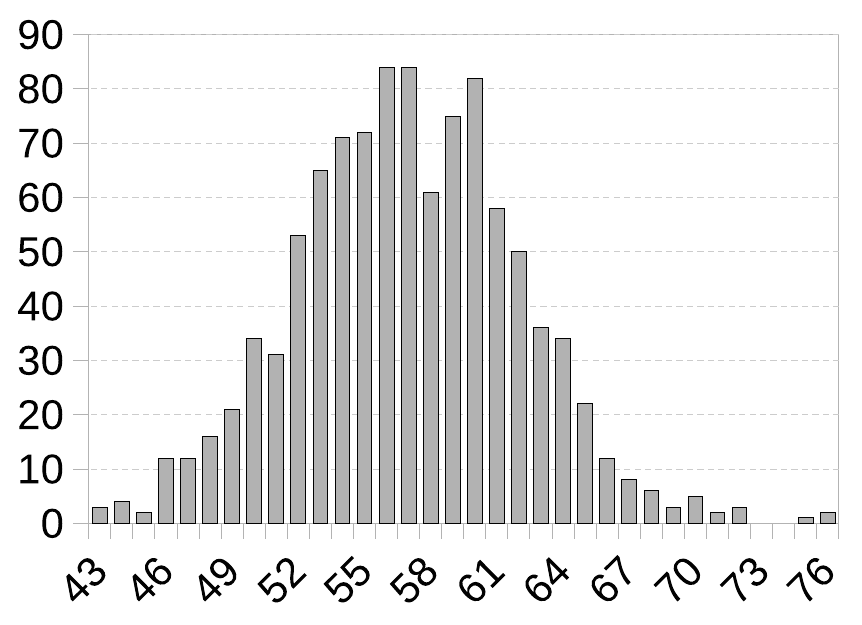}{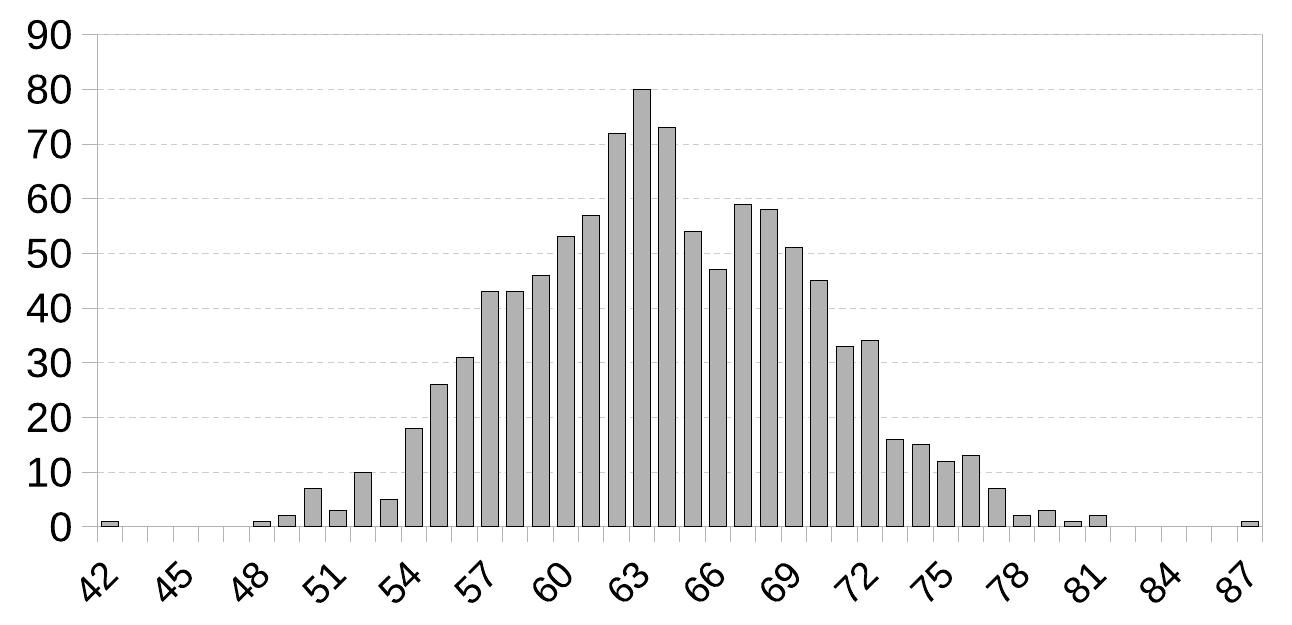}{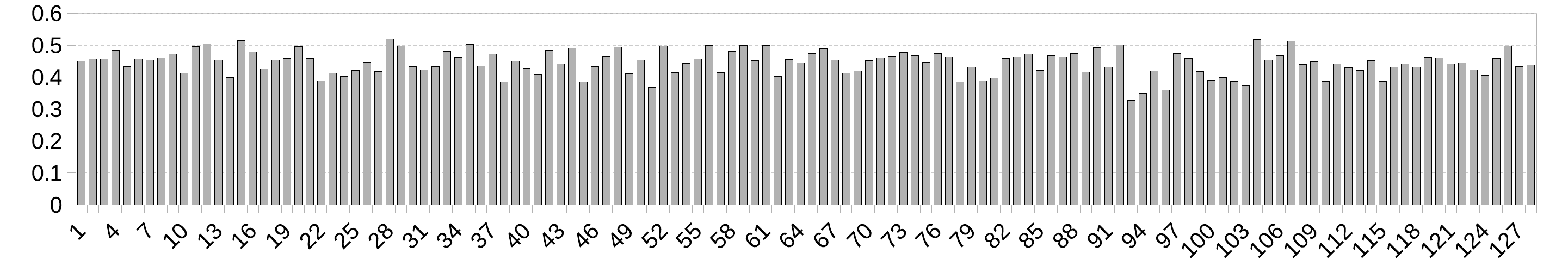}

\textbf{Misses, predicted:} Average = 56.981, Min = 43, Max = 76.

\textbf{Misses, random:} Average = 63.967, Min = 42, Max = 87.

To get this ``achievement'', the total number of train samples, as well as the batch, were made 8 times bigger.

{\ }

\resultitem
Even a single byte of the hash is difficult for NN to match.

\textbf{CHF:} Keccak-1600-256, 2 rounds, 64-bit message. Mask: bits 0--7.

\textbf{NN:} 3 layers, 512 cells each. 4096 samples, 256 in batch.

\traintestcharts{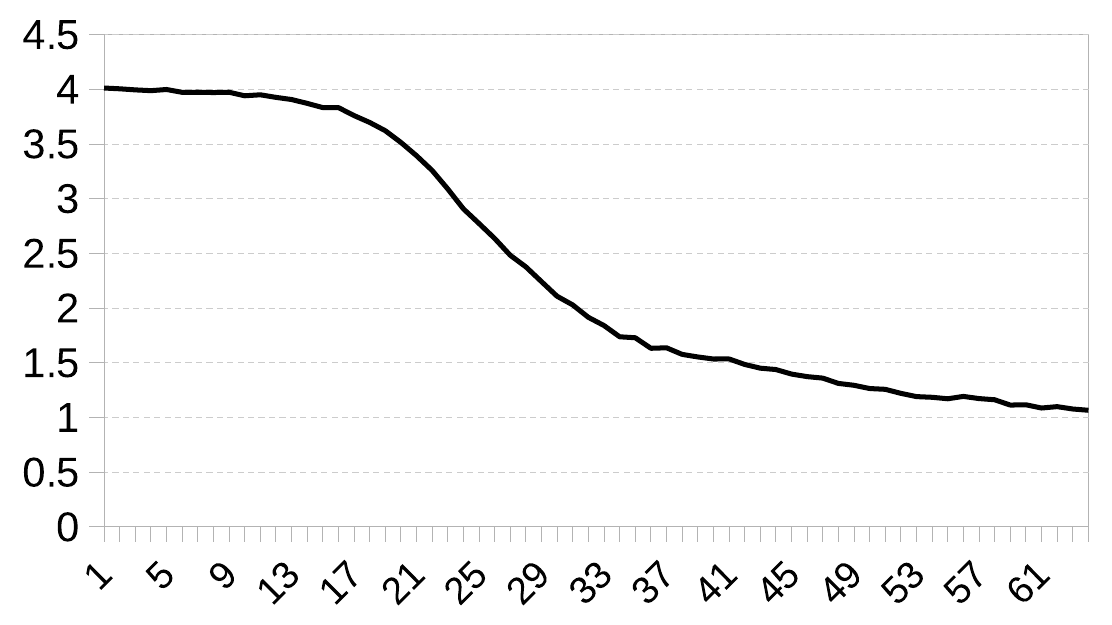}{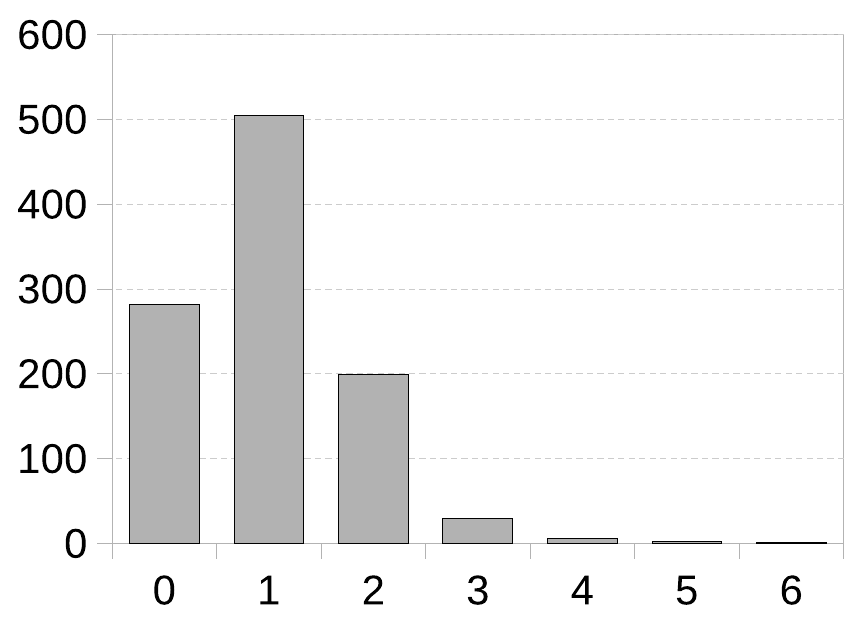}{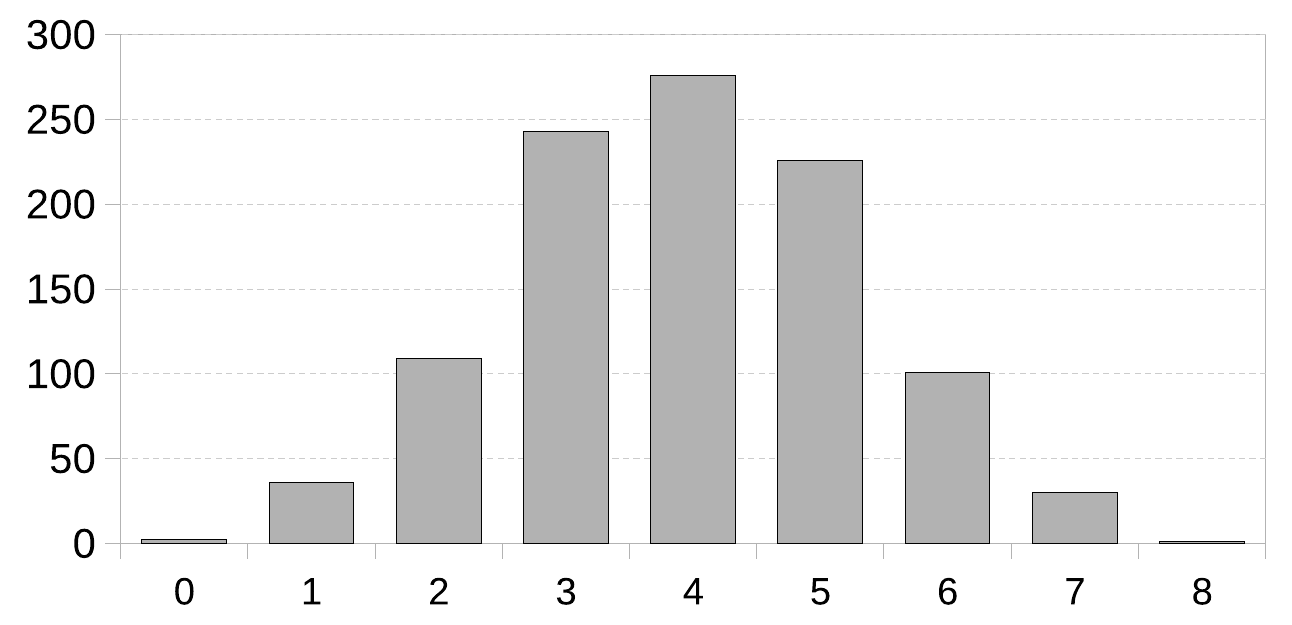}{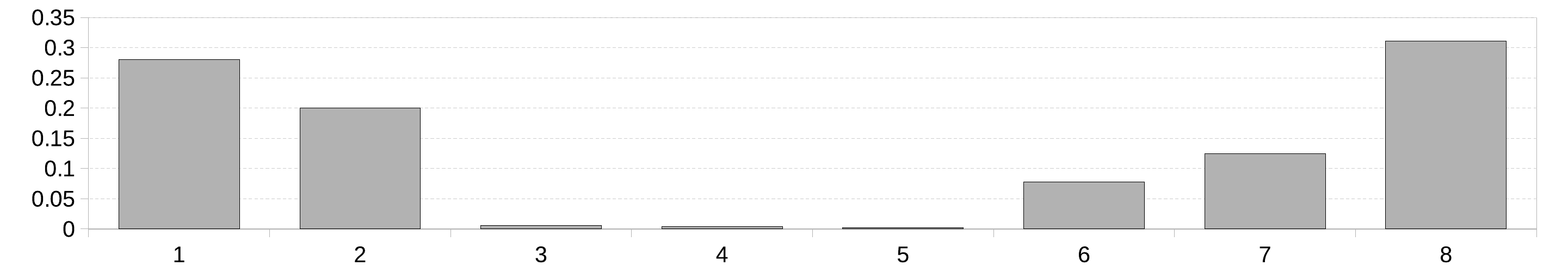}

\textbf{Misses, predicted:} Average = 1.006, Min = 0, Max = 6.

\textbf{Misses, random:} Average = 3.946, Min = 0, Max = 8.

The number of train samples, the batch, and the number of epochs were doubled.

Similarly to Res. \ref{resKec128delay} and \ref{resKec128lanedelay},
NN initially delays before it starts to seemingly increase the accuracy:
up to 8--9th epoch there are no signs of the forthcoming ``success''.
This behaviour is known to raise expectations, ---
maybe we should ``wait a little longer and see how NN makes it'', though the asymptote of accuracy is still positive.
In other words, we would like to be sure that NN from Res. \ref{resKec256nolearn} doesn't just need more time.

When the mask is 0--31 bits (first double word of the hash), this NN after 64 epochs is able to match 20 bits of the hash
in average, while for random messages it's 16 bits, and the loss remains quite close to 16 for the first 12--13 epochs.

{\ }

\resultitem
As usual, short messages are much easier.

\textbf{CHF:} Keccak-1600-256, 2 rounds, 16-bit message. Mask: full.

\textbf{NN:} 2 layers, 1024 cells each. 1024 samples, 64 in batch.

\traintestcharts{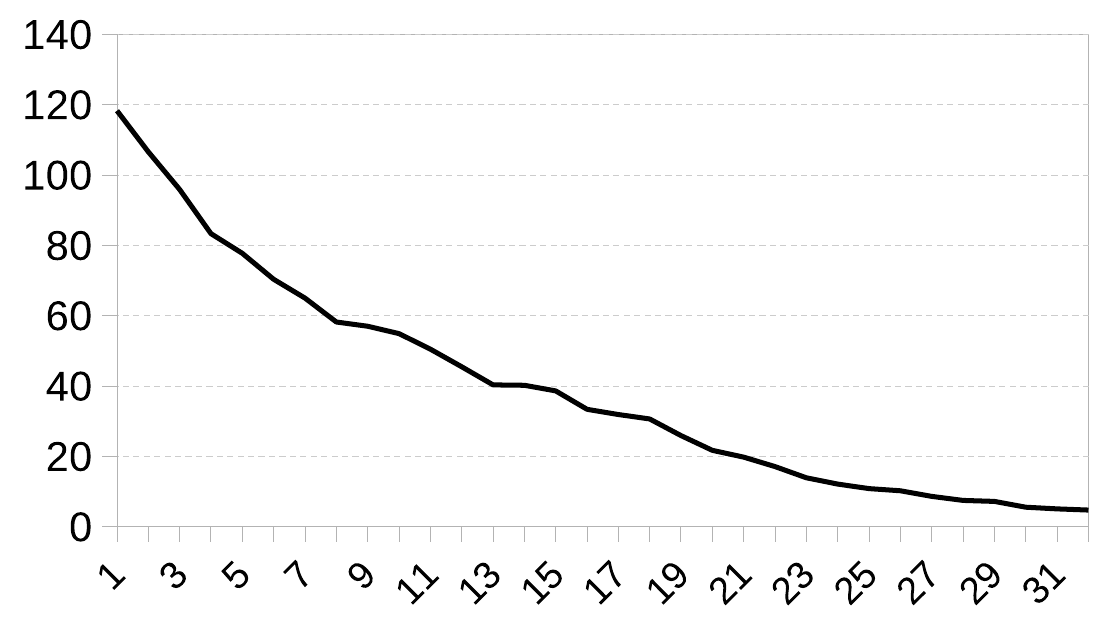}{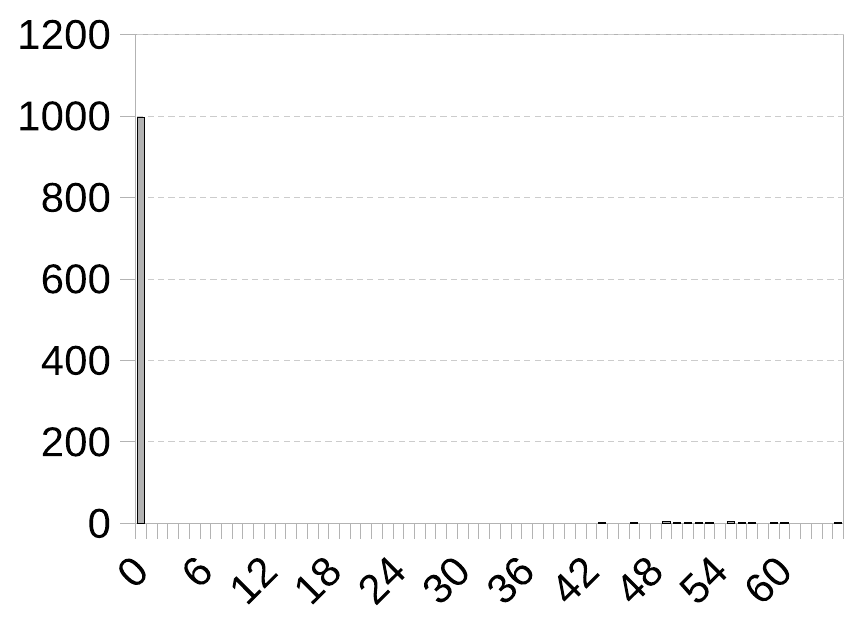}{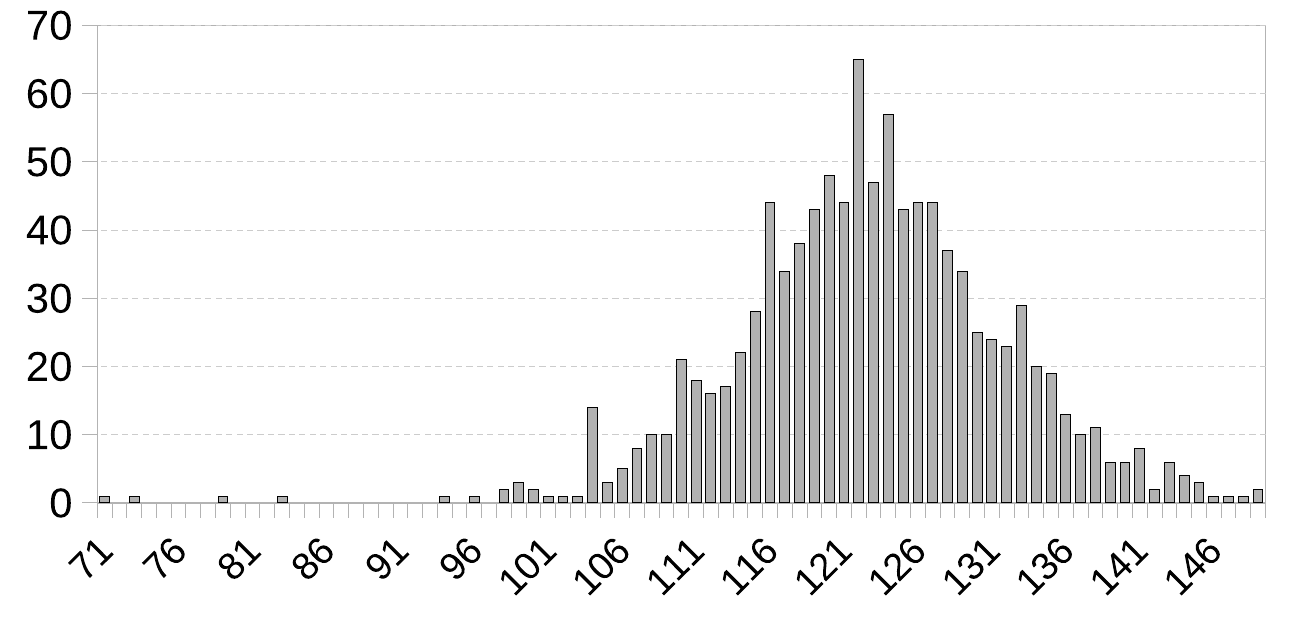}{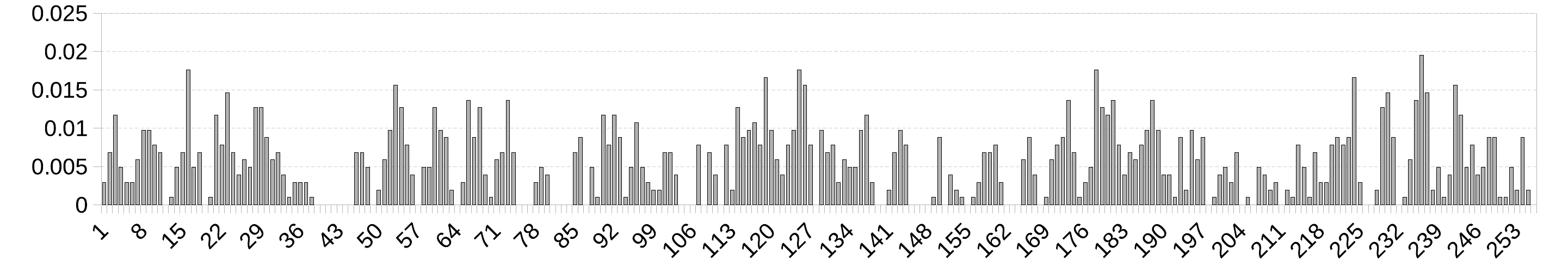}

\textbf{Misses, predicted:} Average = 1.401, Min = 0, Max = 65.

\textbf{Misses, random:} Average = 122.68, Min = 71, Max = 149.

997 hashes of 1024 were exactly inverted. This may look like overfitting, for there are only $2^{16} = 65536$ such messages,
though our training could contain at most half of them.

It is surely much faster and simpler to verify all $2^{16}$ messages for a given hash.

{\ }

\resultitem
The limits of ``easiness'' are quite narrow: only 8 additional bits almost nullify it.

\textbf{CHF:} Keccak-1600-256, 2 rounds, 24-bit message. Mask: full.

\textbf{NN:} 2 layers, 1024 cells each. 1024 samples, 64 in batch.

\traintestcharts{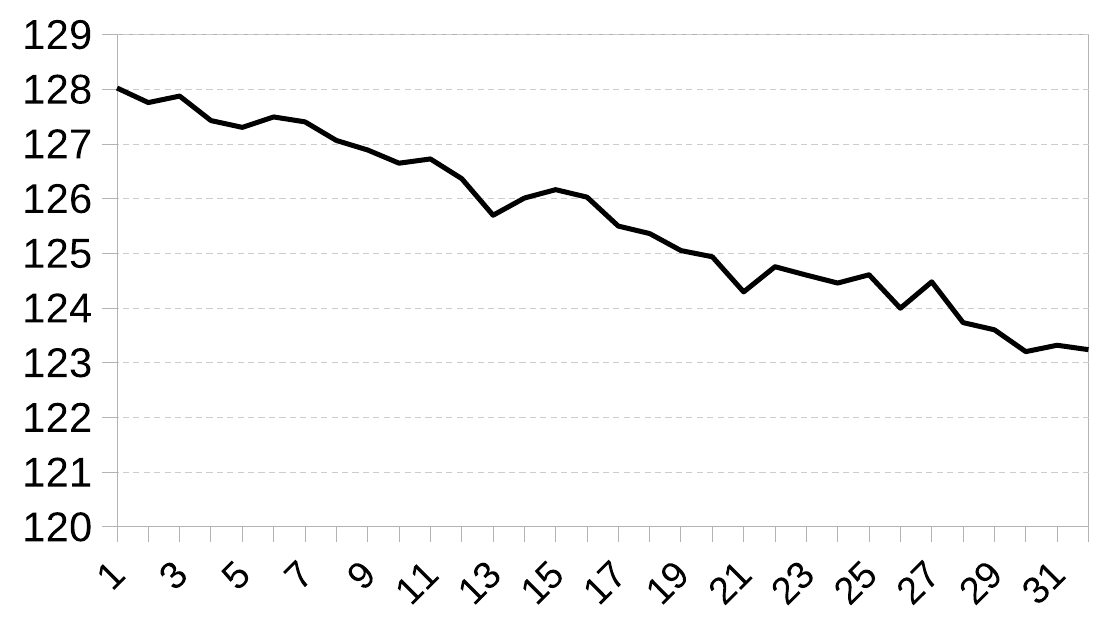}{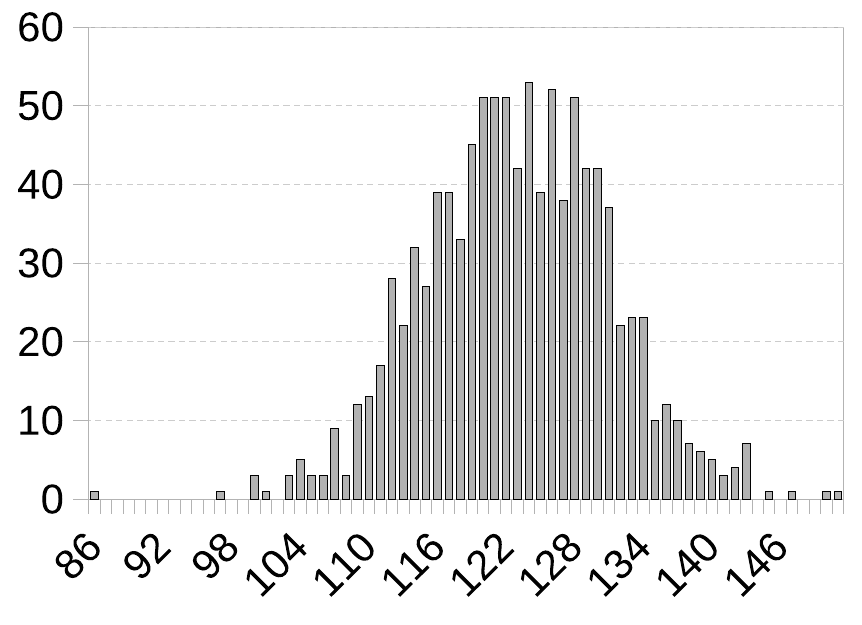}{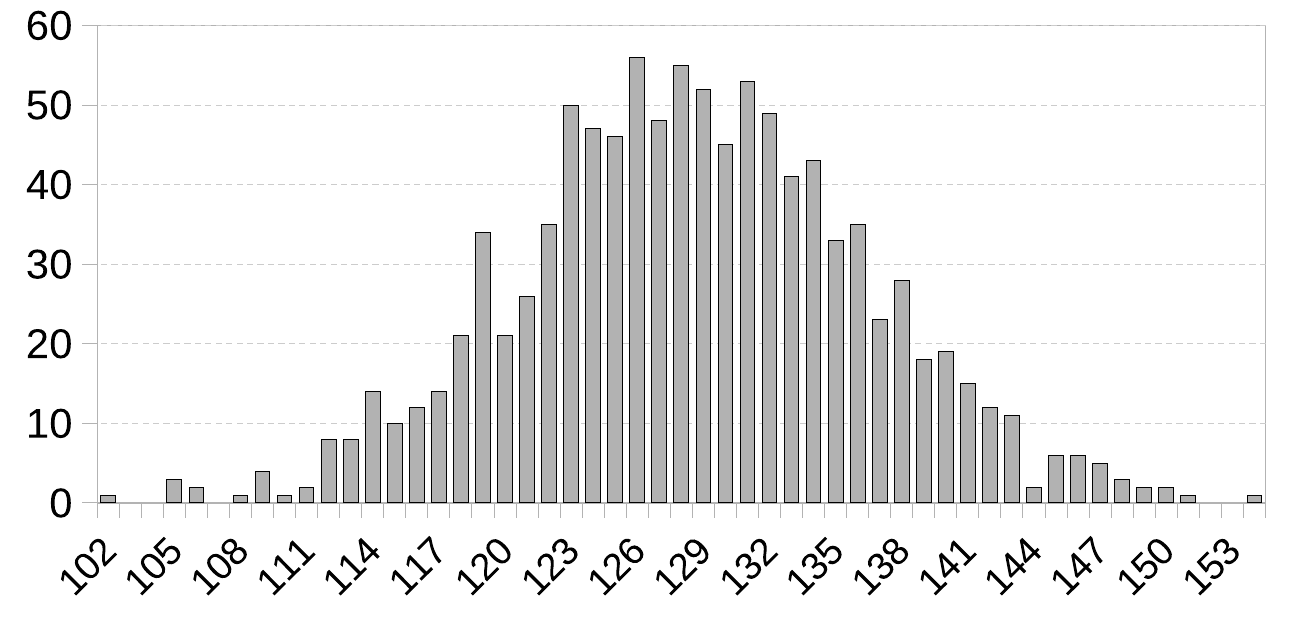}{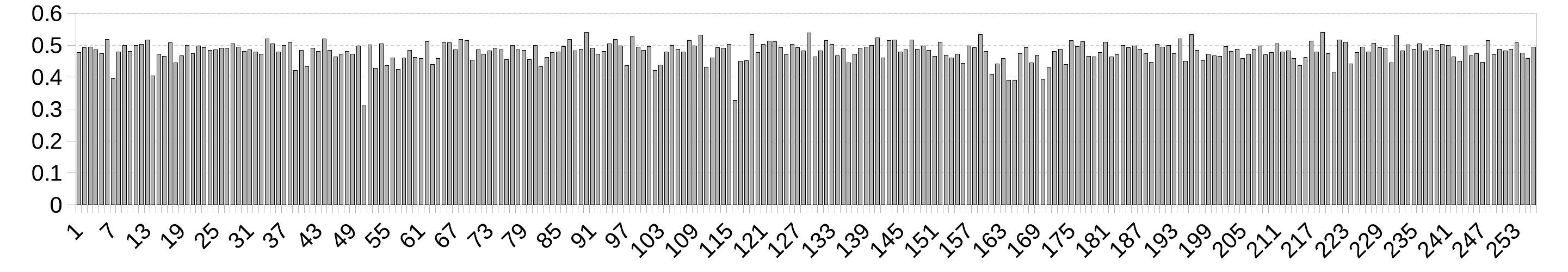}

\textbf{Misses, predicted:} Average = 122.957, Min = 86, Max = 151.

\textbf{Misses, random:} Average = 128.462, Min = 102, Max = 154.

Further training might increase the accuracy a little; for now only 2 bits of the hash have better match than random guess.

{\ }

\resultitem
Now we consider Keccak with weakened rounds, removing some of $\theta$, $\rho$ and $\pi$, $\chi$, $\iota$ mappings
(see \cite[3.2]{fips202}) from each round.
First of all, we remove $\theta$ since it is the main source of diffusion of the whole transformation
(\cite[2.3.2]{bertoni2011}).

\textbf{CHF:} Keccak-1600-256 without $\theta$, 6 rounds, 128-bit message. Mask: full.

\textbf{NN:} 2 layers, 1024 cells each. 1024 samples, 64 in batch.

\traintestcharts{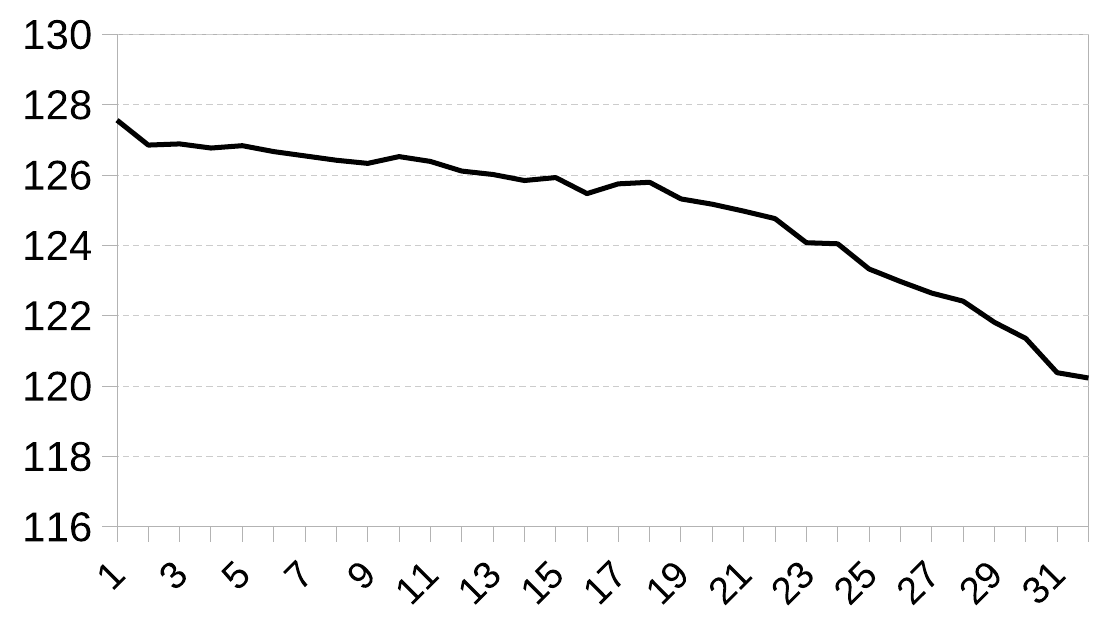}{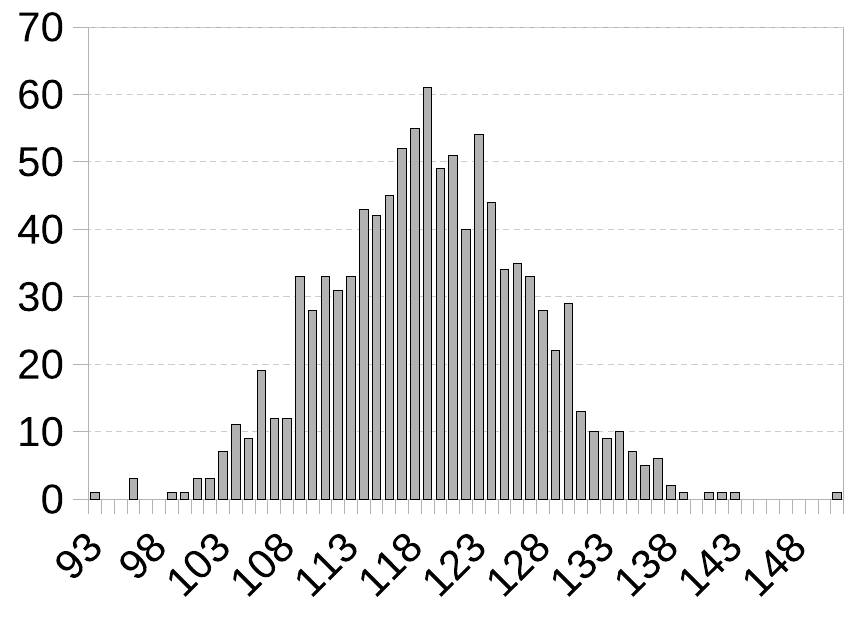}{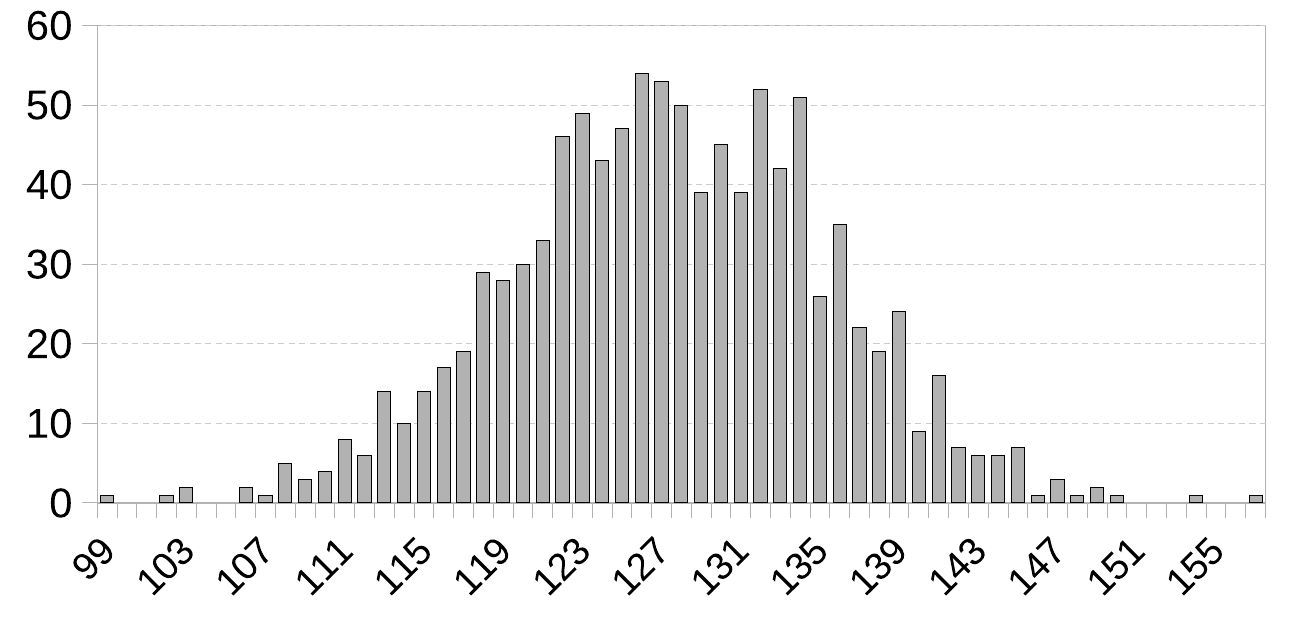}{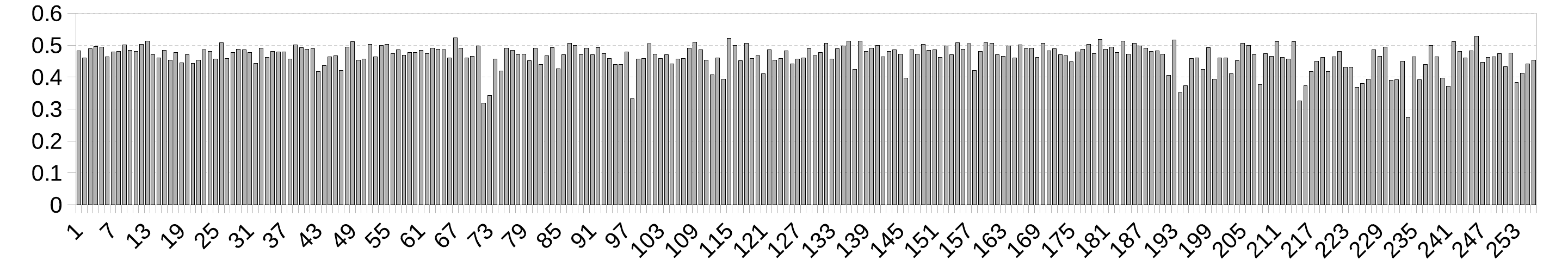}

\textbf{Misses, predicted:} Average = 119.131, Min = 93, Max = 151.

\textbf{Misses, random:} Average = 127.446, Min = 99, Max = 157.

Not much, but NN is able to learn, and there's a ``potential'' for subsequent improvement (the accuracy will increase).

{\ }

\resultitem
Again, the hashes of short messages can be inverted exactly.

\textbf{CHF:} Keccak-1600-256 without $\theta$, 6 rounds, 48-bit message. Mask: full.

\textbf{NN:} 2 layers, 1024 cells each. 1024 samples, 64 in batch.

\traintestcharts{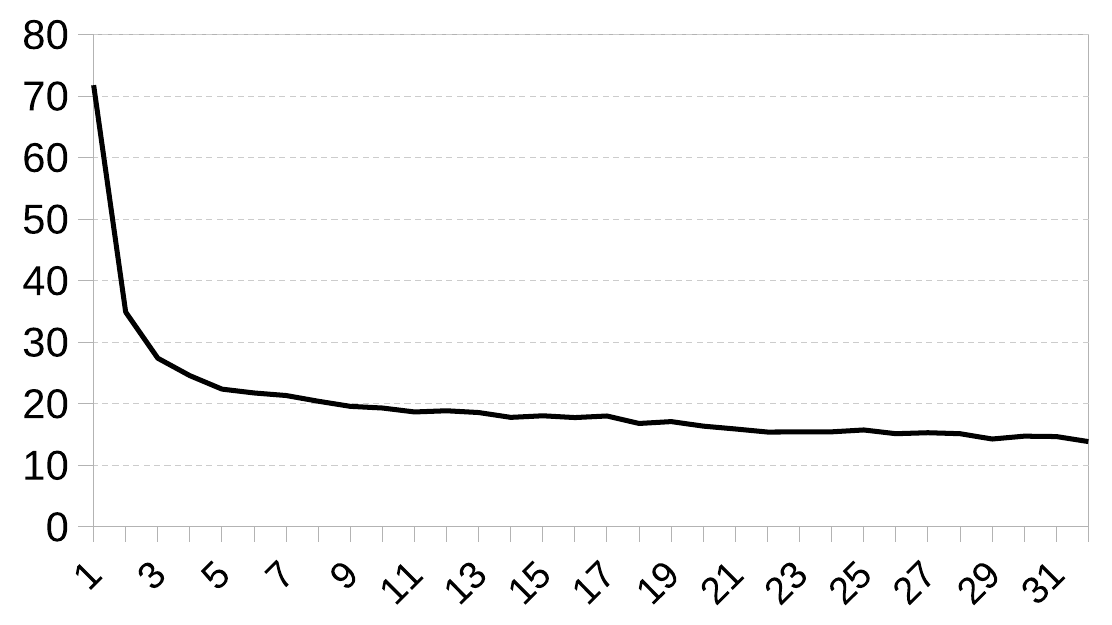}{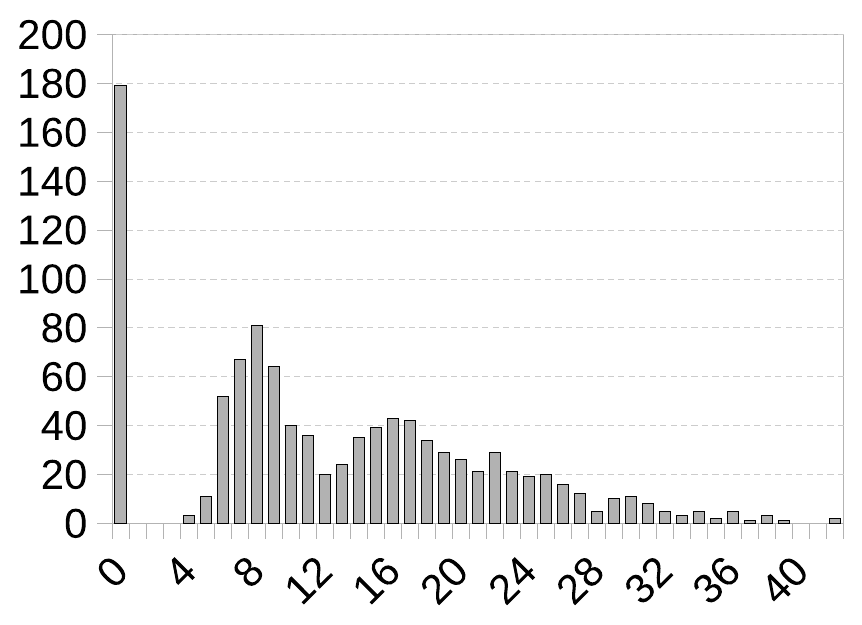}{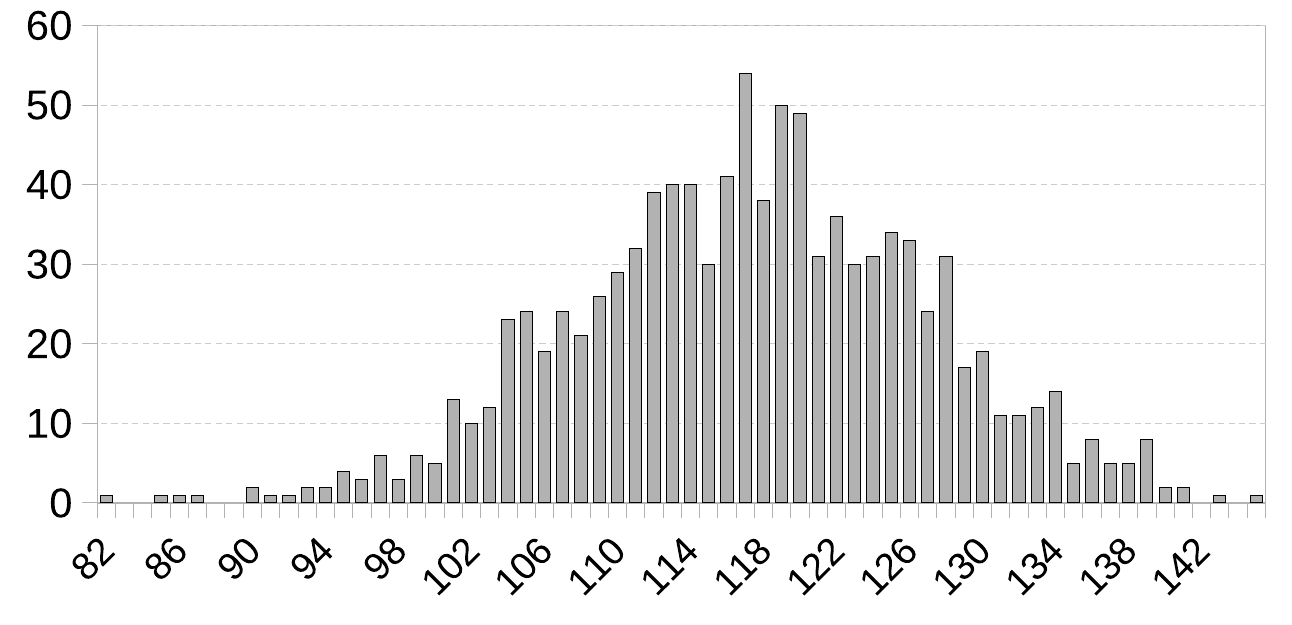}{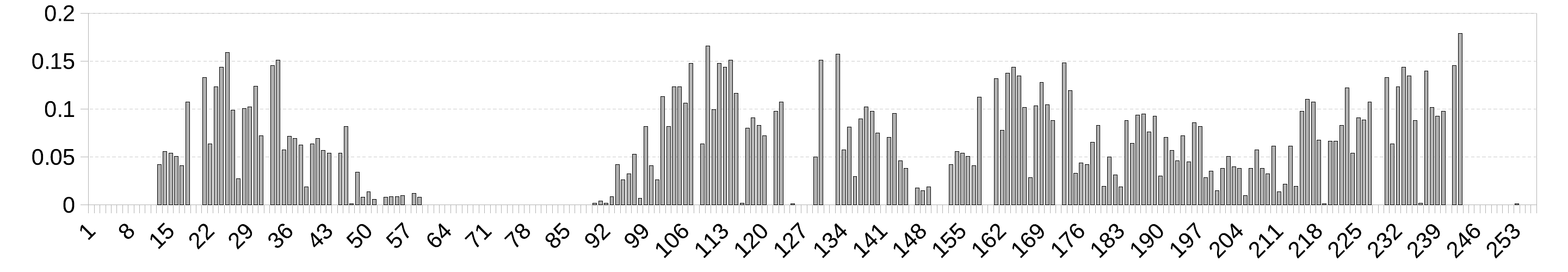}

\textbf{Misses, predicted:} Average = 12.499, Min = 0, Max = 42.

\textbf{Misses, random:} Average = 117.321, Min = 82, Max = 145.

{\ }

\resultitem
On the other hand, $\theta$ alone is not everything...

\textbf{CHF:} Keccak-1600-256 with only $\theta$, 2 rounds, 128-bit message. Mask: full.

\textbf{NN:} 2 layers, 1024 cells each. 1024 samples, 64 in batch.

\traintestcharts{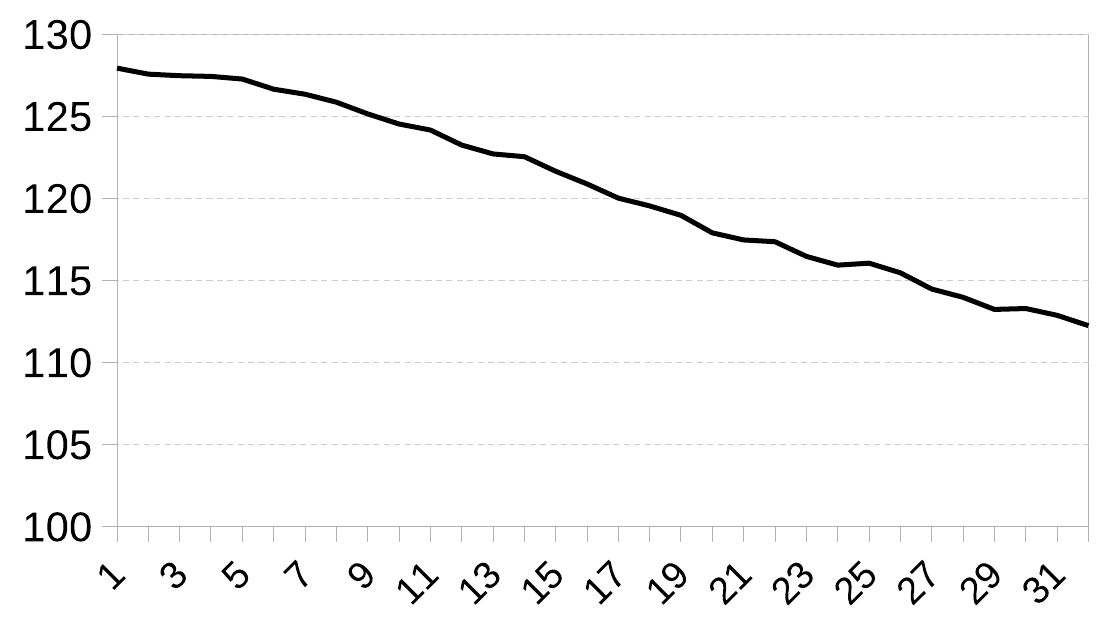}{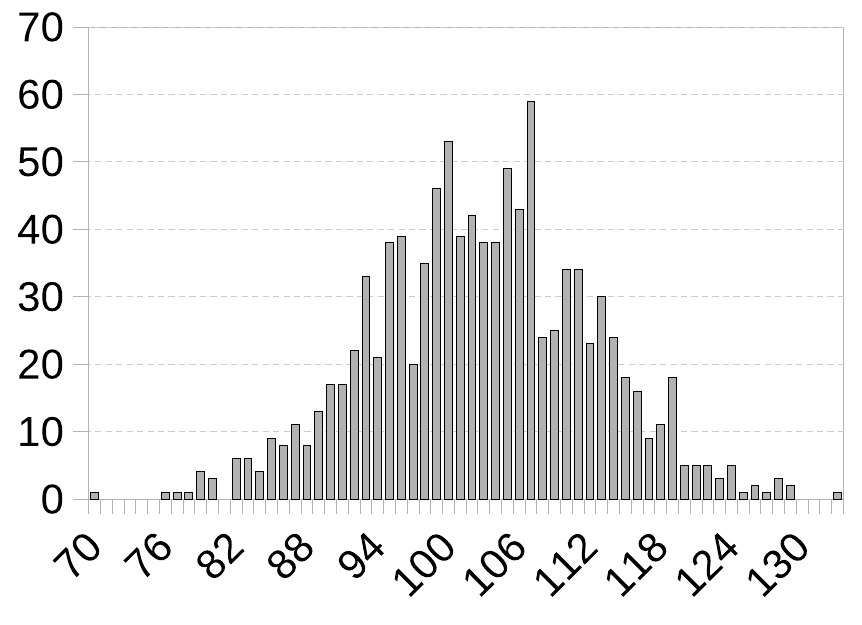}{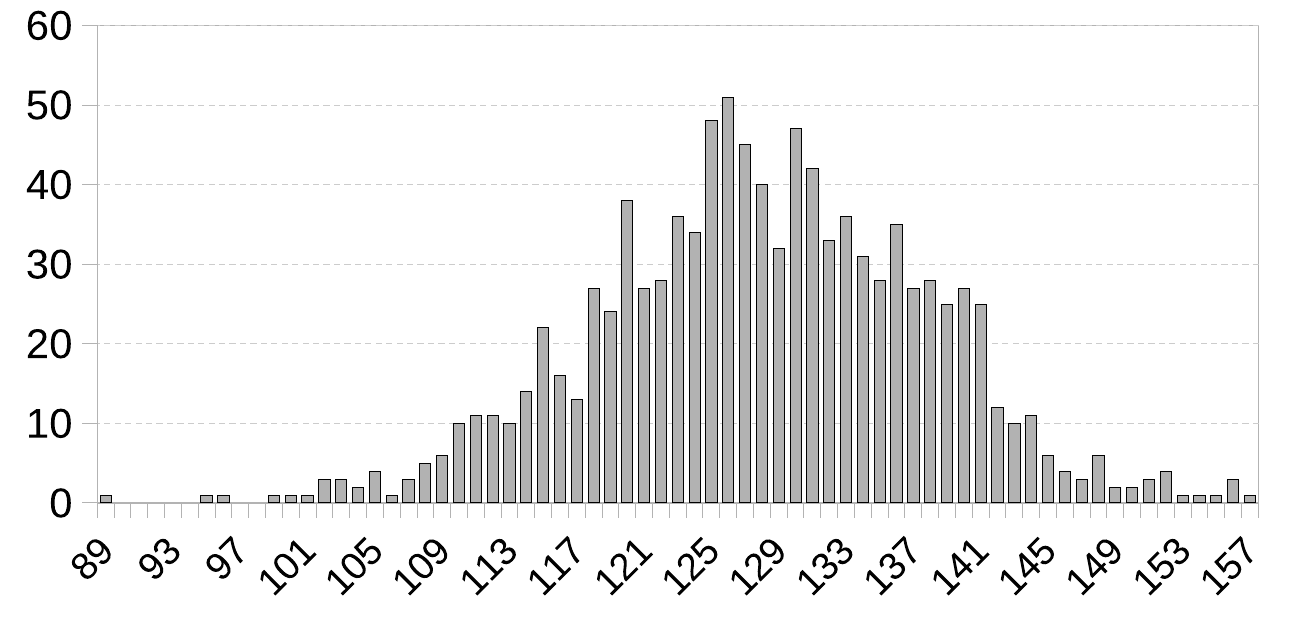}{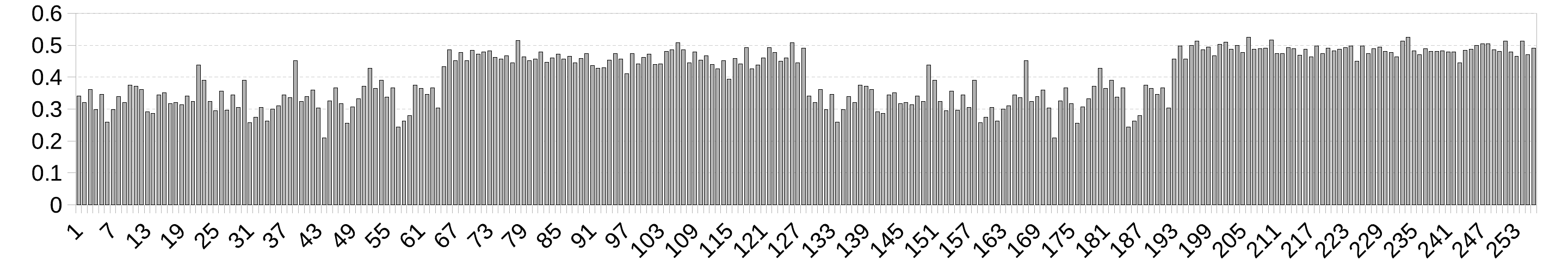}

\textbf{Misses, predicted:} Average = 102.848, Min = 70, Max = 133.

\textbf{Misses, random:} Average = 127.918, Min = 89, Max = 157.

...on yet another hand, $\theta$ is probably the principal obstacle, because \textit{3-round} version with only $\theta$
seemed to be uninvertible by our NNs even for \textit{32-bit} messages.

{\ }

\resultitem
Leaving just $\chi$ --- the only non-linear mapping (\cite[2.3.1]{bertoni2011}) --- produces a ``split''.

\textbf{CHF:} Keccak-1600-256 with only $\chi$, 6 rounds, 64-bit message. Mask: full.

\textbf{NN:} 2 layers, 1024 cells each. 1024 samples, 64 in batch.

\traintestcharts{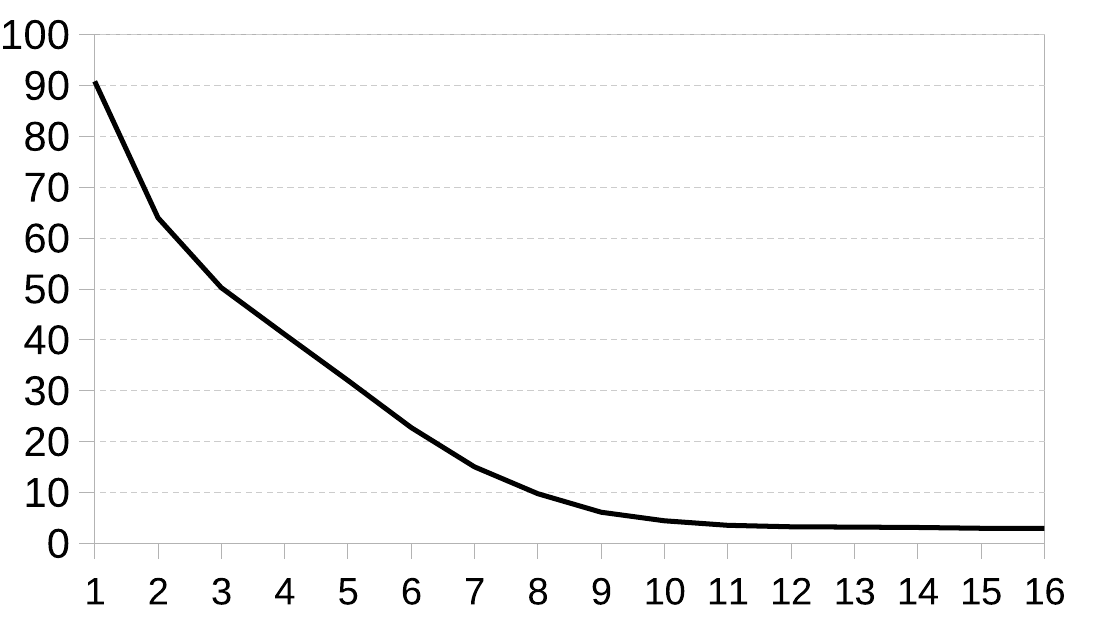}{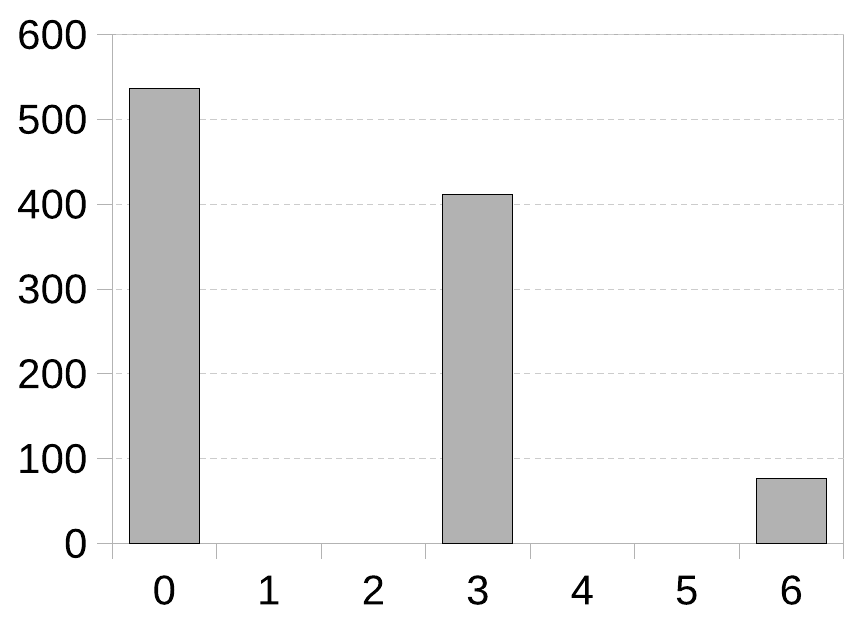}{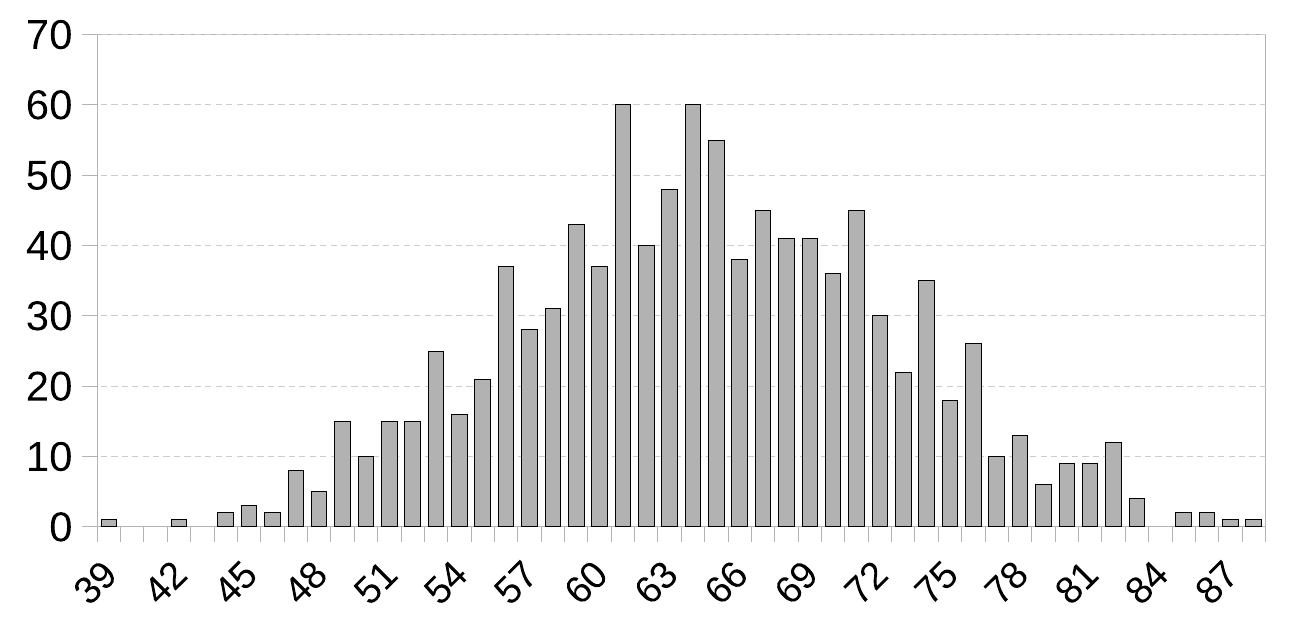}{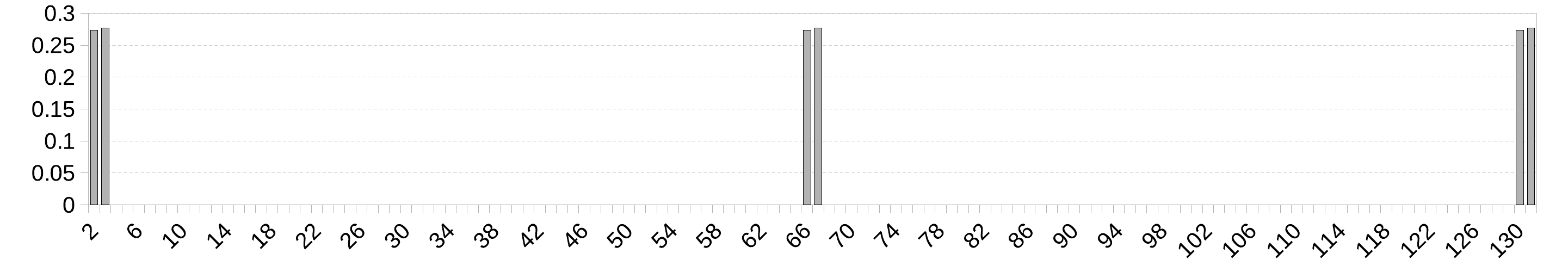}

\textbf{Misses, predicted:} Average = 1.652, Min = 0, Max = 6.

\textbf{Misses, random:} Average = 64.468, Min = 39, Max = 88.

There are 6 bits, 2 in each of 3 groups, about which NN is ``uncertain'' and therefore exactly inverts only half of the test hashes.

{\ }

\resultitem
Finally, the attempt to invert two bytes of 1-round Keccak-$f[200]$ permutation of the 200-bit \textit{state}
(NN gets the state after the permutation and tries to restore the state before it,
absorbing and squeezing phases are omitted); see \cite[2.4]{bertoni2011.csf}, \cite[1.2]{bertoni2011}.

\textbf{CHF:} Keccak-200 state permutation, 1 round, 200-bit message=state. Mask: bits 0--15.

\textbf{NN:} 3 layers, 512 cells each. 4096 samples, 256 in batch.

\traintestcharts{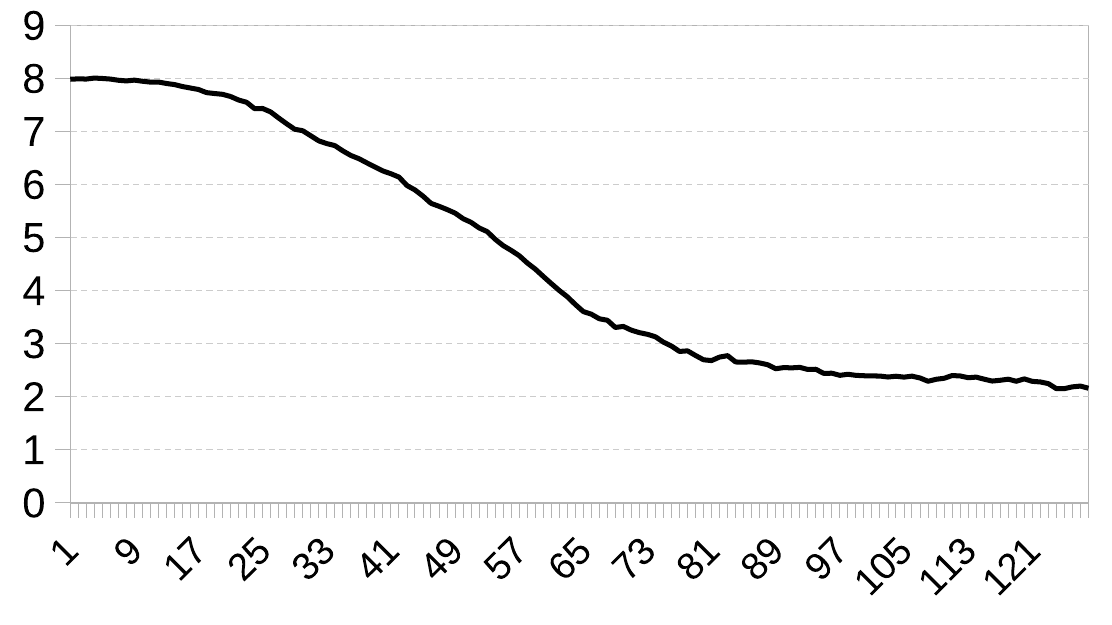}{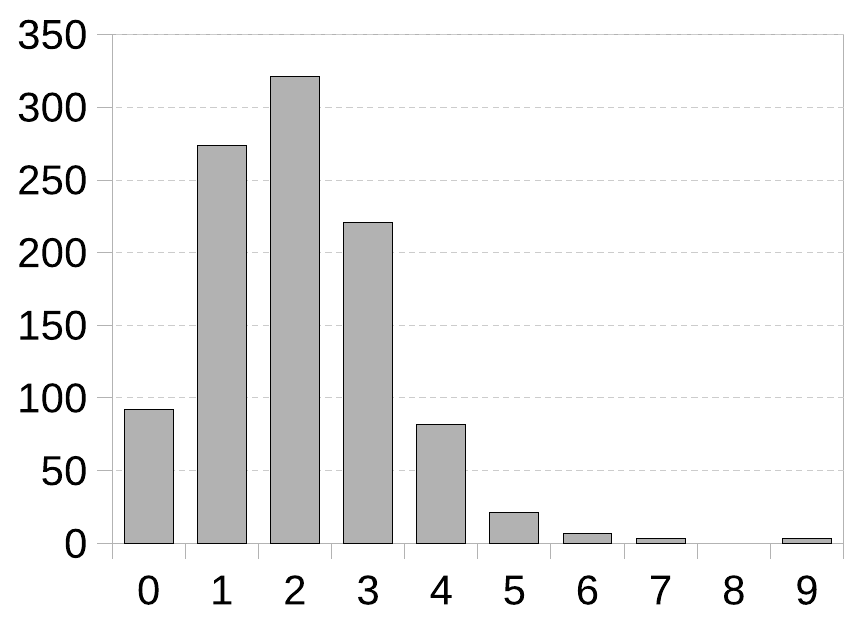}{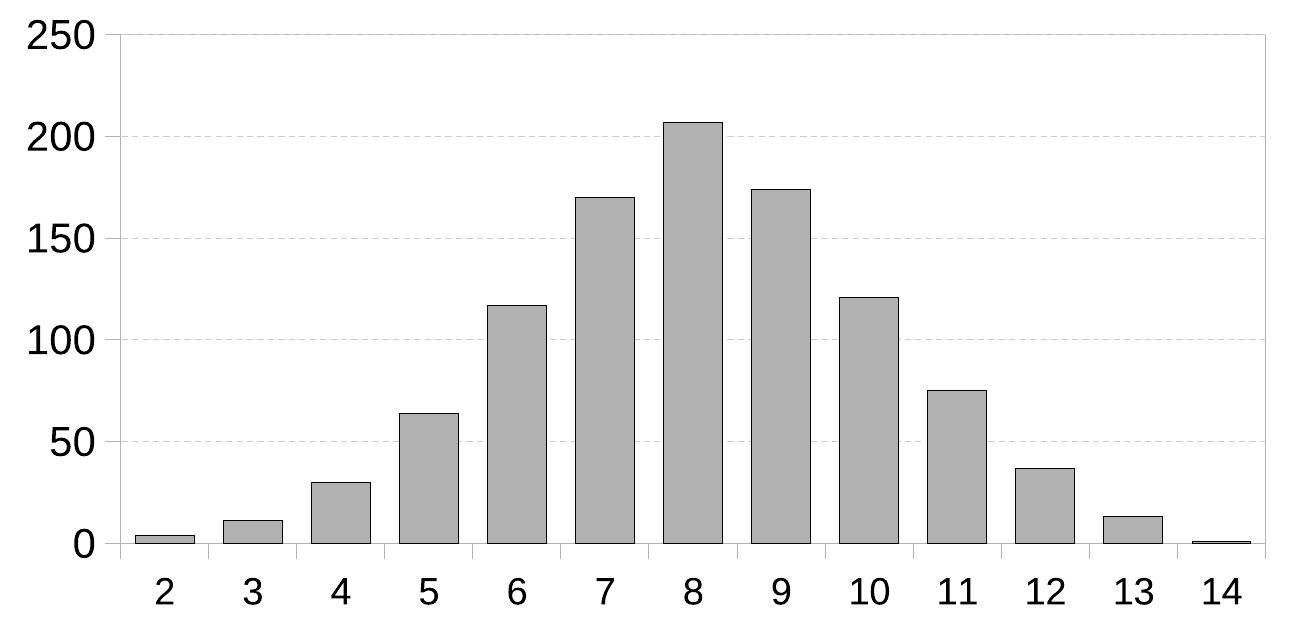}{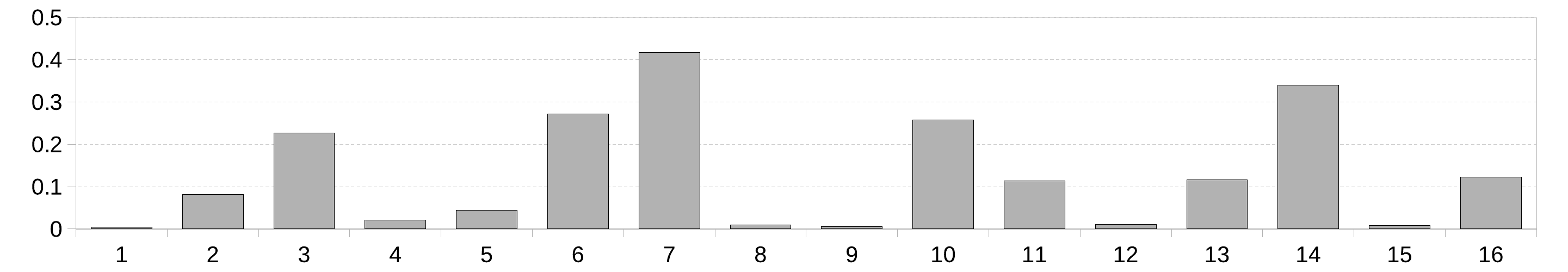}

\textbf{Misses, predicted:} Average = 2.053, Min = 0, Max = 9.

\textbf{Misses, random:} Average = 8.063, Min = 2, Max = 14.

There's the familiar initial delay of the learning,
and for wider masks the difficulty increases; in particular, when the mask is full (0--199),
the results of this NN's training look like Res. \ref{resKec256nolearn}.

The mask bits can be non-adjacent: for the mask $\{ 17+11i \mid i =\ovline{0,15} \}$
the average number of misses after 128 epochs was $\approx 1.3$, and 164 times out of 1024 all 16 bits coincided.

\newpage

\section*{Ancillary files}
\addcontentsline{toc}{section}{Ancillary files}

\forceparindent
The ancillary files to this paper contain the Python implementation of the fuzbit-string CHFs calculation,
NumPy- and backend-based, and the NN training to invert few rounds of these CHFs,
so that the results presented here can be reproduced, verified, and extended easily if needed.

{\ }

\cenlin{\small \texttt{\begin{tabular}{|p{4.5cm}|p{1cm}|p{3.5cm}|p{3.5cm}|}
\hline
FILE & SIZE & SHA2-256 & SHA3-256\\
\hline
\hline
bytehash.py
& 13014
& b68d 3d60 1da4 54b7 a4e9 aaea 39b2 1720 7e45 ed90 b1b5 1177 ee60 4a88 3bef aa25
& 47a2 76dd 722c bde5 4d95 e7de 49c3 57c2 7594 b27b 0655 1780 3c5c f55f 3be2 67cf\\
\hline
fuzzybitsconv.py
& 1085
& 4ad1 b87f 3bb3 1c07 04f3 f2ab 1dcc 06ce 8b9c f834 3204 b33d 33fd 990f 2295 91b4
& e99f dff2 d00f 8196 8d1d f6ef 7824 f5bc ca15 a72f 20d3 ecd3 3db7 8ef3 0d88 f066\\
\hline
fuzzyops\_numpy.py
& 2436
& f20c 98a6 3779 3d18 343a ed40 4cae d16b e9f7 0098 d555 26c2 168e 9c4c 92a9 c2fa
& d3e6 c403 43d4 65d7 c6c2 9bdb 86e0 44db ca5b 84f8 8afa 280a fd21 4288 bf53 f8d5\\
\hline
fuzzyops\_backend.py
& 2484
& b902 25d0 ea38 6954 85fe aa8d 292b 7e7a 2902 2742 dde0 70ad 0490 f55c 773d e623
& b97a 8403 3e5e 2f20 528b 2381 9582 f010 78ef 0cf9 f2a2 642d 8a23 73d9 940e 1614\\
\hline
fuzzyops\_circ\_numpy.py
& 993
& 9e31 1a28 bda4 ab6c 9510 c823 1d46 ab5f cc0d 7377 4ee6 2d02 01d9 f6a3 63ac 3632
& 096f 99c4 9650 a336 8092 a517 cae5 1924 1732 15ef 272d 422c c509 07f8 0740 3b9e\\
\hline
fuzzyops\_circ\_backend.py
& 1156
& c556 d418 f693 c13a 4fe9 cd6b e844 5bd2 4ccc 174c c959 e073 829f 7a80 72f6 dc21
& 0422 c066 5be0 a3d3 c701 641e 76e7 3c79 8ced 3f1b c4ed eafa 7174 0186 eba0 1c8c\\
\hline
fuzzyhash\_numpy.py
& 14645
& 3f6a b4be 17e2 84af 3bdd 8f43 518f 40a8 74d1 4aec 1fab 2ff0 779c c79e 77d2 c549
& f1be f28c f4af 9ee3 16bf db38 cd90 4a98 148c c761 bbb7 750f 12f4 15dc 91fb b4c2\\
\hline
fuzzyhash\_backend.py
& 15372
& 84b9 da7b c5b3 e0c1 f690 1efb 6c2b 1eae 9cb6 456b 9d35 d327 17d0 5b13 c616 3982
& fc7c 1f12 6e95 ac6a 427c 880d 5a3e 838d ebb0 cf53 e514 b514 571f 00b9 2ef3 20c2\\
\hline
learnfuzzyhashinversion.py
& 10758
& b049 fb12 56b1 9bd2 0730 d261 497a d587 d1ae 2f6e be7d 7140 e01a 16db 07f0 c9c0
& ee6b f77e 420a 7580 2491 6eaf fcdf ca5b 4e65 0fbd b552 8ee6 1628 9eeb 2434 4275\\
\hline
readme.txt
& 1419
& d727 9406 b436 1eb1 8baf f843 5649 c928 c715 4573 c2c6 052e 5900 d30c eec1 0efc
& 07d5 a664 8f71 c5bd 953e 9ebe 445f 7b04 52b6 6c1c feeb 6816 2132 9491 ee6c c84a\\
\hline
\end{tabular}}}

\end{document}